\documentclass{amsart}

\usepackage{amsmath}
\usepackage{amsfonts}
\usepackage{amssymb}
\usepackage{stmaryrd}
\usepackage{diagrams}
\usepackage{verbatim}
\usepackage{epic}

\usepackage{times}
\usepackage{cmll}
\usepackage{pst-tree}
\usepackage[dvips]{epsfig}

\usepackage[capbesideposition=inside,facing=yes,capbesidesep=quad]{floatrow}

\newtheorem{thm}{Theorem}
\newtheorem{prop}{Proposition}
\newtheorem{lem}{Lemma}
\newtheorem{cor}{Corollary}
\newtheorem{example}{Example}

\newtheorem{remark}{Remark}

\newtheorem{fact}{Fact}

\newtheorem{defin}{Definition}

\newtheorem{notats}{Notations}{\bfseries}{\itshape}

\newenvironment{minilist}{\begin{list}{$\bullet$}{
  \setlength{\parsep}{0pt}
  \setlength{\topsep}{0pt}
  \setlength{\itemsep}{-\parsep}
  \setlength{\labelsep}{0.4em}
  \setlength{\leftmargin}{1.3em}}}{\end{list}}

\newcommand{\LL}{\textnormal{\textsf{LL}}}

\newcommand{\cosize}[1]{\textit{cosize}(#1)}

\newcommand{\D}{D}

\newcommand{\sem}[1]{\llbracket #1\rrbracket}
\newcommand{\Nat}{\ensuremath{\mathbb{N}}}

\psset{treemode=U,levelsep=5mm}

\def\restriction#1#2{\mathchoice
              {\setbox1\hbox{${\displaystyle #1}_{\scriptstyle #2}$}
              \restrictionaux{#1}{#2}}
              {\setbox1\hbox{${\textstyle #1}_{\scriptstyle #2}$}
              \restrictionaux{#1}{#2}}
              {\setbox1\hbox{${\scriptstyle #1}_{\scriptscriptstyle #2}$}
              \restrictionaux{#1}{#2}}
              {\setbox1\hbox{${\scriptscriptstyle #1}_{\scriptscriptstyle #2}$}
              \restrictionaux{#1}{#2}}}
\def\restrictionaux#1#2{{#1\,\smash{\vrule height .8\ht1 depth .85\dp1}}_{\,#2}} 

\newcommand{\dom}[1]{\textsf{dom}(#1)}
\newcommand{\codom}[1]{\textsf{codom}(#1)}
\newcommand{\im}[1]{\textsf{im}(#1)}
\newcommand{\integer}[1]{\ulcorner #1 \urcorner}
\newcommand{\setofpairs}[1]{\mathfrak{P}_2(#1)}
\newcommand{\setfparts}[1]{\mathfrak{P}_\textsf{fin}(#1)}
\newcommand{\multi}[1]{[#1]}
\newcommand{\setfmulti}[1]{\mathfrak{M}_\textsf{fin}(#1)}
\newcommand{\emptysequence}{\varepsilon}

\newcommand{\Ports}{\textbf{Ports}}

\newcommand{\typesoflinks}{\mathcal{T}}
\newcommand{\tens}{\otimes}
\newcommand{\one}{1}
\newcommand{\bottom}{\bot}
\newcommand{\cod}{\oc}
\newcommand{\contr}{\wn}

\newcommand{\Links}{\textbf{Cells}}
\newcommand{\type}[1]{\textsf{t}(#1)}
\newcommand{\arity}[1]{\textsf{a}(#1)}
\newcommand{\links}[1]{\mathcal{C}(#1)}
\newcommand{\tenslinks}[1]{\mathcal{C}^\tens(#1)}
\newcommand{\parrlinks}[1]{\mathcal{C}^\parr(#1)}
\newcommand{\multlinks}[1]{\mathcal{C}^\textsf{m}(#1)}
\newcommand{\onelinks}[1]{\mathcal{C}^\one(#1)}
\newcommand{\botlinks}[1]{\mathcal{C}^\bottom(#1)}
\newcommand{\contrlinks}[1]{\mathcal{C}^?(#1)}
\newcommand{\LinksofPorts}[1]{\mathbb{C}(#1)}
\newcommand{\ports}[1]{\mathcal{P}(#1)}
\newcommand{\mapleftports}[1]{\textsf{P}^\textsf{left}(#1)}
\newcommand{\maprightports}[1]{\textsf{P}^\textsf{right}(#1)}
\newcommand{\mapauxports}[1]{\textsf{P}^\textsf{aux}(#1)}
\newcommand{\mappriports}[1]{\textsf{P}^\textsf{pri}(#1)}
\newcommand{\portsoflink}[1]{\textsf{P}_\textsf{c}(#1)}
\newcommand{\auxports}[1]{\mathcal{P}^\textsf{aux}(#1)}
\newcommand{\priports}[1]{\mathcal{P}^\textsf{pri}(#1)}
\newcommand{\paxnumber}[1]{\#(#1)}

\newcommand{\weakenings}[1]{\mathcal{C}^{?\textsf{w}}(#1)}
\newcommand{\derelictions}[1]{\mathcal{C}^{?\textsf{d}}(#1)}
\newcommand{\contractionspax}[1]{\mathcal{C}^{?\textsf{c}_{\textsf{auxd}}}(#1)}
\newcommand{\contractions}[1]{\mathcal{C}^{?\textsf{c}_{\textsf{b}}}(#1)}
\newcommand{\bangs}[1]{\mathcal{C}^!(#1)}
\newcommand{\auxdoors}[1]{\textsf{Auxdoors}(#1)}
\newcommand{\linksofports}[2]{\mathcal{C}(#1)(#2)}

\newcommand{\PPLPS}{\textbf{PPLPS}}
\newcommand{\PortsofPPLPS}[1]{\mathbb{P}(#1)}
\newcommand{\conclusions}[1]{\mathcal{P}^{\textsf{f}}(#1)}
\newcommand{\edges}[1]{\mathcal{W}(#1)}
\newcommand{\terminallinks}[1]{\mathcal{C}^\textsf{t}(#1)}
\newcommand{\axioms}[1]{\textsf{Ax}(#1)}
\newcommand{\terminalaxioms}[1]{\textsf{Ax}^\textsf{t}(#1)}
\newcommand{\isolatedaxioms}[1]{\textsf{Ax}^\textsf{i}(#1)}
\newcommand{\below}[1]{\textsf{b}_{#1}}
\newcommand{\PPLPSind}{\textbf{PPLPS}_\textsf{ind}}

\newcommand{\PLPS}{\textbf{PLPS}}
\newcommand{\conclusionunder}[2]{\textsf{c}(#1)(#2)}
\newcommand{\depth}[2]{\textit{depth}(#1)(#2)}
\newcommand{\emptyPLPS}{\emptyset\textbf{-}\PLPS}
\newcommand{\axPLPS}{\textsf{ax}\textbf{-}\PLPS}
\newcommand{\multPLPS}{\textsf{mult}\textbf{-}\PLPS}
\newcommand{\unitPLPS}{\textsf{unit}\textbf{-}\PLPS}
\newcommand{\weakPLPS}{?_{\textsf{w}}\textbf{-}\PLPS}
\newcommand{\derPLPS}{?_{\textsf{d}}\textbf{-}\PLPS}
\newcommand{\contrPLPS}{?_{\textsf{$c_{\textsf{b}}$}}\textbf{-}\PLPS}
\newcommand{\cboxPLPS}{\textbf{?-box-}\PLPS}

\newcommand{\contrunitPLPS}{\textbf{?unit-}\PLPS}
\newcommand{\bangunitPLPS}{\textbf{!unit-}\PLPS}

\newcommand{\LPS}{\textbf{LPS}}

\newcommand{\PS}{\textbf{PS}}
\newcommand{\LPSofPPS}[1]{\textsf{LPS}(#1)}

\newcommand{\mapbangauxd}[1]{\textsf{b}(#1)}
\newcommand{\portsofbox}[1]{\textsf{B}(#1)}
\newcommand{\boxofbang}[1]{\overline{\textsf{B}}(#1)}

\newcommand{\pInj}{\textbf{pInj}}

\newcommand{\enleverunecouche}[1]{\overline{#1}}
\newcommand{\PSind}{\textbf{PS}_\textsf{ind}}

\newcommand{\enleverunecoucheunitsweakenings}[2]{{#1}_{[#2]}}

\newcommand{\mes}[1]{\textit{mes}(#1)}

\newcommand{\PLPSind}{\textbf{PLPS}_\textsf{ind}}

\newcommand{\LPSind}{\textbf{LPS}_\textsf{ind}}

\newcommand{\dig}{\textsf{dig}}

\newcommand{\groupoidpM}{\textbf{pM}}

\newcommand{\groupoidD}{\textbf{D}}
\newcommand{\groupoidsD}{\textbf{sD}}
\newcommand{\groupoidsDM}{\textbf{sDM}}
\newcommand{\groupoidpD}{\textbf{pD}}
\newcommand{\groupoidM}{\textbf{M}}

\newcommand{\groupoidsM}{\textbf{sM}}
\newcommand{\groupoidpsM}{\textbf{psM}}
\newcommand{\groupoidppsM}{\textbf{ppsM}}
\newcommand{\groupoidBij}{\textbf{Bij}}

\newcommand{\bridge}{\textit{B}}

\newcommand{\sm}[1]{\llbracket #1 \rrbracket}

\newcommand{\omegaPPLPSind}{\mathbf{\omega}\textbf{PPLPS}_\textsf{ind}}

\title{The relational model is injective for Multiplicative Exponential Linear Logic (without weakenings)}

\author{Daniel de Carvalho}
\address{Daniel de Carvalho\\ Laboratoire d'Informatique de Paris-Nord - Universit\'e Paris~13}
\author{Lorenzo Tortora de Falco}
\address{Lorenzo Tortora de Falco\\ Dipartimento di Filosofia - Roma~III}

\begin{document}

\begin{abstract}
We show that for Multiplicative Exponential Linear Logic (without weakenings) the syntactical equivalence relation on proofs induced by cut-elimination coincides with the semantic equivalence relation on proofs induced by the multiset based relational model: one says that the interpretation in the model (or the semantics) is injective. We actually prove a stronger result: two cut-free proofs of the full multiplicative and exponential fragment of linear logic whose interpretations coincide in the multiset based relational model are the same ``up to the connections between the doors of exponential boxes''. 
\end{abstract}

\maketitle

\section{Introduction}

Separation is an important mathematical property, and several theorems are often referred to as ``separation theorems''. In theoretical computer science, one of the most well-known examples of separation theorem is B\"ohm's theorem (\cite{Bohmtheorem}) for pure $\lambda$-calculus: if $t,t'$ are two distinct closed $\beta\eta$-normal terms, then there exists a context $C[\ ]$ s.t.\ $C[t]\simeq_\beta 0$ and $C[t']\simeq_\beta 1$. Another way of stating the theorem is to say that it is possible to define an order relation (i.e.\ a $T_0$ topology) on the $\beta\eta$-equivalence classes of (normalizable) $\lambda$-terms. Later on, this kind of question has been studied by Friedman and Statman for the simply typed $\lambda$-calculus (\cite{Stat?}), leading to what is often called ``typed B\"ohm's theorem'' (see also~\cite{phdjoly},~\cite{dosenpietric} for sharper formulations). We believe that if no other result of this kind has been produced for a long time, it is due to the absence of interesting logical systems where proofs could be represented in a nice ``canonical'' way. 

The situation radically changed in the nineties, mainly due to Linear Logic (LL~\cite{ll}), a refinement of intuitionistic (and classical) logic characterized by the introduction of new connectives (the exponentials) which give a \emph{logical}  status to the operations of erasing and copying (corresponding to the \emph{structural rules} of logic): this change of viewpoint had striking consequences in proof-theory, like the introduction of proof-nets, a geometric way of representing computations. In the framework of proof-nets, the separation property can be studied: the first work on the subject is~\cite{llbohm} where the authors deal with the translation in LL of the pure $\lambda$-calculus; it is a key property of ludics (\cite{locus}) and has been studied more recently for the intuitionistic multiplicative fragment of LL (\cite{Matsuoka07}) and for differential nets (\cite{MazPag07lpar}). For Parigot's $\lambda\mu$-calculus, see~\cite{lmbohm} and~\cite{saurin05}.

Still in LL's framework, a semantic approach to the question of separation is developped in~\cite{phdtortora} and~\cite{injectcoh}, where the (very natural) question of ``injectivity'' of the semantics is adressed: do the equivalence relation on proofs defined by the cut-elimination procedure and the one defined by a given denotational model (sometimes/always) coincide? When the answer is positive one says that the model is \emph{injective} (it separates syntactically different proofs). Indeed, two proofs are ``syntactically'' equivalent when (roughly speaking) they have the same cut-free form (in a confluent and weakly normalizing system), and they are ``semantically'' equivalent in a given denotational model  (a semantics of proofs in logical terms) when they have the same \emph{interpretation}. It is worth noticing that the study of both these equivalence relations is at the heart of the whole research area between proof-theory and theoretical computer science: cut-elimination is a crucial property of logical systems since Gentzen (with a renewal of interest in this property after the discovery of the Curry-Howard correspondence: a proof is a program whose execution corresponds to applying the cut-elimination procedure to the proof) and the general goal of denotational semantics is to give a ``mathematical'' counterpart to syntactical devices such as proofs and programs, bringing to the fore their essential properties. 
The basic pattern is to associate with every formula/type an object of some category and with every proof/program a morphism of this category (its interpretation).

The works~\cite{phdtortora} and~\cite{injectcoh} give partial results and counterexamples to the question of injectivity, mainly for the (multiset based) coherent model: in particular the counterexamples show that this model is not injective for multiplicative and exponential LL ($MELL$). Also, it was conjectured that the (multiset based) relational model is injective for $MELL$, but despite many efforts (\cite{phdtortora},~\cite{injectcoh},~\cite{Boudesunifying},~\cite{Pag07mscs},~\cite{MazPag07lpar},~\cite{PagTas09lics}...) all the attempts to prove the conjecture failed up to now: no real progress has been done since~\cite{injectcoh}, where a proof of injectivity of the relational model is given for a fragment of $MELL$\footnote{Precisely, for the $(?\wp)$LL fragment given by 
$A::= \: X \: | \: ?A \wp A \: | \: A \wp ?A \: | \: A \wp A \: | \: A \tens A \:| \: !A $
.}. Game semantics is much closer to syntax than relational and coherent semantics, and positive answers have been obtained for little fragments like the multiplicative fragment $MLL$ or the fragment corresponding to the $\lambda$-calculus (\cite{ajm},\cite{ho}), but also for the polarized fragment of LL (\cite{synsempol}).

We prove here that for $MELL$ without weakenings (and without the multiplicative unit $\bot$) relational semantics is injective (Corollary~\ref{cor:injectNoW}). This tremendous improvement w.r.t.\ the previous situation is an immediate consequence of a much stronger result: in the full $MELL$ fragment (with units) two proof-nets $R$ and $R'$ with the same interpretation are the same ``up to the connections between the doors of exponential boxes'' (we say they have the same LPS: Theorem~\ref{theorem:injectLPS} and Corollary~\ref{cor:injectLPS}). This result can be expressed in terms of differential nets (\cite{dif}): two cut-free proof-nets with different LPS have different Taylor expansions. We also believe this work is an essential step towards the proof of the full conjecture.

In the style of~\cite{lics06} and~\cite{carvalhopaganitortora08} we work in an untyped framework; we do not define (proof-)nets nor cut-elimination but only cut-free proof-structures (PS, Definition~\ref{def:PS}): we prove that two PS with the same interpretation have the same LPS (Corollary~\ref{cor:injectLPS}). A (proof-)net (as defined in~\cite{carvalhopaganitortora08}) is a particular case of PS so that the result holds for untyped (so as for typed) $MELL$ (proof-)nets (Remark~\ref{rem:InjectStandardMELL}). Since we want to prove that two PS are isomorphic in Theorem~\ref{theorem:injectLPS}, it is mandatory to have a (simple and clear) notion of isomorphism between PS (Definition~\ref{def:isoPPLPSind})\footnote{We actually use in our theorem an even subtler notion: the one of isomorphism between $k$-experiments of indexed LPS (Definition~\ref{def:isoExpInd}).}, and this is why in Section~\ref{sect:syntax} we give a very sharp description of the syntax in the style of interaction nets (\cite{pnin},~\cite{phdmazza}): we cannot only rely 
on a graphic intuition. The notion of Linear Proof-Structure (LPS), which comes from~\cite{injectcoh}, is our main syntactical tool: with every (proof-)net $R$ of (say)~\cite{carvalhopaganitortora08} is associated a LPS, which is obtained from $R$ by forgetting some informations about $R$'s exponential boxes, namely which auxiliary doors correspond to which $!$-link (using standard LL's terminology); this is particularly clear in Definition~\ref{def:PS} of PS: a PS is a LPS and a function allowing to recover boxes. Recovering this function from the interpretation of a PS is the only missing point in the proof of the full conjecture, but a simple remark shows that the function can be recovered from the LPS when the PS is a connected graph: this yields injectivity for $MELL$ without weakenings and $\bot$ (Corollary~\ref{cor:injectNoW}). 
In Section~\ref{sect:experiments}, we introduce a domain $D$ to interpret PS which is exactly the one already defined in~\cite{carvalhopaganitortora08}. Like in~\cite{injectcoh}, we use here experiments (introduced in~\cite{ll}) which can be thought as objects in between syntax and semantics and are related to type derivations in the $\lambda$-calculus (\cite{Carvalho:ExecutionTime}). Experiments are functions defined on (proof-)nets allowing to compute the interpretation pointwise: the set of \emph{results} of all the experiments of a given (proof-)net is its interpretation\footnote{The result of an experiment $e$ is the image of the conclusions of the (proof-)net through the function $e$; so that contrary to an experiment its result is a truly semantic object.}. Usually an experiment $e$ of a (proof-)net $R$ is a labeling of $R$ at depth $0$ and a function associating with every $!$-link $l$ of $R$ a set of experiments of the content of the box associated with $l$. We noticed that a particular kind of experiment called \emph{$k$-experiment} (Definition~\ref{def:experiment}) can be defined directly on LPS (boxes are not needed). We conclude Section~\ref{sect:experiments} by stating our results and reducing the problem of injectivity to Proposition~\ref{prop : KeyProposition}, which is proven in Section~\ref{sect:InjectivityLPS}. The paper ends with a technical appendix, containing some obvious definitions and the formal details of some constructions previously used.

\begin{figure}
\fcapside{\caption{\textbf{Example of PS.} In the standard syntax of~\cite{carvalhopaganitortora08} we have a box with a unique auxiliary door represented by the port $p_2$ (the dashed arrow allows to determine the doors of the box) and a dereliction link (the port $p_1$); the conclusions of the auxiliary door and the dereliction are then contracted.}\label{fig:NoUniqueExp} 
}
{\includegraphics[width=6cm,height=3cm]{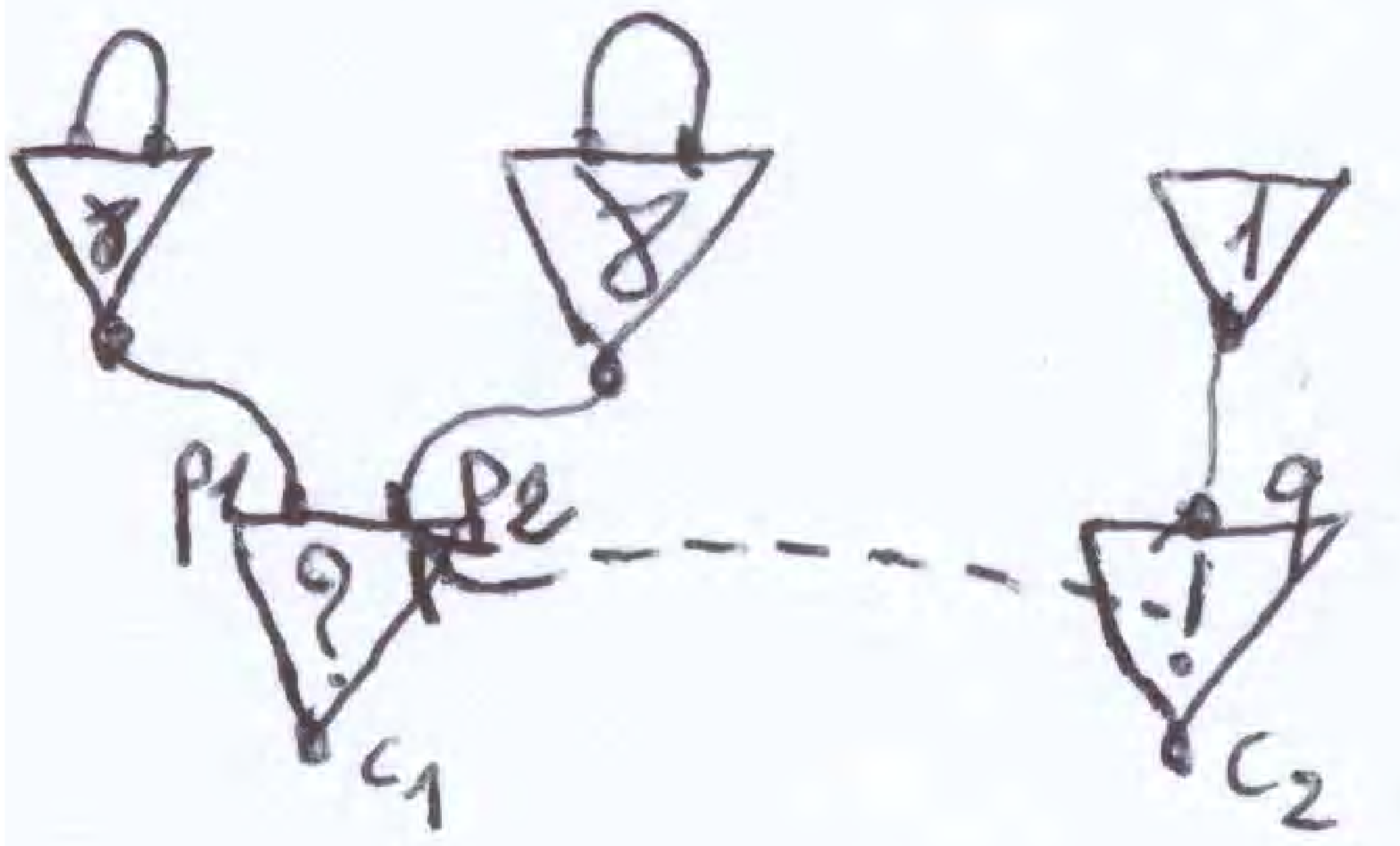}}
\end{figure}

In~\cite{injectcoh}, a single (well-chosen!)  point of the interpretation of a proof-net allowed to ``rebuild'' the entire proof-net (in some particular cases and for coherent semantics). Something similar happens in this paper, with a notable difference that makes everything much more complicated: in~\cite{injectcoh} the well-chosen point of the interpretation of a proof-net allowed not only to rebuild the proof-net but also the experiment having this point as result. This is not the case here, where the well-chosen points of the interpretation of a PS are atomic injective $k$-points (Definition~\ref{def:injectivek-point}): we show using Figure~\ref{fig:NoUniqueExp} that there exist different experiments having as result the same atomic injective $k$-point. 
We can define two experiments $e_1$ and $e_2$ of the PS $R$ represented in Figure~\ref{fig:NoUniqueExp} in such a way that $e_1(p_1)= \multi{\zeta_1}$, $e_2(p_1)= \multi{\zeta_2}$, $e_1(p_2)= \multi{\zeta_2, \zeta_3, \zeta_4}$ and $e_2(p_2)=\multi{\zeta_1, \zeta_3, \zeta_4}$, where $\zeta_j = (-, \gamma_j, \gamma_j)$ and the $\gamma_j$ are distinct atoms. The two (different) experiments have the same result, which is an atomic and injective $3$-point. Let us conclude by mentioning the main novelties in our proof:
\begin{itemize}
 \item 
the use of injective experiments in a completely different sense than in~\cite{injectcoh}: intuitively, our injective $k$-experiments associate with an axiom link with depth $d$, $k^d$ different labels, while the injective $k$-obsessional experiments of~\cite{injectcoh} associate a unique label with such an axiom link (see Remark~\ref{rem:OldNewInjExp}). A crucial aspect of our new injective $k$-experiments is that they can be recognized by their results (Definition~\ref{def:injectivek-point}), and this was not the case for \emph{relational} injective $k$-obsessional experiments
\item
the use of an equivalence relation on experiments: the idea is that the two experiments of the PS of Figure~\ref{fig:NoUniqueExp} previously defined are ``the same'' experiment, and we should not try to distinguish them (or choose one of the two). Indeed $e_1$ and $e_2$ are the same ``up to the labels of the axiom links'': a precise definition of this equivalence is given and is a key ingredient in the proof of Proposition~\ref{prop : KeyProposition}.
\end{itemize}
Summing up, we show that if the interpretation of the PS $R$ contains an atomic injective $k$-point, then every $R'$ with the same interpretation as $R$ has the same LPS as $R$ (Corollary~\ref{cor:injectLPS}); and contrary to~\cite{injectcoh} we do not know the experiment which produced this point.

\bigskip

\textbf{Conventions. }
We use the notation $[ \: ]$ for multisets while the notation $\{ \: \}$ is, as usual, for sets. For any set $A$, we denote by $\setfmulti{A}$ the set of finite multisets $a$ whose support, denoted by $ \textit{Supp}(a) $, is a subset of $ A $. 
The pairwise union of multisets given by term-by-term addition of multiplicities is denoted by a $+$ sign and, following this notation, the generalized union is denoted by a $\sum$ sign. The neutral element for this operation, the empty multiset, is denoted by $\multi{ \: }$. For $k \in \mathbb{N}$ and $a$ multiset, we denote by $k\cdot a$ the multiset defined by $\textit{Supp}(k\cdot a)=\textit{Supp}(a)$ and for every $\alpha\in\textit{Supp}(a)$, $(k\cdot a)(\alpha)=k a(\alpha)$.

For any $k \in \mathbb{N}$, we set $\integer{k} = \{ 1, \ldots, k \}$. For any set $A$, we denote by $\mathfrak{P}(A)$ the powerset of $A$ and by $\setofpairs{A}$ the set 
$\{ \{ a, b \} \in \mathfrak{P}(A) \: / \: a,b \in A \textrm{ and } a \not=b \} $. A function $f:A\to B$ has domain $A=\dom{f}$, codomain $B=\codom{f}$, image $\im{f}=\{f(a)/\ a\in A\}$; 
we denote by $_{A'|}{f}_{|B'}$ the restriction of $f$ to the domain $A'$ and to the codomain $B'$ and by 
$\mathfrak{P}(f):\mathfrak{P}(A)\to\mathfrak{P}(B)$ the function wich associates with $X\subseteq A$ the set $\{f(x)/\ x\in X\}$. We denote by $\emptysequence$ the unique element of ${\integer{k}}^0$ for any $k \in \mathbb{N}$ and by $A \uplus B$ the disjoint union of the sets $A$ and $B$. 

\bigskip

\section{Syntax}\label{sect:syntax}

\subsection{Cells and Ports}

We introduce cells and ports, which intuitively correspond to ``links with their premises and conclusions'' in the theory of linear logic proof-nets (\cite{ll}, \cite{hilbert}, \cite{phdtortora}, \ldots). Our presentation is in the style of interaction nets (\cite{pnin}, \cite{phdmazza}), where principal (resp.\ auxiliary) ports correspond to the conclusions (resp.\ the premises) of the links and axiom links of the usual syntax become wires (see Definition~\ref{def:axioms}). We deal with (the analogue of) unary $!$-links, while $?$-links can have an arbitrary number of premises. More precisely, we set $\typesoflinks = \{ \tens, \parr, \one, \bottom, \cod, \contr \}$ and we define $\Links$ and $\Ports$ as follows, where the function $\arity{\mathbb{C}}$ associates with a given cell $l$ its arity $\arity{\mathbb{C}}(l)$.

\begin{defin}
Let $\Links$ be the set of pairs $\mathbb{C}=(t, a)$ such that
\begin{minilist}
\item $t$ is a function such that $\dom{t}$ is finite and $\codom{t} = \typesoflinks$;
\item $a$ is a function $\dom{t} \rightarrow \mathbb{N}$ such that 
$a(l) = \left\lbrace \begin{array}{ll} 
2 & \textrm{if $t(l) \in \{ \tens, \parr \}$} \\
0 & \textrm{if $t(l) \in \{ \one, \bottom \}$} \\
1 & \textrm{if $t(l) = \cod$.}
\end{array} \right. $
\end{minilist}
We set $\type{\mathbb{C}} = t$ and $\arity{\mathbb{C}} = a$.
\end{defin}

\begin{notats}
For any $\mathbb{C} \in \Links$, we set $\links{\mathbb{C}} = \dom{\type{\mathbb{C}}}$, $\tenslinks{\mathbb{C}} = \{ l \in \links{\mathbb{C}} \: / \: \type{\mathbb{C}}(l) = \tens \}$, $\parrlinks{\mathbb{C}} = \{ l \in \links{\mathbb{C}} \: / \: \type{\mathbb{C}}(l) = \parr \}$, 
$\multlinks{\mathbb{C}} = \tenslinks{\mathbb{C}} \cup \parrlinks{\mathbb{C}}$, $\onelinks{\mathbb{C}} = \{ l \in \links{\mathbb{C}} \: / \: \type{\mathbb{C}}(l) = \one \}$, $\botlinks{\mathbb{C}} = \{ l \in \links{\mathbb{C}} \: / \: \type{\mathbb{C}}(l) = \bottom \}$, $\contrlinks{\mathbb{C}} = \{ l \in \links{\mathbb{C}} \: / \: \type{\mathbb{C}}(l) = \contr \}$ and $\bangs{\mathbb{C}} = \{ l \in \links{\mathbb{C}} \: / \: \type{\mathbb{C}}(l) = \cod \}$.
\end{notats}

\begin{defin}
For any $\mathbb{C}, \mathbb{C'} \in \Links$, we write $\varphi_0 : \mathbb{C} \simeq \mathbb{C'}$ if, and only if, 
$\varphi_0$ is a bijection from $\links{\mathbb{C}}$ to $\links{\mathbb{C'}}$ such that 
the following diagram commutes:
\begin{diagram}
\typesoflinks & \lTo^{\type{\mathbb{C}}} & \links{\mathbb{C}} & \rTo^{\arity{\mathbb{C}}} & \mathbb{N} \\
& \luTo_{\type{\mathbb{C'}}} & \dTo^{\varphi_0} & \ruTo_{\arity{\mathbb{C'}}} \\
& & \links{\mathbb{C'}}
\end{diagram}
\end{defin}

\begin{defin}\label{def:Ports}
Let $\Ports$ be the set of 6-tuples $\mathbb{P} =(\mathbb{C}, \mathcal{P}, \textsf{P}_\textsf{c}, \textsf{P}^\textsf{pri}, \textsf{P}^\textsf{left}, \#)$ such that
\begin{minilist}
\item $\mathbb{C} \in \Links$; the elements of $\links{\mathbb{C}}$ are \emph{the cells} of $\mathbb{P}$;
\item $\mathcal{P}$ is a finite set whose elements are \emph{the ports} of $\mathbb{P}$;
\item $\textsf{P}_\textsf{c}$ is a function $\links{\mathbb{C}} \rightarrow \mathfrak{P}(\mathcal{P})$ such that
\begin{minilist}
\item for any $l_1, l_2 \in \links{\mathbb{C}}$, we have $\textsf{P}_\textsf{c}(l_1) \cap \textsf{P}_\textsf{c}(l_2) \not= \emptyset \Rightarrow l_1 = l_2$;
\item and for any $l \in \links{\mathbb{C}}$, we have $\textit{Card}(\textsf{P}_\textsf{c}(l)) = \arity{\mathbb{C}}(l) + 1$;
\end{minilist}
the elements of $\textsf{P}_\textsf{c}(l)$ are the ports of $l$;
\item $\textsf{P}^\textsf{pri}$ is a function $\links{\mathbb{C}} \rightarrow \mathcal{P}$ such that for any $l \in \links{\mathbb{C}}$, we have $\textsf{P}^\textsf{pri}(l) \in \textsf{P}_\textsf{c}(l)$; $\textsf{P}^\textsf{pri}(l)$ is \emph{the principal port} of $l$. A port of $l$ different from $\textsf{P}^\textsf{pri}(l)$ is an \emph{auxiliary port};
\item $\textsf{P}^\textsf{left}$ is a function $\multlinks{\mathbb{C}} \rightarrow \mathcal{P}$ such that for any $l \in \multlinks{\mathbb{C}}$, we have $\textsf{P}^\textsf{left}(l) \in \textsf{P}_\textsf{c}(l) \setminus \{ \textsf{P}^\textsf{pri}(l) \}$. 
\item $\#$ is a function $\bigcup_{l \in \contrlinks{\mathbb{C}}} (\textsf{P}_\textsf{c}(l) \setminus \{ \textsf{P}^\textsf{pri}(l) \}) \rightarrow \mathbb{N}$.
\end{minilist}
We set $\ports{\mathbb{P}}=\mathcal{P}$, $\LinksofPorts{\mathbb{P}}=\mathbb{C}$, $\portsoflink{\mathbb{P}} = \textsf{P}_\textsf{c}$, $\mappriports{\mathbb{P}}=\textsf{P}^\textsf{pri}$, $\mapleftports{\mathbb{P}} = \textsf{P}^\textsf{left}$, $\paxnumber{\mathbb{P}} = \#$, 
$\links{\mathbb{P}}=\links{\LinksofPorts{\mathbb{P}}}$, $\tenslinks{\mathbb{P}} = \tenslinks{\LinksofPorts{\mathbb{P}}}$, $\parrlinks{\mathbb{P}}=\parrlinks{\LinksofPorts{\mathbb{P}}}$,
$\multlinks{\mathbb{P}}=\multlinks{\LinksofPorts{\mathbb{P}}}$, $\onelinks{\mathbb{P}}=\onelinks{\LinksofPorts{\mathbb{P}}}$, $\botlinks{\mathbb{P}}=\botlinks{\LinksofPorts{\mathbb{P}}}$, 
$\bangs{\mathbb{P}}=\bangs{\LinksofPorts{\mathbb{P}}}$ and $\contrlinks{\mathbb{P}}=\contrlinks{\LinksofPorts{\mathbb{P}}}$. We set $\type{\mathbb{P}} = \type{\LinksofPorts{\mathbb{P}}}$ and $\arity{\mathbb{P}} = \arity{\LinksofPorts{\mathbb{P}}}$.

For any $\mathcal{P}_0 \subseteq \ports{\mathbb{P}}$, we set $\linksofports{\mathbb{P}}{\mathcal{P}_0} = \{ l \in \links{\mathbb{P}} \: / \: (\exists p \in \mathcal{P}_0) \: p \in \portsoflink{\mathbb{P}}(l) \}$.
\end{defin}

\begin{remark}\label{rem:explainingPorts}
(i) Intuitively, $\mathbb{P}\in\Ports$ corresponds to what is called ``a set of links'' in the usual syntax of~\cite{injectcoh}. Notice that the functions $\textsf{P}^\textsf{pri}$ and $\textsf{P}^\textsf{left}$ of Definition~\ref{def:Ports} induce the function $\mapauxports{\mathbb{P}}:\links{\LinksofPorts{\mathbb{P}}} \rightarrow \mathfrak{P}(\ports{\mathbb{P}})$ defined by $\mapauxports{\mathbb{P}}(l) = \portsoflink{\mathbb{P}}(l) \setminus \{ \mappriports{\mathbb{P}}(l) \}$ and the function $\maprightports{\mathbb{P}}:\multlinks{\mathbb{P}} \rightarrow \ports{\mathbb{P}}$ defined by $\{\maprightports{\mathbb{P}}(l)\} = \mapauxports{\mathbb{P}}(l) \setminus \{ \mapleftports{\mathbb{P}}(l) \}$: $\mappriports{\mathbb{P}}$ and $\mapauxports{\mathbb{P}}$ allow to distinguish the principal ports (conclusions in~\cite{injectcoh}) from the auxiliary ports (premises in~\cite{injectcoh}), while for multiplicative cells the functions $\mapleftports{\mathbb{P}}$ and $\maprightports{\mathbb{P}}$ allow to distinguish the left auxiliary port (left premise in~\cite{injectcoh}) from the right one. We denote by $\priports{\mathbb{P}}$ (resp.\ $\auxports{\mathbb{P}}$) the set of principal (resp.\ auxiliary) ports of $\mathbb{P}$.

(ii) There is however a notable difference w.r.t.~\cite{pnin} in the way we handle boxes in our \PS{} (Definition~\ref{def:PS}): here the function $\#$ plays a crucial role. If $p\in\mapauxports{\mathbb{P}}(l)$ for some $l\in\contrlinks{\mathbb{P}}$, then the integer $\paxnumber{\mathbb{P}}(p)$ is in the syntax of~\cite{injectcoh} the number of auxiliary doors of boxes of the exponential branch corresponding to $p$. For instance, for the $\mathbb{P}$ in Figure \ref{fig:NoUniqueExp}, we have $\paxnumber{\mathbb{P}}(p_1) = 0$ and $\paxnumber{\mathbb{P}}(p_2) = 1$. In the spirit of \LL, we split the set $\contrlinks{\mathbb{P}}$ into the four following disjoint sets: 
\begin{minilist}
\item
$\weakenings{\mathbb{P}}=\{ l \in \contrlinks{\mathbb{P}} \: / \: \arity{\mathbb{P}}(l) = 0 \}$ which (in~\cite{injectcoh}) corresponds to the set of weakening links of $\mathbb{P}$
\item
$\derelictions{\mathbb{P}}=\{l \in \contrlinks{\mathbb{P}} \: / \: \arity{\mathbb{P}}(l)=1 \textrm{ and } \mathfrak{P}(\paxnumber{\mathbb{P}})(\mapauxports{\mathbb{P}(l)})= \{ 0 \}\}$, which (in~\cite{injectcoh}) corresponds to the set of dereliction links of $\mathbb{P}$
\item
$\contractions{\mathbb{P}}=\{ l \in \contrlinks{\mathbb{P}} \: / \: \arity{\mathbb{P}}(l) > 1 \textrm{ and } 
(\exists p \in\mapauxports{\mathbb{P}}(l)) \: \paxnumber{\mathbb{P}}(p) = 0 \}$, which (in~\cite{injectcoh}) corresponds to the set of contraction links of $\mathbb{P}$ having at least the conclusion of one dereliction link among their premises
 \item 
$\contractionspax{\mathbb{P}}=\{l\in\contrlinks{\mathbb{P}} \: / \: \arity{\mathbb{P}}(l) \geq 1\textrm{ and }(\forall p \in \mapauxports{\mathbb{P}}(l))\: \paxnumber{\mathbb{P}}(p)>0\}$, which (in~\cite{injectcoh}) corresponds to the set of contraction links having only conclusions of auxiliary doors of boxes among their premises.
\end{minilist}
The \emph{auxiliary ports of the $?$-cells} of $\mathbb{P}$ are the ports belonging to the set $\textsf{Aux}^?(\mathbb{P}) = \bigcup_{l \in \contrlinks{\mathbb{P}}} \mapauxports{\mathbb{P}}(l)$, while the \emph{auxiliary doors} of $\mathbb{P}$ are the elements of $\auxdoors{\mathbb{P}} = \{ p \in \textsf{Aux}^?(\mathbb{P}) \: / \: \paxnumber{\mathbb{P}}(p)>0\}$.
\end{remark}

\begin{defin}\label{def:isoPorts}
Let $\mathbb{P}, \mathbb{P'} \in \Ports$ and let $\varphi$ be a pair $(\varphi_\mathcal{C}, \varphi_\mathcal{P})$ with $\varphi_\mathcal{C} : \links{\mathbb{P}} \simeq \links{\mathbb{P'}}$ 
and $\varphi_\mathcal{P}$ a bijection $\ports{\mathbb{P}} \simeq \ports{\mathbb{P'}}$. 
For writing $\varphi : \mathbb{P} \simeq \mathbb{P'}$, we require that the following diagrams commute: 
\begin{diagram}
\links{\mathbb{P}} & \rTo^{\varphi_\mathcal{C}} & \links{\mathbb{P'}} & & \links{\mathbb{P}} & \rTo^{\varphi_\mathcal{C}} & \links{\mathbb{P'}} & & \multlinks{\mathbb{P}} & \rTo^{\varphi_\mathcal{C}} & \multlinks{\mathbb{P'}} \\
\dTo_{\portsoflink{\mathbb{P}}} & & \dTo_{\portsoflink{\mathbb{P'}}} & & \dTo_{\mappriports{\mathbb{P}}} & & \dTo_{\mappriports{\mathbb{P'}}} & & \dTo_{\mapleftports{\mathbb{P}}} & & \dTo_{\mapleftports{\mathbb{P'}}}\\
\mathfrak{P}(\ports{\mathbb{P}}) & \rTo_{\mathfrak{P}(\varphi_\mathcal{P})} & \mathfrak{P}(\ports{\mathbb{P'}}) & & \mathcal{P} & \rTo_{\varphi_\mathcal{P}} & \mathcal{P'} & & \mathcal{P} & \rTo_{\varphi_\mathcal{P}} & \mathcal{P'}
\end{diagram}
If these diagrams commute, then we have $\im{{\varphi_\mathcal{P}}_{|\textsf{Aux}^?(\mathbb{P})}} = \textsf{Aux}^?(\mathbb{P'})$. Hence we can consider $\varphi' = _{\textsf{Aux}^?(\mathbb{P'})|}{\varphi_\mathcal{P}}_{|\textsf{Aux}^?(\mathbb{P})}$. We then require moreover that $\paxnumber{\mathbb{P'}} \circ \varphi' = \paxnumber{\mathbb{P}}$.

For any $\mathbb{P}, \mathbb{P'} \in \Ports$, for any $\varphi = (\varphi_\mathcal{C}, \varphi_\mathcal{P}) : \mathbb{P} \simeq \mathbb{P'}$, we set $\ports{\varphi} = \varphi_\mathcal{P}$ and $\links{\varphi} = \varphi_\mathcal{C}$.
\end{defin}

We now introduce two sizes on elements of $\Ports$ which will be used in the sequel: an integer and an ordered pair (pairs are lexicographically ordered).

\begin{defin}\label{def:measure}
Let $\mathbb{P} \in \Ports$. We set $\cosize{\mathbb{P}} = \max \{ \arity{\mathbb{P}}(l) \: / \: l \in \contrlinks{\mathbb{P}} \}$ and 
$\mes{\mathbb{P}} = (\sum_{l \in \contrlinks{\mathbb{P}}} \arity{\mathbb{P}}(l),$ $\textit{Card}(\ports{\mathbb{P}}) + \sum_{p \in \auxdoors{\mathbb{P}}} \paxnumber{\mathbb{P}}(p))$.
\end{defin}

\subsection{Pre-Linear Proof-Structures (PLPS)}

With $\PPLPS$ (Pre-Pre-Linear Proof-Structures) we shift from ``sets of cells'' (elements of $\Ports$) to graphs, and this amounts to give the rules allowing to connect the ports of the different cells. We give conditions on the set of wires of our graphs: condition~\ref{item:1} implies that three ports cannot be connected by two wires, condition~\ref{item:2} implies that auxiliary ports can never be conclusions of PPLPS (see Definition~\ref{def:axioms}), condition~\ref{item:3} implies that when the principal port of a cell is connected to another port this is necessarily a port of some cell, condition~\ref{item:4} corresponds to the fact that PPLPS are cut-free.

The reader acquainted with the theory of linear logic proof-nets might be interested in the reasons why our structures (PPLPS and later PLPS and PS) never contain cuts. There are essentially two reasons:
\begin{enumerate}
\item
(cut-free) PS are enough for our purpose, since the property we want to prove (injectivity) deals with cut-free proofs: once a precise notion of ``identity'' (or better said isomorphism) between cut-free PS is given (see Definition~\ref{def:isoPPLPSind}), if we prove that two different PS have different interpretations, then injectivity is proven (w.r.t.\ the chosen interpretation) whatever system of proofs one considers, provided the notion of cut-free proof of this system coincides with the one of PS\footnote{We already mentioned in the introduction that a standard cut-free proof-net (as defined for example in~\cite{injectcoh} or in~\cite{carvalhopaganitortora08}) is a particular case of PS.}.
\item
We can thus avoid a technical problem related to the presence of cuts in untyped proof-structures: it might happen that applying a cut-elimination step to an untyped proof-structure which ``contains a cycle'' (meaning that it does not satisfy the proof-net correctness criterion) yields a graph without cuts but containing ``vicious cycles'' (a premise of some link is also its conclusion: see the discussion before Definition~\ref{def:PLPS} of PLPS). It is precisely to avoid this problem that in~\cite{carvalhopaganitortora08} we decided to restrict to nets (proof-structures ``without cycles'' i.e.\ satisfying the correctness criterion).
\end{enumerate}

\begin{defin}\label{definition : PPLPS}
Let $\PPLPS$ be the set of pairs $\Phi=(\mathbb{P}, \mathcal{W})$ with $\mathbb{P} \in \Ports$ and $\mathcal{W} \subseteq \setofpairs{\ports{\mathbb{P}}}$ such that
\begin{enumerate}
\item\label{item:1} 
for any $w, w' \in \mathcal{W}$, we have $(w \cap w' \not= \emptyset \Rightarrow w=w')$;
\item \label{item:2} 
for any $p \in \ports{\mathbb{P}} \setminus \priports{\mathbb{P}}$, there exists $q \in \ports{\mathbb{P}}$ such that $\{ p, q \} \in \mathcal{W}$;
\item\label{item:3} 
for any $p \in \ports{\mathbb{P}} \setminus \im{\portsoflink{\mathbb{P}}}$, there exists $q \in \ports{\mathbb{P}} \setminus \priports{\mathbb{P}}$ s.t. $\{ p, q \} \in \mathcal{W}$;
\item\label{item:4}  
for any $w \in \mathcal{W}$, there exists $p \in w$ such that $p \notin \priports{\mathbb{P}}$.
\end{enumerate}
We set $\PortsofPPLPS{\Phi} = \mathbb{P}$ and $\edges{\Phi} = \mathcal{W}$. The elements of $\ports{\PortsofPPLPS{\Phi}}$ are \emph{the ports of $\Phi$} and those of $\edges{\Phi}$ are \emph{the wires} of $\Phi$.
\end{defin}

We now introduce precisely axioms and conclusions of a PPLPS $\Phi$; a consequence of our definition is that a conclusion $p$ of $\Phi$ is either the principal port of some cell or an axiom port.

\begin{defin}\label{def:axioms}
For any $\Phi \in \PPLPS$, we set:
\begin{minilist}
 \item 
$\conclusions{\Phi} = \{ p \in \ports{\PortsofPPLPS{\Phi}} \: / \: p \notin \im{\portsoflink{\PortsofPPLPS{\Phi}}} \cap \bigcup \edges{\Phi} \} $; the elements of $\conclusions{\Phi}$ are the \emph{free ports} or the \emph{conclusions} of $\Phi$
\item
$\terminallinks{\Phi} = \{ l \in \links{\PortsofPPLPS{\Phi}} \: / \: \mappriports{\PortsofPPLPS{\Phi}}(l) \in \conclusions{\Phi} \}$; the elements of $\terminallinks{\Phi}$ are the \emph{terminal cells} of $\Phi$
\item
$\axioms{\Phi} = \{ \{ p, q \} \in \edges{\Phi} \: / \: p, q \notin \priports{\PortsofPPLPS{\Phi}} \}$; the wire $\{ p, q \} \in\axioms{\Phi}$ is \emph{an axiom} of $\phi$ and the ports $p$ and $q$ are \emph{axiom ports}
\item
$\terminalaxioms{\Phi} = \{ w \in \axioms{\Phi} \: / \: (\exists p \in w) p \in \conclusions{\Phi} \}$ 
and $\isolatedaxioms{\Phi} = \{ w \in \axioms{\Phi} \: / \: (\forall p \in w) \: p \in \conclusions{\Phi} \}$; the wires of $\terminalaxioms{\Phi}$ (resp.\ $\isolatedaxioms{\Phi}$) are the \emph{terminal axioms} (resp.\ the \emph{isolated axioms}) of $\Phi$.
\end{minilist}
\end{defin}

\begin{defin}
For any $\Phi, \Phi' \in \PPLPS$, we write $\varphi : \Phi \simeq \Phi'$ if, and only if, $\varphi : \PortsofPPLPS{\Phi} \simeq \PortsofPPLPS{\Phi'}$ and for every $\{ p, q \}\in\setofpairs{\PortsofPPLPS{\Phi}}$, we have $\{ p, q \} \in \edges{\Phi}$ iff $\{ \ports{\varphi}(p), \ports{\varphi}(q) \} \in \edges{\Phi'}$.
\end{defin}

Intuitively, an axiom port is ``above'' a unique conclusion. But for general PPLPS this is wrong and we can only say that an axiom port cannot be ``above'' two different conclusions (Lemma~\ref{lemma : <= : unicity}). We thus consider the reflexive and transitive closure $\leq_\Phi$ of the relation $\below{\Phi}$ ``$p$ is immediately below $p'$ in $\Phi$''\footnote{See Definition \ref{definition : <=} of the appendix for a formal definition.} and show that our statement holds provided $\leq_\Phi$ is antisymmetric (Lemma~\ref{lemma:axports-above-unique-conclusion}), that is for PLPS (Definition~\ref{def:PLPS}).

\begin{lem}\label{lemma : <= : unicity}
Let $\Phi \in \PPLPS$. We have $(\forall w \in \axioms{\Phi})$ $(\forall p \in w)$ $(\forall c, c' \in \conclusions{\Phi})$ $((c \leq_\Phi p \textrm{ and } c' \leq_\Phi p) \Rightarrow c = c') $.
\end{lem}

The proof of Lemma \ref{lemma : <= : unicity} is just an application of Facts \ref{fact : tout filtre est une chaine} and \ref{fact : conclusions are minimal}:

\begin{fact}\label{fact : tout filtre est une chaine}
Let $\Phi \in \PPLPS$ and $p,q_1,q_2\in\ports{\PortsofPPLPS{\Phi}}$. If $q_1\leq_\Phi p$ and $q_2 \leq_\Phi p$, then $q_1 \leq_\Phi q_2$ or $q_2  \leq_\Phi q_1$.
\end{fact}

\begin{proof}
If $q_1 \below{\Phi} p$ and $q_2 \below{\Phi} p$, then $q_1 = q_2$.
\end{proof}

\begin{fact}\label{fact : conclusions are minimal}
Let $\Phi \in \PPLPS$. If $c \in \conclusions{\Phi})$ and $p\leq_\Phi c$, then $p=c$.
\end{fact}

\begin{proof}
If $c \in \conclusions{\Phi})$ then $\neg p \below{\Phi} c$ for every $p\in\ports{\Phi}$.
\end{proof}

A PPLPS $\Phi$ can have ``vicious cycles'' like for example a cell $l$ such that $p$ (resp.\ $p'$) is the principal (resp.\ an auxiliary) port of $l$ and $\{p,p'\}$ is a wire of $\Phi$: in~\cite{injectcoh} this corresponds to a link having a premise which is also the conclusion of the link (this does not occur in the typed framework of~\cite{injectcoh} but it cannot be excluded in our untyped framework). Let us stress that such a cycle is called ``vicious'' to distinguish it from the cycles in the so-called correctness graphs, which are related to the issue of sequentialization (see the discussion before Corollary~\ref{cor:injectNoW}). A PLPS is a PPLPS without vicious cycles:

\begin{defin}\label{def:PLPS}
We set $\PLPS = \{ \Phi \in \PPLPS \: / \: \textrm{the relation }\leq_\Phi \textrm{ is antisymmetric} \}$.
\end{defin}

\begin{lem}\label{lemma:axports-above-unique-conclusion}
Let $\Phi \in \PLPS$. We have $(\forall w \in \axioms{\Phi})$ $(\forall p \in w)$ $(\exists ! c \in \conclusions{\Phi}) \: c \leq_\Phi p $.
\end{lem}

\begin{proof}
For the unicity, apply Lemma \ref{lemma : <= : unicity}. For the existence, use the antisymmetry of $\leq_\Phi$ and the following property: we have $(\forall q \in \ports{\PortsofPPLPS{\Phi}})$ $((\forall p \in \ports{\PortsofPPLPS{\Phi}})(p \leq_\Phi q \Rightarrow p=q) \Rightarrow q \in \conclusions{\Phi}) $.
\end{proof}

The depth of a cell $l$ is (in the usual syntax see~\cite{injectcoh}) the number of exponential boxes containing $l$. We have not yet defined our notion of box (Definition~\ref{def:PS}), but since we are cut-free, $l$'s depth can also be defined as the number of doors of boxes below $l$; this makes sense in our framework too thanks to Lemma~\ref{lemma:axports-above-unique-conclusion}. We thus obtain the following definition (where the function $\#$ plays a crucial role, as mentioned in Remark~\ref{rem:explainingPorts}):

\begin{defin}\label{def:depth}
Let $\Phi \in \PLPS$. For any $p \in \ports{\PortsofPPLPS{\Phi}}$:
\begin{minilist}
\item we denote by $\conclusionunder{\Phi}{p}$ the unique $c \in \conclusions{\Phi}$ such that $c \leq_\Phi p$
\item $\depth{\Phi}{p} =$ $\textit{Card}(\{ l \in \bangs{\PortsofPPLPS{\Phi}} \: /$ $\mappriports{\PortsofPPLPS{\Phi}}(l) <_\Phi p\})+$ $\sum_{q \in \auxdoors{\PortsofPPLPS{\Phi}}, q < p} \paxnumber{\PortsofPPLPS{\Phi}}(q)$.
\end{minilist}
The depth of a PLPS $\Phi$ is the maximal depth of its ports and it is denoted by $\textit{depth}(\Phi)$.
\end{defin}

In the sequel, we will apply to $\Phi\in\PLPS$ transformations, depending on its terminal cells: $\Phi$ can of course have different terminal cells, but notice that in case $\Phi\in\cboxPLPS$ defined below, every terminal cell of $\Phi$ belongs to the set $\bangs{\mathbb{P}(\Phi)}\cup\contractionspax{\mathbb{P}(\Phi)}$.

\begin{defin}\label{def:ListPLPS}
We set:
\begin{minilist}
\item 
$\emptyPLPS = \{ \Phi \in \PLPS \: / \: \edges{\Phi} = \emptyset \}$.
\item
$\axPLPS = \{ \Phi \in \PLPS \: / \: \isolatedaxioms{\Phi} \not= \emptyset \}$.
\item
$\multPLPS = \{ \Phi \in \PLPS \: / \: (\exists l \in \terminallinks{\Phi}) \: \type{\PortsofPPLPS{\Phi}}(l) \in \{ \tens, \parr \} \}$.
\item
$\unitPLPS = \{ \Phi \in \PLPS \: / \: (\exists l \in \terminallinks{\Phi}) \: \type{\PortsofPPLPS{\Phi}}(l) \in \{ \one, \bot \} \}$.
\item
$\weakPLPS = \{ \Phi \in \PLPS \: / \: (\exists l \in \terminallinks{\Phi}) \: l \in \weakenings{\PortsofPPLPS{\Phi}} \}$.
\item
$\derPLPS = \{ \Phi \in \PLPS \: / \: (\exists l \in \terminallinks{\Phi}) \: l \in \derelictions{\PortsofPPLPS{\Phi}} \}$.
\item
$\contrPLPS = \{ \Phi \in \PLPS \: / \: (\exists l \in \terminallinks{\Phi}) \: l \in \contractions{\PortsofPPLPS{\Phi}} \}$.
\item $\contrunitPLPS = \{ \Phi \in \PLPS \: / \: (\exists l \in \terminallinks{\Phi} \cap \contrlinks{\PortsofPPLPS{\Phi}}) (\exists p \in \mapauxports{\PortsofPPLPS{\Phi}}(l)) (\paxnumber{\PortsofPPLPS{\Phi}}(p) \geq 1 \textrm{ and } (\forall q \geq_\Phi p) q \notin \bigcup{\axioms{\Phi}}) \} \setminus \contrPLPS$;
\item $\bangunitPLPS = \{ \Phi \in \PLPS \: / \: (\exists l \in \terminallinks{\Phi} \cap \bangs{\PortsofPPLPS{\Phi}}) (\exists p \in \mapauxports{\PortsofPPLPS{\Phi}}(l)) (\forall q \geq_\Phi p) q \notin \bigcup{\axioms{\Phi}}) \}$;
\item
$\cboxPLPS = \PLPS \setminus (\emptyPLPS \cup \axPLPS \cup \multPLPS \cup \unitPLPS \cup \weakPLPS \cup \derPLPS \cup \contrPLPS \cup \contrunitPLPS \cup \bangunitPLPS)$.
\end{minilist}
\end{defin}

Later on we will ``eliminate a terminal cell $l$'' from (some particular) PLPS: this is immediate when $l\in\weakenings{\PortsofPPLPS{\Phi}}$ or $\type{\PortsofPPLPS{\Phi}}(l)\in\{ \one, \bot \}$ since there is nothing ``above'' $l$. In case $\type{\PortsofPPLPS{\Phi}}(l)\in\{ \tens, \parr, \cod\}$ or $l\in\derelictions{\PortsofPPLPS{\Phi}}$, ``to eliminate $l$'' is intuitively clear, that is why we do not give the formal definition\footnote{See Definition \ref{def:eliminate-l} in the appendix for such a definition.}.

The peculiarity of the PLPS elements of $\contrunitPLPS\cup\bangunitPLPS$ is that they contain ``isolated subgraphs'': if ``above'' an auxiliary port $p$ of $l\in\bangs{\PortsofPPLPS{\Phi}} \cup \contrlinks{\PortsofPPLPS{\Phi}}$ there are no axioms, then the subgraph ``above'' $p$ is isolated. In presence of ``isolated subgraphs'', we can apply to the PLPS $\Phi$ the following transformations \emph{without damage} (Fact~\ref{fact : Phi LPS => Phi[l] LPS}) and shrinking the measure $\mes{\PortsofPPLPS{\Phi}}$ of $\Phi$ (see Definition~\ref{def:measure}), which will be used in the proof of Proposition~\ref{prop : KeyProposition}. For any $\Phi \in \PLPS$, for any $l \in \terminallinks{\Phi} \cap (\bangs{\PortsofPPLPS{\Phi}} \cup \contrlinks{\PortsofPPLPS{\Phi}})$, we denote by $\enleverunecoucheunitsweakenings{\Phi}{l}$ the PLPS obtained as follows: 
\begin{itemize}
\item if $l \in\bangs{\PortsofPPLPS{\Phi}}$, then we distinguish between two cases:
\begin{itemize}
\item if $\{ p \in \bigcup \axioms{\Phi} \: / \: p \geq_\Phi \mappriports{\PortsofPPLPS{\Phi}}(l) \} \not= \emptyset$, then $\enleverunecoucheunitsweakenings{\Phi}{l} = \Phi$;
\item otherwise, we remove $l$;
\end{itemize}
\item if $l \in \contrlinks{\PortsofPPLPS{\Phi}}$, $\enleverunecoucheunitsweakenings{\Phi}{l}$ is $\Phi$, except when there exists $q\in\mapauxports{\PortsofPPLPS{\Phi}}(l)$ such that $\paxnumber{\PortsofPPLPS{\Phi}}(q) \geq 1$ and $\{ p \in \bigcup \axioms{\Phi} \: / \: p \geq_\Phi q \} = \emptyset$: in that case $\enleverunecoucheunitsweakenings{\Phi}{l}$ is $\Phi$ where for every such $q$ one has $\paxnumber{\PortsofPPLPS{\enleverunecoucheunitsweakenings{\Phi}{l}}} (q) = \paxnumber{\PortsofPPLPS{\Phi}}(q)-1$.
\end{itemize}

\subsection{Linear Proof-Structures (LPS)}\label{subsection : LPS}

In a (cut-free) Proof-Structure of~\cite{injectcoh}, the depth of an axiom link is easily defined as the number of boxes in which the link is contained. In our framework this notion makes sense only when the two ports of an axiom have the same depth (Definition~\ref{def:depth}). This condition is not fulfilled by every PLPS: when this is the case we have a LPS.

\begin{defin}\label{def:LPS}
A LPS is a PLPS $\Phi$ such that $(\forall \{ p_1, p_2 \} \in \axioms{\Phi})$ $\depth{\Phi}{p_1} = \depth{\Phi}{p_2}$. We denote by $\LPS$ the set of LPS.\footnote{Our notion of \emph{LPS} has not to be confused with what is sometimes called ``the linearization of a proof-net'': the ``linearization'' forgets the auxiliary doors, and obviously there are some PS that have the same ``linearization'' but different LPS.}
\end{defin}

\begin{figure}
\fcapside{\caption{\textbf{Example of LPS.} 
Let $\Psi_2 \in \PPLPS$ as beside and such that $\paxnumber{\PortsofPPLPS{\Psi_2}}(p_1) = 1 = \paxnumber{\PortsofPPLPS{\Psi_2}}(p_2)$. Then we have  $\Psi_2 \in \cboxPLPS \cap \LPS$. 
}\label{example : LPS}
}
{\includegraphics[width=6cm, height=1.8cm]{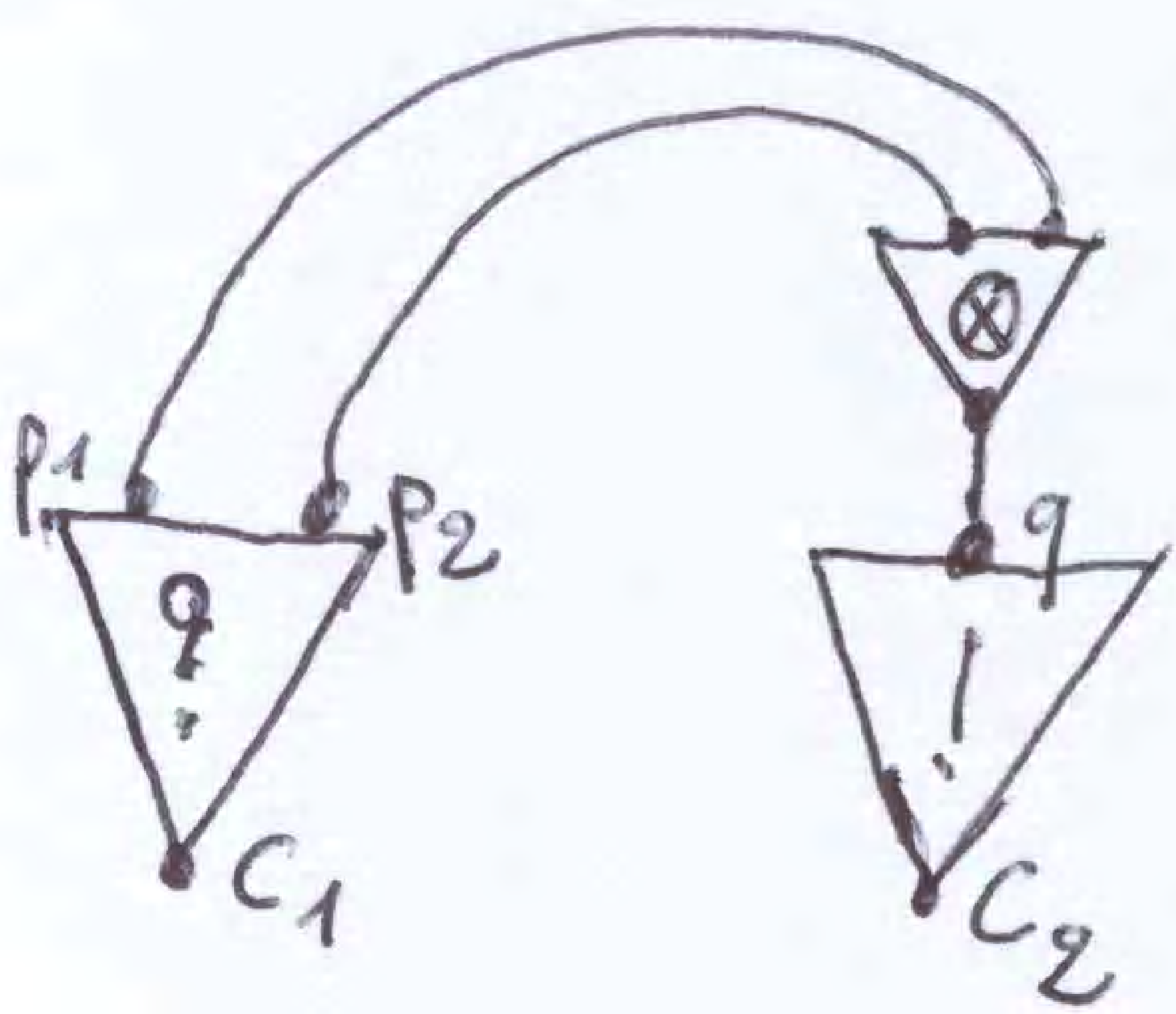}}
\end{figure}

\begin{fact}\label{fact : no terminal axioms in cbox-LPS}
For any $\Phi \in \cboxPLPS \cap \LPS$, we have $\terminalaxioms{\Phi} = \emptyset$.
\end{fact}

\begin{proof}
Let $\{p,q\}\in\axioms{\Phi}$, suppose $p\in\conclusions{\Phi}$ and let $c_q$ be the unique conclusion below $q$ (Definition~\ref{def:depth}): by Definition~\ref{def:depth} $\depth{\Phi}{p}=0$. Since $\Phi\not\in\axPLPS$ we have $q\neq c_q$ and thus $c_q$ is not an axiom port: in this case $c_q$ is the principal port of some cell $l$ of $\Phi$. By Definition~\ref{def:ListPLPS} this means that $l\in\bangs{\mathbb{P}(\Phi)}\cup\contractionspax{\mathbb{P}(\Phi)}$, which entails that $\depth{\Phi}{q}>0$, thus contradicting Definition~\ref{def:LPS} of LPS.
\end{proof}

A consequence of Fact~\ref{fact : no terminal axioms in cbox-LPS} is that in case $\Phi\in\cboxPLPS\cap\LPS$ all $\Phi$'s conclusions are  principal ports of some cells of the set $\bangs{\mathbb{P}(\Phi)}\cup\contractionspax{\mathbb{P}(\Phi)}$; in the syntax of~\cite{injectcoh} this corresponds to a proof-structure $\Phi$ with no links at depth $0$ except boxes and contraction links. We call $\overline{\Phi}$ the LPS obtained from such a $\Phi$ by decreasing $\Phi$'s depth by $1$, which can be easily done since $\Phi\in\cboxPLPS\cap\LPS$.\footnote{See Definition \ref{definition : enleverunecouche} in the appendix for a formal definition.}

\begin{fact}\label{fact : Phi LPS => Phi[l] LPS}
For any $\Phi \in \LPS$, for any $l \in \terminallinks{\Phi} \cap (\bangs{\PortsofPPLPS{\Phi}} \cap \contrlinks{\PortsofPPLPS{\Phi}})$, we have $\enleverunecoucheunitsweakenings{\Phi}{l} \in \LPS$.
\end{fact}

\begin{proof}
We have $\axioms{\enleverunecoucheunitsweakenings{\Phi}{l}} = \axioms{\Phi}$ and for any $\{ p, q \} \in \axioms{\Phi}$, $\depth{\Phi}{p} = \depth{\enleverunecoucheunitsweakenings{\Phi}{l}}{p}$.
\end{proof}

\subsection{Proof-Structures (PS)}

Intuitively, what is still missing in $\Phi \in \LPS$ to be a (cut-free) Proof-Structure in the standard sense (\cite{injectcoh}) is the connection between the doors of exponential boxes (once this information has been correctly produced, it automatically yields boxes). We then introduce a function $b$ associating with every $v\in\bangs{\PortsofPPLPS{\Phi}}$ a set of auxiliary doors of $\PortsofPPLPS{\Phi}$: this is precisely what was missing, provided certain conditions are satisfied (Definition~\ref{def:PS}). In particular, one asks that with every $v\in\bangs{\PortsofPPLPS{\Phi}}$ is associated a Proof-Structure: this is the usual notion of exponential box (see for example~\cite{hilbert}). In our framework, in order to define the Proof-Structure associated with $v$\footnote{We use the fact $v$'s box is itself a Proof-Structure in Definition~\ref{def:expPS}.}, we first build a PLPS $\Phi_{v}$ by taking ``everything what is above $v$ and the doors associated by $b$ with $v$'' and add a dereliction under every ``auxiliary conclusion''; doing this we take care to change the value of $\#$ on the auxiliary doors. We then remove $v$ (using Definition~\ref{def:eliminate-l}); finally we define from $b$ the new function $b_{v}$: 

\begin{defin}\label{def:PS}
A \emph{Proof-Structure (PS)} is a pair $R=(\Phi, b)$ where $\Phi \in \LPS$ and $b$ is a function $\bangs{\PortsofPPLPS{\Phi}} \rightarrow \mathfrak{P}(\auxdoors{\PortsofPPLPS{\Phi}})$ such that for any $p \in \auxdoors{\PortsofPPLPS{\Phi}}$, $\paxnumber{\PortsofPPLPS{\Phi}}(p) = \textit{Card} \{l\in\bangs{\PortsofPPLPS{\Phi}}\: / \: p \in b(l)\}$. Proof-Structures are defined by induction on the number of $!$-cells: we ask that with every $v\in\bangs{\PortsofPPLPS{\Phi}}$ is associated a PS called \emph{the box} of $v$ (denoted by $\boxofbang{R}(v)$)\footnote{Two examples of boxes are in Figures~\ref{fig:Example1} and~\ref{fig:PSandLPS}.}, and defined from the following subset $B_{v}$ of $\ports{\PortsofPPLPS{\Phi}}$: $$B_{v}= \{ q \in \ports{\PortsofPPLPS{\Phi}} \: / \: (\exists p \in \mapauxports{\PortsofPPLPS{\Phi}}(v) \cup b(v)) \: p \leq_\Phi q \}.$$

We ask that for $v,v'\in\bangs{\PortsofPPLPS{\Phi}}$ either $B_{v}\cap B_{v'}=\emptyset$ or $B_{v}\subseteq B_{v'}$ or $B_{v'}\subseteq B_{v}$\footnote{\label{footnote : nesting condition}This is the usual nesting condition of the definition of proof-net: two boxes are either disjoint or contained one in the other.}.

In order to define $\boxofbang{R}(v)$ one first defines $\Psi\in\PLPS$, starting from two sets $\mathcal{L}_0$ and $\mathcal{P}_0$ and from two bijections $p_1 : \mathcal{L}_0 \simeq b(v)$ and $p_0 : \mathcal{L}_0 \simeq \mathcal{P}_0$, by setting:
\begin{itemize}
 \item 
$\dom{\type{\PortsofPPLPS{\Psi}}} = \mathcal{L}_0 \uplus (\linksofports{\PortsofPPLPS{\Phi}}{B_{v}}\setminus  \linksofports{\PortsofPPLPS{\Phi}}{b(v)})$; $\type{\PortsofPPLPS{\Psi}}(l)=\contr$ for every $l\in\mathcal{L}_0$
\item
$\arity{\PortsofPPLPS{\Psi}}(l) = \left\lbrace \begin{array}{ll} 1 & \textrm{if $l \in \mathcal{L}_0$;} \\ \arity{\PortsofPPLPS{\Phi}}(l) & \textrm{otherwise;} \end{array} \right.$
\item
$\ports{\PortsofPPLPS{\Psi}} = (B_{v}\cup\{\mappriports{\PortsofPPLPS{\Phi}}(v)\}) \uplus \mathcal{P}_0$;
\item
$\portsoflink{\PortsofPPLPS{\Psi}}(l) = \left\lbrace \begin{array}{ll} \portsoflink{\PortsofPPLPS{\Phi}}(l) & \textrm{if $l \notin \mathcal{L}_0$;} \\ \{ p_1(l), p_0(l) \} & \textrm{if $l \in \mathcal{L}_0$;} \end{array} \right.$
\item
$\mappriports{\PortsofPPLPS{\Psi}}(l) = \left\lbrace \begin{array}{l} \mappriports{\PortsofPPLPS{\Phi}}(l) \textrm{ if $l \notin \mathcal{L}_0$;} \\ p_0(l) \textrm{ if $l \in \mathcal{L}_0$;} \end{array} \right.$
\item
$\mapleftports{\PortsofPPLPS{\Psi}} = \restriction{\mapleftports{\PortsofPPLPS{\Phi}}} 
{\multlinks{\PortsofPPLPS{\Phi}} \cap \linksofports{\PortsofPPLPS{\Phi}}{B_v}}$;
\item
$\paxnumber{\PortsofPPLPS{\Psi}}(p) =
\textsf{Card}\{w\in\bangs{\PortsofPPLPS{\Phi}}\cap\linksofports{\PortsofPPLPS{\Phi}}{B_v}\: / \:w\neq v\textrm{ and }p\in b(w)\}$;
\item
$\edges{\Psi} = \{ \{ p, q \} \in \edges{\Phi} \: / \: p, q \in B_{v} \}$;
\end{itemize}

The box of $v$, denoted by $\boxofbang{R}(v)$, is the pair $(\Phi_v, b_v)$ such that $\Phi_v$ is obtained from $\Psi$ by eliminating the terminal link $v$ (Definition~\ref{def:eliminate-l}) and such that $b_v = \restriction{b}{\bangs{\PortsofPPLPS{\Phi_v}}}$.

We set $\LPSofPPS{R} = \Phi$, $\mapbangauxd{R} = b$ and we will write \emph{the ports of $R$} (resp. \emph{the cells of $R$}) meaning the ports of $\Phi$ (resp. the cells of $\Phi$).
\end{defin}

In order to establish the equality (or better said an isomorphism) between two graphs representing (some kind of) proof we need to say how the conclusions of the two graphs correspond one another: we thus introduce the notion of indexed PPLPS (resp.\ PLPS, LPS, PS).

\begin{defin}
We denote by $\PPLPSind$ the set of pairs $(\Phi, \textsf{ind})$ such that 
$\Phi \in \PPLPS$ and $\textsf{ind}$ is a bijection $\conclusions{\Phi} \simeq \integer{\textit{Card}(\conclusions{\Phi})}$.

We set $\PSind = \{ (R, \textsf{ind}) \: / \: R \in \PS \textrm{ and } (\LPSofPPS{R}, \textsf{ind}) \in \PPLPSind \}$.
\end{defin}

\begin{defin}\label{def:isoPPLPSind}
For any $(\Phi, \textsf{ind}), (\Phi', \textsf{ind'}) \in \PPLPSind$, we write $\varphi : (\Phi, \textsf{ind}) \simeq (\Phi', \textsf{ind'})$ if, and only if, there exists $\varphi : \Phi \simeq \Phi'$ such that $\textsf{ind'} \circ \conclusions{\varphi} = \textsf{ind}$, 
where $\conclusions{\varphi}$ denotes the bijection $_{\conclusions{\Phi'}|}{\ports{\varphi}}_{|\conclusions{\Phi}} : \conclusions{\Phi} \simeq \conclusions{\Phi'}$.
\end{defin}

\begin{defin}
Let $(R, \textsf{ind}), (R', \textsf{ind'}) \in \PSind$. We write $\varphi : (R, \textsf{ind}) \simeq (R', \textsf{ind'})$ if, and only if, $\varphi: (\LPSofPPS{R}, \textsf{ind})\simeq (\LPSofPPS{R'}, \textsf{ind'})$ and the following diagram commutes\footnote{Recall that the notation $\mathcal{C}(\varphi)$ refers to Definition~\ref{def:isoPorts} and that for a function $f$ the notation $\mathfrak{P}(f)$ is among the ones introduced in the conventions at the beginning of this section.}:
\begin{diagram}
\bangs{\PortsofPPLPS{\LPSofPPS{R}}} & \rTo^{\textsf{b}(R)} & \mathfrak{P}(\auxdoors{\PortsofPPLPS{\LPSofPPS{R}}})\\
\dTo^{\links{\varphi}} & & \dTo_{\mathfrak{P}(\ports{\varphi})}\\
\bangs{\PortsofPPLPS{\LPSofPPS{R'}}} & \rTo_{\textsf{b}(R')} & \mathfrak{P}(\auxdoors{\PortsofPPLPS{\LPSofPPS{R'}}})
\end{diagram}
\end{defin}

\begin{defin}
Let $R = (\Phi, \textsf{ind}) \in \PLPSind$ 
and let $l \in \terminallinks{\Phi} \cap (\bangs{\Phi} \cup \contrlinks{\Phi})$. We set $\enleverunecoucheunitsweakenings{R}{l} = (\enleverunecoucheunitsweakenings{\Phi}{l}, \enleverunecoucheunitsweakenings{\textsf{ind}}{l})$, where $\enleverunecoucheunitsweakenings{\textsf{ind}}{l}(p) = \textsf{ind}(\conclusionunder{\Phi}{p})$ for $p \in \conclusions{\enleverunecoucheunitsweakenings{\Phi}{l}}$.
\end{defin}

\begin{defin}\label{definition : enleverunecoucheind}
Let $(\Phi, \textsf{ind}) \in \LPSind$ such that $\Phi \in \cboxPLPS$. We set $\enleverunecouche{(\Phi, \textsf{ind})} = (\enleverunecouche{\Phi}, \enleverunecouche{\textsf{ind}})$, where $\enleverunecouche{\Phi}$ has been defined in Subsection~\ref{subsection : LPS}\footnote{and, more formaly in Definition \ref{definition : enleverunecouche} of the appendix} and $\enleverunecouche{\textsf{ind}}(p) = \textsf{ind}(\conclusionunder{\Phi}{p})$.
\end{defin}

\section{Experiments}\label{sect:experiments}

In~\cite{lics06} and~\cite{carvalhopaganitortora08} experiments are defined in an untyped framework; we follow here the same approach in our Definition~\ref{def:expPS}. Experiments allow to compute the semantics of proof-nets (more generally of proof-structures):  the \emph{interpretation} $\sm{\pi}$ of a proof-net $\pi$ is the set of the results of $\pi$'s experiments, and the same happens in our framework for PS (Definition~\ref{definition : interpretation}). Like in~\cite{carvalhopaganitortora08}, in the following definition the set $\{+,-\}$ is used in order to ``semantically distinguish'' cells of type $\tens$ from cells of type $\parr$, which is mandatory in an untyped framework (as already discussed and used in~\cite{carvalhopaganitortora08}). The function $(\ )^{\bot}$ (which is the semantic version of linear negation) flips polarities (see Definition~\ref{def:ortogonal} of the appendix for the details).

\begin{defin}\label{definition : D}
We fix a set $A$ which does not contain any couple nor any $3$-tuple and such that $\ast\not\in A$; we call \emph{atoms} the elements of $A$. By induction on $n$ we define $D_n$:
$D_0 = A\cup(\{+,-\}\times\{\ast\})$ and $D_{n+1} = D_0\cup(\{+,-\}\times D_n\times D_n)$ $\cup (\{ +, - \} \times \setfmulti{D_n})$. 
We set $D = \bigcup_{n\in\Nat}D_n$.
\end{defin}

We need in the sequel the notion of injective $k$-point of $D^{< \omega}$, and for $E \in\mathfrak{P}(D^{< \omega})$ the notion of $E$-atomic element:

\begin{defin}\label{def:injectivek-point}
Given $k\in\Nat$, we say that $r\in D^{< \omega}$ is a \emph{$k$-point} when if $(+,[\alpha_1,\ldots, \alpha_m])$ occurs in $r$\footnote{See Definition \ref{definition : appears}\newcounter{AB}\setcounter{AB}{\value{footnote}} of the appendix for a formal definition of this expression.}, then $m=k$.

We say that $r\in D^{< \omega}$ is \emph{injective} when for every $\gamma \in A$, either $\gamma$ does not occur in $r$\footnotemark[\value{AB}] or there are exactly two occurrences of $\gamma$ in $r$\footnotemark[\value{AB}].

Given $E \in\mathfrak{P}(D^{< \omega})$, we say that $r\in E$ is \emph{$E$-atomic} when for every $r'\in E$ and every substitution\footnote{A subsitution is a function $\sigma: D\rightarrow D$ induced by a function $\sigma^{A}:A\rightarrow D$ (see Definition~\ref{def:substitution} of the appendix for the details).} $\sigma$ such that $\sigma(r')=r$ one has $\sigma(\gamma)\in A$ for every $\gamma \in A$ that occurs in $r'$. For $E \in\mathfrak{P}(D^{< \omega})$, we denote by ${E}_\textit{At}$ the subset of $E$ consisting of the $E$-atomic elements.
\end{defin}

\begin{remark}\label{rem:OldNewInjExp}
The notion of $k$-point is reminiscent of the notion of ``result of a $k$-obsessional experiment'' (\cite{injectcoh}), and it is also used in~\cite{lics06}. Notice however that the notion of injective point \emph{is not} related to what is called in~\cite{injectcoh} a result of an injective $k$-obsessional experiment: we keep the idea that all positive multisets have the same size, but we are very far from obsessionality. In some sense we do here exactly the opposite than obsessional experiments do: a $k$-obsessional experiment takes $k$ copies of the same ($k$-obsessional) experiment every time it crosses a box, while the intuition here is that injective $k$-points are results of experiments obtained by taking $k$ \emph{pairwise different} ($k$-)experiments every time a box is crossed.
\end{remark}

We now adapt to our framework the definition of experiment (given in~\cite{ll}; see also~\cite{phdtortora},~\cite{injectcoh},~\cite{carvalhopaganitortora08} for alternative definitions), the key tool to define the interpretation of a PS. Intuitively, an experiment of a PS $\Phi$ is a labeling of its ports by elements of $D$: this works perfectly well in the multiplicative fragment of LL (see for example~\cite{correctcoh}), but of course for PS with depth greater than zero things become a bit more complicated. One can either say that an experiment is defined only on ports $p$ such that $\depth{\Phi}{p}=0$ and that with every $!$-cell with depth zero is associated a multiset of experiments of its box (allowing to define the labels of the ports with depth zero): this is the choice made in~\cite{lics06} and~\cite{carvalhopaganitortora08}. Or one can follow (as we are going to do here in the spirit of~\cite{phdtortora} and~\cite{injectcoh}) the intuition that even with ports $p$ such that $\depth{\Phi}{p}>0$, an experiment associates labels, but not necessarily a unique label for every port (they might be several or none): formally it will associates with $p$ a multiset of elements of $D$ (and thus with every $!$-cell a multiset of multisets of experiments). Of course the two definitions associate the same interpretation with a given PS (Definition~\ref{definition : interpretation}).

\begin{defin}\label{def:expPS}
An experiment $e$ of a PS $R=(\Phi,b)$ is given by a function $\ports{\PortsofPPLPS{\Phi}} \rightarrow\mathcal{M}_{\textrm{fin}}(D)$\footnote{The elements of $e(p)$ are often called \emph{the labels} of $p$. Notice that $e(p)\not\in D$.} and for every $v\in\bangs{\PortsofPPLPS{\Phi}}$ a finite multiset of finite multisets of experiments of $v$'s box (i.e. $\boxofbang{R}(v)$) $e(v)=\multi{\multi{e^{1}_1,\ldots,e^{1}_{n_{1}}},\ldots,\multi{e^{l_{v}}_1,\ldots,e^{l_{v}}_{n_{l_{v}}}}}$, where $l_v\geq 0$ and $n_{i}\geq 0$ for every $1\leq i\leq l_v$. Experiments are defined by induction on $\textsf{depth}(\Phi)$ and we ask that $\textsf{Card}(e(v))=1$ for $v\in\bangs{\PortsofPPLPS{\Phi}}$ such that $\depth{\Phi}{\mappriports{\PortsofPPLPS{\Phi}}(v)}=0$ and that $\textsf{Card}(e(p))=1$ for $p\in\ports{\PortsofPPLPS{\Phi}} \setminus \auxdoors{\PortsofPPLPS{\Phi}}$ such that $\depth{\Phi}{p}=0$. For ports at depth $0$ the following conditions hold:
\begin{itemize}
\item
for any $\{ p, q \} \in \axioms{\Phi}$, we have $\alpha=\beta^\perp$, where $e(p) = [\alpha]$ and $e(q)=[\beta]$;
\item
for any $l \in \tenslinks{\PortsofPPLPS{\Phi}}$, we have $e(\mappriports{\PortsofPPLPS{\Phi}}(l)) = [(+,\alpha,\beta)]$, where $e(\mapleftports{\PortsofPPLPS{\Phi}}(l))=[\alpha]$ and $e(\maprightports{\PortsofPPLPS{\Phi}}(l))=[\beta]$;
\item
for any $l \in \parrlinks{\PortsofPPLPS{\Phi}}$, we have $e(\mappriports{\PortsofPPLPS{\Phi}}(l)) = [(-,\alpha,\beta)]$, where $e(\mapleftports{\PortsofPPLPS{\Phi}}(l))=[\alpha]$ and $e(\maprightports{\PortsofPPLPS{\Phi}}(l))=[\beta]$;
\item
for any $l \in \onelinks{\PortsofPPLPS{\Phi}}$, we have $e(\mappriports{\PortsofPPLPS{\Phi}}(l)) = [(+, \ast)]$;
\item
for any $l \in \botlinks{\PortsofPPLPS{\Phi}}$, we have $e(\mappriports{\PortsofPPLPS{\Phi}}(l)) = [(-, \ast)]$;
\item
for any $l\in\contrlinks{\PortsofPPLPS{\Phi}}$, we have $e(\mappriports{\PortsofPPLPS{\Phi}}(l) = [(-, \sum_{p \in \mapauxports{\PortsofPPLPS{\Phi}}(l)} e(p))]$;
\item
for any $\{p,q\}\in\edges{\Phi}\setminus\axioms{\Phi}$, we have $e(p)=e(q)$.
\end{itemize}
If $\textsf{depth}(\Phi)=0$, the definition is already complete. Otherwise for every $v\in\bangs{\PortsofPPLPS{\Phi}}$ such that $\depth{\Phi}{\mappriports{\PortsofPPLPS{\Phi}}(v)}=0$ we know the multiset $[e_1,\ldots,e_{n_v}]$ of experiments of $v$'s box such that $e(v)=[[e_1,\ldots,e_{n_v}]]$ and we know for every port $p$ of $\Phi$ which is also a port of $\overline{B}(R)(v)$ the multiset $e_i(p)$ (for $i\in\{1,\ldots,n_v\}$). Then we set
\begin{itemize}
\item $e(\mappriports{\PortsofPPLPS{\Phi}}(v))=[(+,\sum_{i \in \{1,\ldots,n_v\}} e_i(p))]$, where $p$ is the unique free port of $\overline{B}(R)(v)$ such that $\mappriports{\PortsofPPLPS{\Phi}}(v) \leq_\Phi p$;\footnote{Let $\{ q_{v} \}= \mapauxports{\PortsofPPLPS{\Phi}}(v)$; then for some port $q_{v}'$ of $\Phi$ we have $\{q_{v},q_{v}'\}\in\edges{\Phi}$. If $\{q_{v},q_{v}'\}\in\axioms{\Phi}$ (resp.\ $\{q_{v},q_{v}'\}\not\in\axioms{\Phi}$), then $q_{v}$ (resp.\  $q_{v}'$) is the unique free port $p$ of $\overline{B}(R)(v)$ such that $\mappriports{\PortsofPPLPS{\Phi}}(v) \leq_\Phi p$.}
\item $e(p)=\sum_{i \in \{1,\ldots,n_v\}} e_i(p)$ 
for every port $p$ of $\Phi$ which is also a port of $\overline{B}(R)(v)$;\footnote{We are using here the nesting condition of Definition~\ref{def:PS} : see Footnote~\ref{footnote : nesting condition}.}
\item
$e(w)=\sum_{i \in \{1,\ldots,n_v\}} e_i(w)$ for every $!$-cell $w$ of $\Phi$ which is also a cell of $\overline{B}(R)(v)$.\footnotemark[\value{footnote}]
\end{itemize}
\end{defin}

\begin{example}\label{example:experiment-Intro}
Consider the PS $R$ of Figure~\ref{fig:NoUniqueExp} and the box $\Psi$ of its unique $!$-cell $v$ represented in Figure~\ref{fig:Example1}.
\begin{figure}
\fcapside{\caption{\textbf{The box $\Psi$ of the unique $!$-cell of the PS $R$ of Figure~\ref{fig:NoUniqueExp}.}\label{fig:Example1}
}
}
{\includegraphics[width=5cm,height=1.5cm]{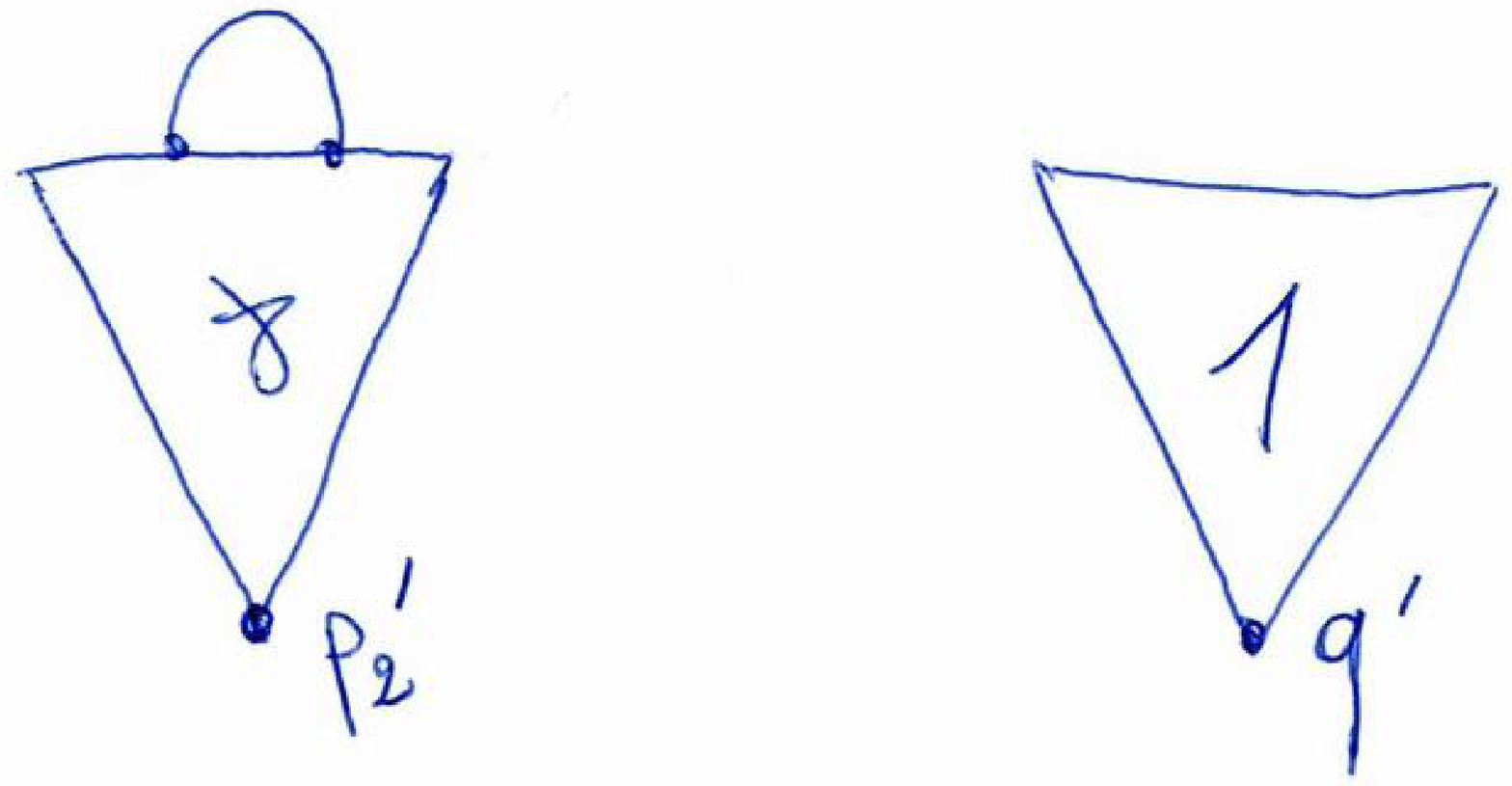}}
\end{figure}
We can define two experiments $e_1$ and $e_2$ of $\Psi$ by choosing $\gamma_1,\gamma_2\in D$: we obtain $e_i(p'_2)=[(-,\gamma_i,\gamma_i^{\bot})]$ and $e_i(q')=[(+,\ast)]$ where $\{q,q'\},\{p_{2},p_{2}'\}\in\edges{\LPSofPPS{R}}$. By choosing $\alpha\in D$, we have an experiment $e$ of $R$ such that $e(p_1)=[(-,\alpha,\alpha^{\bot})]$, $e(p_{2}')=e(p_2)=[(-,\gamma_1,\gamma_1^{\bot}),(-,\gamma_2,\gamma_2^{\bot})]$, $e(c_1)=[(-,[(-,\gamma_1,\gamma_1^{\bot}),(-,\gamma_2,\gamma_2^{\bot})])]$, $e(q')=e(q)=[(+,\ast),(+,\ast)]$, $e(c_2)=[(+,[(+,\ast),(+,\ast)])]$, and $e(v)=[[e_1,e_2]]$.
\end{example}

\begin{defin}
Let $(R, \textsf{ind}) \in \PSind$, let $e$ be an experiment of $R$, let $n=\textit{Card}(\conclusions{R})$ and let $r \in D^n$. We say that $(e, r)$ is an experiment of $(R, \textsf{ind})$ and that $r$ is \emph{the result} of $(e, r)$ if and only if $r = (x_1,\ldots,x_n)$, where $x_i$ is the unique element of the multiset $e \circ \textsf{ind}^{-1}(i)$.
\end{defin}

\begin{defin}\label{definition : interpretation}
If $(R, \textsf{ind}) \in \PSind$, we define \emph{the interpretation of $(R, \textsf{ind})$} as the set $\sm{(R, \textsf{ind})}=\{ r\in D^{\textit{Card}(\conclusions{R})}\: / \: r \textrm{ is the result of an experiment of }(R, \textsf{ind}) \}$.
\end{defin}

The crucial result proven in~\cite{ll} is that if $\pi'$ is a proof-net obtained by applying to $\pi$ some steps of cut-elimination, then $\sm{\pi}=\sm{\pi'}$. Since any cut-free untyped net of~\cite{carvalhopaganitortora08} (and thus any proof-net of, for example,~\cite{injectcoh}) is a PS, in order to prove injectivity for the nets of~\cite{carvalhopaganitortora08} (and thus for the usual proof-nets of, for example,~\cite{injectcoh}) it is enough to prove that two PS with the same interpretation are the same (Corollary~\ref{cor:injectConnected} and Corollary~\ref{cor:injectNoW}).

\subsection{Experiments of PLPS}

In general, if we want to know whether a point is the result of any experiment, it is not enough to know the LPS of the (proof-)net: we have to know ``the connection between the doors of the box''. But if one takes $k$ copies every time one crosses a box, then it is enough: results of $k$-experiments can be defined directly on LPS. This yields the notion of $k$-experiment of a LPS (Definition~\ref{def:experiment}). Actually $k$-experiments are defined ``up to the names of the atoms'' and we thus introduce sequences of indexes: the intuition is that for $\gamma\in A$ and $s\in\mathbb{N}^n$, $(\gamma,s)$ is one of the $k^n$ copies of $\gamma$.



For any $n \in \mathbb{N}$, we define $A'_n$ as follows: $A'_n = \left\lbrace \begin{array}{ll} A & \textrm{if $n=0$;} \\ 
A \times \mathbb{N}^n & \textrm{otherwise.} \end{array} \right.$ We set $A' = \bigcup_{n \in \mathbb{N}} A'_n$. 

We denote by $|\ |$ the function $A' \rightarrow A$ defined by $|\delta| = \left\lbrace \begin{array}{ll} \delta & \textrm{if $\delta \in A$;} \\ \gamma & \textrm{if $\delta = (\gamma, s) \notin A$;} \end{array} \right.$ 

and by $\textsf{loc}$ the function $A' \rightarrow \mathbb{N}^\omega$ defined by $\textsf{loc}(\delta) = \left\lbrace \begin{array}{ll} \emptysequence & \textrm{if $\delta \in A$;} \\ s & \textrm{if $\delta = (\gamma, s) \notin A$.} \end{array} \right.$

\begin{defin}
We set $\dig(\emptysequence) = id_{A'}$ and, for any $s \in \mathbb{N}^{< \omega} \setminus \{ \emptysequence \}$, we denote by $\dig(s)$ the function $A' \rightarrow A'$ defined by $\dig(s)(\delta) = (|\delta|, \textit{conc}(\textsf{loc}(\delta), s))$, where $\textit{conc}$ is the function $\mathbb{N}^{< \omega} \times \mathbb{N}^{< \omega} \rightarrow \mathbb{N}^{< \omega}$ defined by \\$\textit{conc}((d_1, \ldots, d_m), (d'_1, \ldots, d'_{m'})) = (d_1, \ldots, d_m, d'_1, \ldots, d'_{m'})$.
\end{defin}

A construction similar to the one used to define $D$ from $A$ allows to define $D'$ from $A'$: intuitively, an element of $D'$ is an element of $D$ where every atom is followed by a sequence of integers. Notice that since $A\subseteq A'$ one has $D\subseteq D'$, and this will be used in Definition~\ref{def:experiment} (last item) of experiment of a PLPS.

\begin{defin}
By induction on $n$ we define $D'_n$: $D'_0 =  A'\cup(\{+,-\}\times\{\ast\})$ and $D'_{n+1} = D'_0 \cup (\{ +, - \} \times D'_n \times D'_n)$ $\cup (\{ +, - \} \times \setfmulti{D'_n})$. We set $D' = \bigcup_{n \in \mathbb{N}} D'_n $.
\end{defin}

\begin{defin}
We define $\textit{At'}:D' \rightarrow \setfparts{A'}$ the function which associates with $\alpha\in D'$ its atoms, by induction on $\min \{ n \in \mathbb{N} \: / \: \alpha \in D'_n \}$:
\begin{minilist}
\item $\textit{At'}(\delta) = \{ \delta \} $ if $\delta \in A'$;
\item $\textit{At'}(\iota, \ast) = \emptyset$;
\item $\textit{At'}(\iota, \alpha_1, \alpha_2) = \textit{At'}(\alpha_1) \cup \textit{At'}(\alpha_2)$;
\item $\textit{At'}(\iota, \multi{\alpha_1, \ldots, \alpha_m}) = \cup_{j=1}^m \textit{At'}(\alpha_j)$.
\end{minilist}
We still denote by $\textit{At'}$ the function $\setfparts{D'}\rightarrow \setfparts{A'}$ defined by $\textit{At'}(\mathfrak{a})=\bigcup_{\alpha \in \mathfrak{a}} \textit{At'}(\alpha)$; and $\textit{At'}$ will also denote the function ${\setfmulti{D'}}^{< \omega} \rightarrow \setfparts{A'}$ defined by $\textit{At'}(a_1, \ldots, a_n) = \bigcup_{i=1}^n \textit{At'}(\textit{Supp}(a_i))$.
\end{defin}

\begin{defin}\label{definition : actions on points}
The set of partial injections from $A'$ to $A'$ is denoted by $\pInj$.

Let $\tau \in \pInj$. For any $\alpha \in D'$ such that $\textit{At'}(\alpha) \subseteq \dom{\tau}$, we define $\tau \cdot \alpha \in D'$ by induction on $\min \{ n \in \mathbb{N} \: / \: \alpha \in D'_n \}$: 
$$\tau \cdot \alpha = \left\lbrace 
\begin{array}{ll} 
\tau(\delta) & \textrm{if $\alpha = \delta \in A'$;}\\
(\iota, \ast) & \textrm{if $\alpha = (\iota, \ast)$;}\\
(\iota, \tau \cdot \alpha_1, \tau \cdot \alpha_2) & \textrm{if $\alpha = (\iota, \alpha_1, \alpha_2)$;}\\
(\iota, \multi{\tau \cdot \alpha_1, \ldots, \tau \cdot \alpha_m}) & \textrm{if $\alpha = (\iota, \multi{\alpha_1, \ldots, \alpha_m})$.}
\end{array}
\right.$$

For any $a = \multi{\alpha_1, \ldots, \alpha_m} \in \setfmulti{D'}$ such that $\textit{At'}(a) \subseteq \dom{\tau}$, we set $\tau \cdot a = \multi{\tau \cdot \alpha_1, \ldots, \tau \cdot \alpha_m} \in \setfmulti{D'}$. For any $r = (\alpha_1, \ldots, \alpha_n) \in {D'}^{< \omega}$ such that $\textit{At'}(\multi{\alpha_1, \ldots, \alpha_n}) \subseteq \dom{\tau}$, we set $\tau \cdot r = (\tau \cdot \alpha_1, \ldots, \tau \cdot \alpha_n) \in {D'}^{< \omega}$. For any $r = (a_1, \ldots, a_n) \in {\setfmulti{D'}}^{< \omega}$ such that $\textit{At'}(r) \subseteq \dom{\tau}$, we set $\tau \cdot r = (\tau \cdot a_1, \ldots, \tau \cdot a_n) \in {\setfmulti{D'}}^{< \omega}$.
\end{defin}

\begin{defin}
For any $\tau \in \pInj$, for any function $h$ such that $\im{h} \subseteq D'$ and $\textit{At'}(\im{h}) \subseteq \dom{\tau}$, we define $\tau \cdot h:\dom{h} \rightarrow D'$ as follows: $(\tau \cdot h)(x) = \tau \cdot h(x)$.
\end{defin}

The function $\dig_d^k$ associates with $a \in \setfmulti{D'}$ the multiset of the $k^d$ copies of $a$: if for example $a=[\alpha,\beta,\beta]$ for some $\alpha,\beta\in A$, then one has $\dig_1^2(a)=[(\alpha,1),(\alpha,2),(\beta,1),(\beta,2),(\beta,1),(\beta,2)]$. An immediate consequence of the following definition is that for every $a\in\setfmulti{D'}$ and for every integer $d$ one has $\dig_{d+1}^k(a)=\dig_1^k(\dig_d^k(a))$.

\begin{defin}
For any $k, d \in \mathbb{N}$, let $\dig_d^k$ be the function $\setfmulti{D'} \rightarrow \setfmulti{D'}$ defined by $\dig_d^k(a) = \sum_{s \in \integer{k}^d} \sum_{\alpha \in \textsf{Supp}(a)} a(\alpha) \cdot [\dig(s) \cdot \alpha]$.
\end{defin}

We now have all the tools to define (a particular kind of) experiments directly on LPS and not on PS as in the usual setting (Definition~\ref{def:expPS} in our framework). It clearly appears in Subsection~\ref{subsect:MainResult} (and precisely in Fact~\ref{fact:PSandLPS}) how (injective atomic) $k$-experiments of LPS are used in our proof. It is worth noticing that we recover in the framework of LPS the simplicity of the definition of experiment in the multiplicative fragment of linear logic proof-nets (see for example~\cite{correctcoh}and~\cite{Pag07mscs}): despite the presence of exponentials (here $?$-cells and $!$-cells) a $k$-experiment of a PLPS is just a labeling of its ports by elements of $D'$ satisfying some conditions.

\begin{defin}\label{def:experiment}
Let $k \in \mathbb{N}$. For any $\Phi \in \PLPS$, a $k$-experiment $e$ of $\Phi$ is a function $\ports{\PortsofPPLPS{\Phi}} \rightarrow D'$ such that 
\begin{minilist}
\item for any $l \in \tenslinks{\PortsofPPLPS{\Phi}}$, we have $e(\mappriports{\PortsofPPLPS{\Phi}}(l)) = (+, e(\mapleftports{\PortsofPPLPS{\Phi}}(l)), e(\maprightports{\PortsofPPLPS{\Phi}}(l)))$;
\item for any $l \in \parrlinks{\PortsofPPLPS{\Phi}}$, we have $e(\mappriports{\PortsofPPLPS{\Phi}}(l)) = (-, e(\mapleftports{\PortsofPPLPS{\Phi}}(l)), e(\maprightports{\PortsofPPLPS{\Phi}}(l)))$;
\item for any $l \in \onelinks{\PortsofPPLPS{\Phi}}$ (resp. $l \in \botlinks{\PortsofPPLPS{\Phi}}$), we have $e(\mappriports{\PortsofPPLPS{\Phi}}(l)) = (+, \ast)$ (resp. $e(\mappriports{\PortsofPPLPS{\Phi}}(l)) = (-, \ast)$);
\item for any $l \in \bangs{\PortsofPPLPS{\Phi}}$, we have $e(\mappriports{\PortsofPPLPS{\Phi}}(l)) = (+, \sum_{p \in \mapauxports{\PortsofPPLPS{\Phi}}(l)} \dig_1^{k}(\multi{e(p)}))$\footnote{Notice that $\sum_{p \in \mapauxports{\PortsofPPLPS{\Phi}}(l)} \dig_1^{k}(\multi{e(p)})=\dig_1^{k}(\multi{e(p)})$ where $\{p\}=\mapauxports{\PortsofPPLPS{\Phi}}(l)$.};
\item for any $l \in \contrlinks{\PortsofPPLPS{\Phi}}$, we have $e(\mappriports{\PortsofPPLPS{\Phi}}(l) = (-, \sum_{p \in \mapauxports{\PortsofPPLPS{\Phi}}(l)} \dig_{\paxnumber{\PortsofPPLPS{\Phi}}(p)}^k(\multi{e(p)}))$;
\item and for any $\{ p, q \} \in \edges{\Phi}$, we have $e(p) = \left\lbrace \begin{array}{l} e(q)^\perp \textrm{ with  }e(p) \in D,\textrm{ if $\{ p, q \} \in \axioms{\Phi}$;} \\ e(q) \textrm{ otherwise.} \end{array} \right.$\footnote{$\delta^\perp$ is obtained from $\delta\in D'$ by substituting every occurrence of $+$ (resp.\ $-$) by $-$ (resp.\ $+$): see Definition~\ref{def:ortogonal} of the appendix for the details.}
\end{minilist}
\end{defin}

\begin{defin}
Let $k \in \mathbb{N}$, let $\Phi \in \PLPS$. Let $e$ be any $k$-experiment of $\Phi$. 

We say that $e$ is \emph{atomic} if for any $w \in \textsf{Ax}(\Phi)$, for any $p \in w$, we have $e(p) \in A$. 

We say that $e$ is \emph{injective} if for any $w, w' \in \textsf{Ax}(\Phi)$, for any $p \in w, p' \in w'$, we have $\textit{At'}(e(p)) \cap \textit{At'}(e(p')) \not= \emptyset \Rightarrow w = w'$.
\end{defin}

\begin{defin}
Let $k \in \mathbb{N}$. For any $\Phi, \Phi' \in \PLPS$, for any $k$-experiment $e$ of $\Phi$, for any $k$-experiment $e'$ of $\Phi'$, an iso $\varphi : e \simeq e'$ is an iso $\varphi : \Phi \simeq \Phi'$ such that for any $p \in \ports{\PortsofPPLPS{\Phi}}$, we have $e(p) = e'(\ports{\varphi}(p))$.
\end{defin}

\begin{defin}
Let $k \in \mathbb{N}$. Let $(\Phi, \textsf{ind}) \in \PLPSind$. Let $e$ be a $k$-experiment of $\Phi$ and let $r \in (D')^{\textit{Card}(\conclusions{\Phi})}$. We say that $(e, r)$ is a $k$-experiment of $(\Phi, \textsf{ind})$ and that $r$ is the result of $(e,r)$ iff $r = e \circ \textsf{ind}^{-1}$.
\end{defin}

\begin{example}\label{example : experiment}
Let $\Psi_2$ be as in Figure~\ref{example : LPS} and let $\textsf{ind}_2(c_1) =$ $1$ and $\textsf{ind}_2(c_2) =$ $2$. Let $\gamma_1, \gamma_2 \in A$. Let $a_1 =[(\gamma_1,1),(\gamma_1,2),(\gamma_1,3),(\gamma_2,1),(\gamma_2,2),(\gamma_2,3)]$ and\\
$a_2 =[(+,(\gamma_1,1),(\gamma_2,1)),(+,(\gamma_1,2),(\gamma_2,2)), (+,(\gamma_1,3),(\gamma_2,3))]$. Then $r_2 =$ $((-, a_1),$ $(+, a_2))$ is the result of the injective atomic $3$-experiment $e_{2}$ of $(\Psi_2, \textsf{ind}_2)$ such that $e_{2}(p_1)=\gamma_{2}$ and $e_{2}(p_{2})=\gamma_{1}$. Notice that once we have chosen the labels of $p_{1}$ and $p_{2}$ and the integer $k$ (here $k=3$), the $k$-experiment of $\Psi_2$ is entirely determined.
\end{example}

\begin{remark}\label{remark:k-expinducedby1-exp}
As mentioned in Example~\ref{example : experiment}, once an integer $k\geq 1$ and the labels of the axiom ports of the LPS $\Phi$ are chosen, the $k$-experiment of $\Phi$ is entirely determined. In particular, given a $1$-experiment $e_{1}$ of $\Phi$, for every $k\geq 1$ there exists a unique $k$-experiment $e_k$ associating with the axiom ports of $\Phi$ the same labels as $e_{1}$. Clearly, $e_{1}$ is atomic (resp.\ injective) iff $e_k$ is atomic (resp.\ injective).
\end{remark}

We are going to prove a sequence of facts concerning experiments and their results. The first one allows to ``exchange'' two indexes (elements of $\integer{k}$) without changing the result of a given experiment: thanks to this property we'll be able (in Fact~\ref{fact : morphism Q}) to exchange two ``copies'' of $\alpha\in a$ for some multiset $a$ of $D'$.

\begin{fact}\label{fact : result invariant by permutations of dig}
Let $k \in \mathbb{N}$. Let $(\Phi, \textsf{ind}) \in \PLPSind$. Let $(e, r)$ be a $k$-experiment of $(\Phi, \textsf{ind})$. Let $d \in \mathbb{N}$. Let $j_1, j_2 \in \integer{k}$. Let $\rho \in \pInj$ defined by setting \\
$$\rho(\delta) = \left\lbrace \begin{array}{l} 
\dig(s)(\dig(j_2)(\delta_0)) \textrm{ if $\delta = \dig(s)(\dig(j_1)(\delta_0))$ with $s \in {\integer{k}}^d$ and $\delta_0 \in A'$;}\\
\dig(s)(\dig(j_1)(\delta_0)) \textrm{ if $\delta = \dig(s)(\dig(j_2)(\delta_0))$ with $s \in {\integer{k}}^d$ and $\delta_0 \in A'$;}\\
\delta \textrm{ otherwise.}
\end{array} \right.$$
Then we have $\rho \cdot r = r$.
\end{fact}

\begin{proof}
By induction on $\textit{Card}(\links{\PortsofPPLPS{\Phi}})$.
\end{proof}

We now show how one can obtain a $k$-experiment of $\enleverunecouche{\Phi}$ from a $k$-experiment of the LPS $\Phi$, which will be useful in the case $\Phi\in\cboxPLPS$ of the proof of Proposition~\ref{prop : KeyProposition}.

\begin{fact}\label{fact : enleverunecouche a un resultat}
Let $k \in \mathbb{N}$. Let $(\Phi, \textsf{ind}) \in \LPSind$ such that $\Phi \in \cboxPLPS$ and let $(e, r)$ be a $k$-experiment of $(\Phi, \textsf{ind})$. Then there exists a unique $k$-experiment $(\enleverunecouche{e}, \enleverunecouche{r})$ of $\enleverunecouche{(\Phi, \textsf{ind})} = (\enleverunecouche{\Phi}, \enleverunecouche{\textsf{ind}})$ such that
\begin{minilist}
\item for any $p \in (\ports{\PortsofPPLPS{\Phi}} \setminus \conclusions{\Phi}) \cap \ports{\PortsofPPLPS{\enleverunecouche{\Phi}}}$, we have $\enleverunecouche{e}(p) = e(p)$;
\item if $r(i) = (+, a)$, then there exists $\alpha \in D'$ such that $\enleverunecouche{r}(i) = \alpha$ and $a = \sum_{j=1}^k \dig(j) \cdot \multi{\alpha}$; if $r(i) = (-, a)$, then there exists $b \in \setfmulti{D'}$ such that $\enleverunecouche{r}(i) = (-, b)$ and $a = \sum_{j=1}^k \dig(j) \cdot b$.
\end{minilist}

Moreover, if $e$ is atomic (resp. injective), then $\enleverunecouche{e}$ is atomic (resp. injective).
\end{fact}

\begin{proof}
For any $l \in \contractionspax{\PortsofPPLPS{\Phi}} \cap \terminallinks{\Phi}$, we have $e(\mappriports{\PortsofPPLPS{\Phi}}(l))=\sum_{p \in \mapauxports{\PortsofPPLPS{\Phi}}(l)} \dig_{\paxnumber{\PortsofPPLPS{\Phi}}(p)}^k([e(p)]) =\sum_{j=1}^k \dig(j) \cdot \sum_{p \in \mapauxports{\PortsofPPLPS{\Phi}}(l)} \dig_{\paxnumber{\PortsofPPLPS{\Phi}}(p)-1}^k([e(p)])$. For any $l \in \bangs{\PortsofPPLPS{\Phi}} \cap \terminallinks{\Phi}$, we have $e(\mappriports{\PortsofPPLPS{\Phi}}(l))=\sum_{p \in \mapauxports{\PortsofPPLPS{\Phi}}(l)} \dig_1^k([e(p)]) = \sum_{j=1}^k \dig(j) \cdot \multi{e(q)}$, where $\{ q \} = \mapauxports{\PortsofPPLPS{\Phi}}(l)$.
\end{proof}

For every $\rho\in\pInj$ (Definition~\ref{definition : actions on points}) and for every $\alpha\in D'$, when $\textit{At'}(\alpha) = \emptyset$, one has $\rho \cdot \alpha=\alpha$. We will use in the sequel (in particular in subsections~\ref{subsect:expunit} and~\ref{subsect:cbox}) the remark that any multiset $b\in \setfmulti{D'}$ can be decomposed into a (possibly empty) multiset $b^\textit{At}$ in which atoms occur and a (possibly empty) multiset $b^\ast$ in which no atom occurs: $b=b^\textit{At}+b^\ast$, where $b^\textit{At}$ and $b^\ast$ are precisely defined as follows.

\begin{defin}
For any $D_0 \subseteq D'$, we set ${D_0}^{\textit{At}} = \{ \alpha \in D_0 \: / \: \textit{At'}(\alpha) \not= \emptyset \}$ and ${D_0}^\ast = \{ \alpha \in D_0 \: / \: \textit{At'}(\alpha) = \emptyset \}$.

For any $a \in \setfmulti{D'}$, we set $a^\textit{At} = \restriction{a}{{\textit{Supp}(a)}^\textit{At}}$ and $a^\ast = \restriction{a}{{\textit{Supp}(a)}^\ast}$.
\end{defin}

The following Fact~\ref{fact : bangunit : syntaxe} and Fact~\ref{fact : contrunit : syntaxe} are similar in spirit to Fact~\ref{fact : enleverunecouche a un resultat}: they allow to obtain a $k$-experiment $\enleverunecoucheunitsweakenings{e}{l_0}$ of $\enleverunecoucheunitsweakenings{\Phi}{l_0}$ from a $k$-experiment $e$ of $\Phi\in \LPS$, and they will be used in the cases $\Phi \in\bangunitPLPS$ and $\Phi \in\contrunitPLPS$ of the proof of Proposition~\ref{prop : KeyProposition}. In both the facts the hypothesis $a^\ast\neq \multi{}$ (for $a\in\setfmulti{D'}$ such that $e(p)=(\iota,a)$ with $p$ port of $\Phi$) is crucial: it implies that ``above'' $p$ there is an ``isolated subgraph'', which allows to apply the transformations defined in Section~\ref{sect:syntax}, thus shrinking the measure of $\Phi$.

\begin{fact}\label{fact : bangunit : syntaxe}
Let $k \in \mathbb{N}$. Let $R = (\Phi, \textsf{ind}) \in \LPSind$ and let $(e, r)$ be a $k$-experiment of $(\Phi, \textsf{ind})$. 
Let $l_0 \in\bangs{\PortsofPPLPS{\Phi}}\cap\terminallinks{\Phi}$ and $\beta \in D'$ such that $e( \mappriports{\PortsofPPLPS{\Phi}}(l_0)) = (+, \dig_1^k (\multi{\beta}))$ and ${(\dig_1^k (\multi{\beta}))}^\ast \not= \multi{}$. 
Then $\mes{\PortsofPPLPS{\enleverunecoucheunitsweakenings{\Phi}{l_0}}} < \mes{\PortsofPPLPS{\Phi}}$ and there exists a unique $k$-experiment $(\enleverunecoucheunitsweakenings{e}{l_0}, \enleverunecoucheunitsweakenings{r}{l_0})$ of $\enleverunecoucheunitsweakenings{R}{l_0}$ such that
\begin{minilist}
\item for any $p \in (\ports{\PortsofPPLPS{\Phi}} \setminus \conclusions{\Phi}) \cap \ports{\PortsofPPLPS{\enleverunecoucheunitsweakenings{\Phi}{l_0}}}$, we have $\enleverunecoucheunitsweakenings{e}{l_0}(p) = e(p)$;
\item 
$$\enleverunecoucheunitsweakenings{r}{l_0}(i) = \left\lbrace \begin{array}{ll} r(i) & \textrm{if $i \not= \textsf{ind}(\mappriports{\PortsofPPLPS{\Phi}}(l_0))$;} \\ \beta & \textrm{if $i = \textsf{ind}(\mappriports{\PortsofPPLPS{\Phi}}(l_0))$.} \end{array} \right.$$
\end{minilist}
Moreover, if $e$ is atomic (resp. injective), then $\enleverunecoucheunitsweakenings{e}{l_0}$ is atomic (resp. injective).
\end{fact}

\begin{proof}
We set $\enleverunecoucheunitsweakenings{e}{l_0}(p) = e(p)$ for any $p \in \ports{\PortsofPPLPS{\enleverunecoucheunitsweakenings{\Phi}{l_0}}}$.
\end{proof}

\begin{fact}\label{fact : contrunit : syntaxe}
Let $k \in \mathbb{N}$. Let $R = (\Phi, \textsf{ind}) \in \LPSind$ such that $\Phi \notin \contrPLPS$ and let $(e, r)$ be a $k$-experiment of $R$. 
Let $l_0 \in\contrlinks{\PortsofPPLPS{\Phi}}\cap\terminallinks{\Phi}$ and $b \in \setfmulti{D'}$ such that $e( \mappriports{\PortsofPPLPS{\Phi}}(l_0)) = (-, \dig_1^k (b))$ and ${(\dig_1^k (b))}^\ast \not= \multi{}$. 
Then $\mes{\PortsofPPLPS{\enleverunecoucheunitsweakenings{\Phi}{l_0}}} < \mes{\PortsofPPLPS{\Phi}}$ and there exists a unique $k$-experiment $(\enleverunecoucheunitsweakenings{e}{l_0}, \enleverunecoucheunitsweakenings{r}{l_0})$ of $\enleverunecoucheunitsweakenings{R}{l_0}$ such that
\begin{minilist}
\item for any $p \in (\ports{\PortsofPPLPS{\Phi}} \setminus \conclusions{\Phi}) \cap \ports{\PortsofPPLPS{\enleverunecoucheunitsweakenings{\Phi}{l_0}}}$, we have $\enleverunecoucheunitsweakenings{e}{l_0}(p) = e(p)$;
\item 
$$\enleverunecoucheunitsweakenings{r}{l_0}(i) = \left\lbrace \begin{array}{ll} r(i) & \textrm{if $i \not= \textsf{ind}(\mappriports{\PortsofPPLPS{\Phi}}(l_0))$;} \\ (-, (\dig_1^k (b))^\textit{At} + b^\ast) & \textrm{if $i = \textsf{ind}(\mappriports{\PortsofPPLPS{\Phi}}(l_0))$.} \end{array} \right.$$
\end{minilist}
Moreover, if $e$ is atomic (resp. injective), then $\enleverunecoucheunitsweakenings{e}{l_0}$ is atomic (resp. injective).
\end{fact}

\begin{proof}
We set 
$\enleverunecoucheunitsweakenings{e}{l_0}(p) = \left\lbrace \begin{array}{ll} e(p) & \textrm{if $p \neq  \mappriports{\PortsofPPLPS{\Phi}}(l_0)$;}\\ 
(-, (\dig_1^k (b))^\textit{At} + b^\ast) & \textrm{if $p = \mappriports{\PortsofPPLPS{\Phi}}(l_0)$.}
\end{array} \right.$
\end{proof}

The following definition extends the notion of isomorphism of $k$-experiments of PLPS to $k$-experiments of indexed PLPS. The proof of Theorem \ref{theorem:injectLPS} will use the obvious fact that, by definition, for any $k$-experiment $(e, r)$ of $(\Phi, \textsf{ind})$, for any $k$-experiment $(e', r')$ of $(\Phi', \textsf{ind'})$, we have $(e, r) \simeq_\textit{At} (e', r') \Rightarrow (\Phi, \textsf{ind}) \simeq (\Phi', \textsf{ind'})$.

\begin{defin}\label{def:isoExpInd}
Let $k \in \mathbb{N}$. Let $(\Phi, \textsf{ind}), (\Phi', \textsf{ind'}) \in \PLPSind$. Let $(e, r)$ be a $k$-experiment of $(\Phi, \textsf{ind})$ and let $(e', r')$ be a $k$-experiment of $(\Phi', \textsf{ind'})$. 
\begin{minilist}
\item We write $\varphi : (e, r) \simeq (e', r')$ if, and only if, $\varphi : e \simeq e'$ and $r = r'$.
\item We write $\varphi : (e, r) \simeq_\textit{At} (e', r')$ if, and only if, there exist $\rho, \rho' \in \pInj$ such that $\varphi : (\rho \cdot e, \rho \cdot r) \simeq (\rho' \cdot e', \rho' \cdot r')$.
\end{minilist}
\end{defin}

Facts~\ref{fact : enleverunecouche a un resultat}, \ref{fact : bangunit : syntaxe} and \ref{fact : contrunit : syntaxe} allow to obtain a $k$-experiment $\enleverunecouche{e}$ of $\enleverunecouche{\Phi}$ and a $k$-experiment $\enleverunecoucheunitsweakenings{e}{l_0}$ of $\enleverunecoucheunitsweakenings{\Phi}{l_0}$ from a $k$-experiment $e$ of a LPS $\Phi$. This will be used in the proof of Proposition~\ref{prop : KeyProposition} to apply the induction hypothesis (since the measure of $\enleverunecouche{\Phi}$ and $\enleverunecoucheunitsweakenings{\Phi}{l_0}$ is strictly smaller than the one of $\Phi$): starting from two experiments $(e,r)$ of $(\Phi,\textsf{ind})$ and $(e',r')$ of $(\Phi',\textsf{ind}')$ such that $r=r'$, we will be able to conclude that $(\enleverunecouche{e},\enleverunecouche{r})\simeq_\textit{At}(\enleverunecouche{e'},\enleverunecouche{r'})$ and $(\enleverunecoucheunitsweakenings{e}{l_0},\enleverunecoucheunitsweakenings{r}{l_0})\simeq_\textit{At}(\enleverunecoucheunitsweakenings{e'}{l'_0},\enleverunecoucheunitsweakenings{r'}{l'_0})$. However, what we want to prove is that $(e,r)\simeq_\textit{At}(e',r')$ (and thus $\Phi\simeq\Phi'$), and for this last step we will use the three following facts concluding this subsection.

\begin{fact}\label{fact : cbox : enleverunecouche de e = enleverunecouche de e' => e = e'}
Let $k \in \mathbb{N}$. Let $R = (\Phi, \textsf{ind}), R' = (\Phi', \textsf{ind'}) \in \LPSind$ such that $\Phi, \Phi' \in \cboxPLPS$ and let $(e, r)$ (resp. $(e', r')$) be a $k$-experiment of $R$ (resp. $R'$). Assume that $(\enleverunecouche{e}, \enleverunecouche{r}) \simeq_\textit{At} (\enleverunecouche{e'}, \enleverunecouche{r'})$. Then we have $(e, r) \simeq_\textit{At} (e', r')$.
\end{fact}

\begin{proof}
Let $\varphi_0=({\varphi_0}_\mathcal{C},{\varphi_0}_\mathcal{P})$, $\rho_0$ and $\rho'_0$ such that $\varphi_0 : (\rho_0 \cdot \enleverunecouche{e}, \rho_0 \cdot \enleverunecouche{r}) \simeq (\rho'_0 \cdot \enleverunecouche{e'}, \rho'_0 \cdot \enleverunecouche{r'})$. 
Let $\psi : \bangs{\PortsofPPLPS{\Phi}} \cap \terminallinks{\Phi} \rightarrow \bangs{\PortsofPPLPS{\Phi'}} \cap \terminallinks{\Phi'}$ defined by $\psi(l_0) = l_0'$ with $\textsf{ind'}(\mappriports{\PortsofPPLPS{\Phi}}(l'_0))=\textsf{ind}(\mappriports{\PortsofPPLPS{\Phi}}(l_0))$.
Then we have $\varphi=(\varphi_\mathcal{C},\varphi_\mathcal{P}) : (\rho \cdot e, \rho \cdot r) \simeq_\textit{At} (\rho' \cdot e', \rho' \cdot r')$, where $\varphi$ is defined as follows:
\begin{itemize}
\item 
$\varphi_\mathcal{C}(l) = \left\lbrace \begin{array}{ll} {\varphi_0}_\mathcal{C}(l) & \textrm{if $l \notin \bangs{\PortsofPPLPS{\Phi}} \cap \terminallinks{\Phi}$;} \\ \psi(l) & \textrm{if $l \in \bangs{\PortsofPPLPS{\Phi}} \cap \terminallinks{\Phi}$;} \end{array} \right.$ 
\item and $\varphi_\mathcal{P}(p) = \left\lbrace \begin{array}{ll} 
{\varphi_0}_\mathcal{P}(p) & \textrm{if there is no $l_0 \in \bangs{\PortsofPPLPS{\Phi}} \cap \terminallinks{\Phi}$ such that $p \in \portsoflink{\PortsofPPLPS{\Phi}} (l_0)$;} \\ 
 \mappriports{\PortsofPPLPS{\Phi'}}(\psi(l_0)) & \textrm{if $p =  \mappriports{\PortsofPPLPS{\Phi}}(l_0)$ with $l_0 \in \bangs{\PortsofPPLPS{\Phi}} \cap \terminallinks{\Phi}$;}\\
q' \textrm{, where} & 
\left\lbrace
\begin{array}{l}
\textrm{$\{ q' \} = \mapauxports{\PortsofPPLPS{\Phi'}}(\psi(l_0))$,}\\ 
\textrm{if $\{ p \} = \mapauxports{\PortsofPPLPS{\Phi}}(l_0)$ with $l_0 \in \bangs{\PortsofPPLPS{\Phi}} \cap \terminallinks{\Phi}$;}
\end{array}
\right.
\end{array} \right.$
\end{itemize}
and $\rho, \rho' \in \pInj$ are defined as follows: \\
$\rho(\delta) = \left\lbrace \begin{array}{ll} \rho_0(\delta) & \textrm{if $\delta \in \textit{At'}(\im{\overline{e}})$;} \\
\dig(j) \cdot {\delta_0}' & \textrm{if $\delta = \dig(j) \cdot \delta_0$, $\delta_0 \in \textit{At'}(\im{\overline{e}})$ and $\dig(j) \cdot \delta_0 \notin \textit{At'}(\im{\overline{e}})$;}
\end{array} \right.$ 
and \\
$\rho'(\delta) = \left\lbrace \begin{array}{ll} \rho'_0(\delta) & \textrm{if $\delta \in \textit{At'}(\im{\overline{e'}})$;} \\
\dig(j) \cdot {\delta_0}' & \textrm{if $\delta = \dig(j) \cdot \alpha$, $\rho_0(\delta_0) = \rho'_0(\alpha)$, 
$\alpha \in \textit{At'}(\im{\overline{e'}})$ and $\dig(j) \cdot \alpha \notin \textit{At'}(\im{\overline{e'}})$;}
\end{array} \right.$ 
where, for any $\delta_0 \in \textit{At'}(\im{\overline{e}})$ such that $\dig(1) \cdot \delta_0 \notin \textit{At'}(\im{\overline{e}})$\footnote{The reader certainly noticed that $\dig(1) \cdot \delta_0 \notin \textit{At'}(\im{\overline{e}})$ iff $\dig(j) \cdot \delta_0 \notin \textit{At'}(\im{\overline{e}})$ for every $j\in\integer{k}$.}, we have chosen $\delta'_0 \in A'$ such that $\dig(1) \cdot \delta'_0, \ldots, \dig(k) \cdot \delta'_0 \notin \textit{At'}(\im{\overline{e}}) \cup \textit{At'}(\im{\overline{e'}})$.
\end{proof}

Like for Fact~\ref{fact : cbox : enleverunecouche de e = enleverunecouche de e' => e = e'}, also in Facts~\ref{fact : bangunit : e[l]=e'[l] => e=e'} and~\ref{fact : contrunit : e[l]=e'[l] => e=e'} some ``new'' substitutions ($\rho, \rho'$ in the proof of Fact~\ref{fact : cbox : enleverunecouche de e = enleverunecouche de e' => e = e'}) have to be constructed from ``existing'' ones ($\rho_0, \rho'_0$ in the proof of Fact~\ref{fact : cbox : enleverunecouche de e = enleverunecouche de e' => e = e'}). However for Facts~\ref{fact : bangunit : e[l]=e'[l] => e=e'} and~\ref{fact : contrunit : e[l]=e'[l] => e=e'} we can just use the existing ones\footnote{With the notations of the proof of Fact~\ref{fact : cbox : enleverunecouche de e = enleverunecouche de e' => e = e'}, we have $\rho=\rho_0$ and $\rho'=\rho'_{0}$.} since there is no difference between the atoms of the experiment $(e,r)$ of $(\Phi, \textsf{ind})$ and the atoms of the experiment $(\enleverunecoucheunitsweakenings{e}{l_0},\enleverunecoucheunitsweakenings{r}{l_0})$ of $(\enleverunecoucheunitsweakenings{\Phi}{l_0},\enleverunecoucheunitsweakenings{\textsf{ind}}{l_0})$: more precisely $\textit{At'}(r)=\textit{At'}(\enleverunecoucheunitsweakenings{r}{l_0})$.

\begin{fact}\label{fact : bangunit : e[l]=e'[l] => e=e'}
Let $k \in \mathbb{N}$. Let $R = (\Phi, \textsf{ind}), R' = (\Phi', \textsf{ind'}) \in \LPSind$ and let $(e, r)$ (resp. $(e', r')$) be a $k$-experiment of $R$ (resp. $R'$). 
Let $l_0 \in\bangs{\PortsofPPLPS{\Phi}}\cap\terminallinks{\Phi}$ and $\beta \in D'$ such that $e(\mappriports{\PortsofPPLPS{\Phi}}(l_0)) = (+, \dig_1^k (\multi{\beta}))$ and ${(\dig_1^k(\multi{\beta}))}^\ast \not= \multi{}$. 
Let $l'_0\in\bangs{\PortsofPPLPS{\Phi'}}\cap\terminallinks{\Phi'}$ be such that $\textsf{ind'}(\mappriports{\PortsofPPLPS{\Phi'}}(l'_0))=\textsf{ind}(\mappriports{\PortsofPPLPS{\Phi}}(l_0))$.
Assume that $(\enleverunecoucheunitsweakenings{e}{l_0}, \enleverunecoucheunitsweakenings{r}{l_0}) \simeq_\textit{At} (\enleverunecoucheunitsweakenings{e'}{l_0'}, \enleverunecoucheunitsweakenings{r'}{l_0'})$. Then we have $(e, r) \simeq_\textit{At} (e', r')$.
\end{fact}

\begin{proof}
Let $\varphi_0=({\varphi_0}_\mathcal{C},{\varphi_0}_\mathcal{P}) : (\enleverunecoucheunitsweakenings{e}{l_0}, \enleverunecoucheunitsweakenings{r}{l_0}) \simeq_\textit{At} (\enleverunecoucheunitsweakenings{e'}{l_0'}, \enleverunecoucheunitsweakenings{r'}{l_0'})$. Then we have $\varphi=(\varphi_\mathcal{C},\varphi_\mathcal{P}) : (e, r) \simeq_\textit{At} (e', r')$, where 
$\varphi_\mathcal{C}(l) = \left\lbrace \begin{array}{ll} {\varphi_0}_\mathcal{C}(l) & \textrm{if $l \not= l_0$;} \\ l_0' & \textrm{if $l = l_0$;} \end{array} \right.$ and 
$$\varphi_\mathcal{P}(p) = \left\lbrace \begin{array}{ll} 
{\varphi_0}_\mathcal{P}(p) & \textrm{if $p \notin\portsoflink{\PortsofPPLPS{\Phi}}(l_0)$;} \\ 
\mappriports{\PortsofPPLPS{\Phi'}}(l_0') & \textrm{if $p = \mappriports{\PortsofPPLPS{\Phi}}(l_0)$;}\\
q' & \textrm{,where $\{ q' \} = \mapauxports{\PortsofPPLPS{\Phi'}}(l_0')$, if $\{ p \} = \mapauxports{\PortsofPPLPS{\Phi}}(l_0)$.}
\end{array} \right.$$
\end{proof}

\begin{fact}\label{fact : contrunit : e[l]=e'[l] => e=e'}
Let $k \in \mathbb{N}$. Let $R = (\Phi, \textsf{ind}), R' = (\Phi', \textsf{ind'}) \in \LPSind$ such that $\Phi, \Phi' \notin \contrPLPS$ and let $(e, r)$ (resp. $(e', r')$) be a $k$-experiment of $R$ (resp. $R'$). 
Let $l_0 \in\contrlinks{\PortsofPPLPS{\Phi}}\cap\terminallinks{\Phi}$ and $b \in \setfmulti{D'}$ such that $e(\mappriports{\PortsofPPLPS{\Phi}}(l_0)) = (-, \dig_1^k (b))$ and ${(\dig_1^k (b))}^\ast \not= \multi{}$. 
Let $l'_0\in\contrlinks{\PortsofPPLPS{\Phi'}}\cap\terminallinks{\Phi'}$ be such that $\textsf{ind'}(\mappriports{\PortsofPPLPS{\Phi'}}(l'_0))=\textsf{ind}(\mappriports{\PortsofPPLPS{\Phi}}(l_0))$.
Assume that $(\enleverunecoucheunitsweakenings{e}{l_0}, \enleverunecoucheunitsweakenings{r}{l_0}) \simeq_\textit{At} (\enleverunecoucheunitsweakenings{e'}{l_0'}, \enleverunecoucheunitsweakenings{r'}{l_0'})$. Then we have $(e, r) \simeq_\textit{At} (e', r')$.
\end{fact}

\begin{proof}
Let $\varphi=(\varphi_\mathcal{C},\varphi_\mathcal{P}) : (\enleverunecoucheunitsweakenings{e}{l_0}, \enleverunecoucheunitsweakenings{r}{l_0}) \simeq_\textit{At} (\enleverunecoucheunitsweakenings{e'}{l_0'}, \enleverunecoucheunitsweakenings{r'}{l_0'})$. Then we have $\varphi : (e, r) \simeq_\textit{At} (e', r')$. Indeed: 
let $b_0 = \sum_{p \in \mapauxports{\PortsofPPLPS{\Phi}}(l_0)} e(p)$ and $b'_0 = \sum_{p' \in \mapauxports{\PortsofPPLPS{\Phi'}}(l_0')} e'(p')$ ; then for any $p \in\mapauxports{\PortsofPPLPS{\Phi}}(l_0)$, we have $e(p) \in \textit{Supp}({b_0}^\ast)$ if, and only if, $e'(\varphi_\mathcal{P}(p)) \in \textit{Supp}({b'_0}^\ast)$, hence $\paxnumber{\PortsofPPLPS{\Phi}}(p) = \paxnumber{\PortsofPPLPS{\Phi'}}(\varphi_\mathcal{P}(p))$.
\end{proof}

\subsection{Main result}\label{subsect:MainResult}

Thanks to the previous sections, we can reduce our main result to the following proposition 
concerning only LPS (and not PS anymore). 
This crucial proposition will be proven by induction on $\mes{\PortsofPPLPS{\Phi}}$, the most delicate cases being $\Phi \in \contrPLPS$ and $\Phi \in \cboxPLPS$.

\begin{prop}\label{prop : KeyProposition}\newcounter{BB}\setcounter{BB}{\value{prop}}
Let $(\Phi, \textsf{ind}), (\Phi', \textsf{ind'}) \in \LPSind$. For any $k > \cosize{\PortsofPPLPS{\Phi}}, \cosize{\PortsofPPLPS{\Phi'}}$, for any $k$-experiment $(e, r)$ of $(\Phi, \textsf{ind})$, for any $k$-experiment $(e', r')$ of $(\Phi', \textsf{ind'})$, $e$ and $e'$ atomic and injective, if there exist $\rho, \rho' \in \pInj$ such that $\rho \cdot r = \rho' \cdot r'$, then $(e, r) \simeq_{\textit{At}} (e', r')$.
\end{prop}

An injective atomic $k$-experiment of an LPS $\Phi$ can be considered as a ``prototype'' of (atomic) $k$-experiment of \emph{any} PS $(\Phi,b)$.\footnote{\label{footnote : k-experiment of PS}Notice that we did not define $k$-experiments of PS but only of LPS: $k$-experiments of nets have been defined in \cite{lics06} and by \emph{(injective) $k$-experiment of a PS} we mean here an experiment having a(n injective) $k$-point as result. A $k$-experiment of a PS $R$ is said to be \emph{atomic} if for any $p \in \bigcup \axioms{\LPSofPPS{R}}$, we have $\textit{Supp}(e(p)) \subseteq A$. \newcounter{BC}\setcounter{BC}{\value{footnote}}} Indeed, every $k$-point of $\sm{(\Phi,b)}_\textit{At}$ can be obtained from the result of an injective atomic $k$-experiment of $\Phi$: to be precise, for $(R, \textsf{ind}) \in \PSind$ we have 
\begin{eqnarray*}
& & \{ r_0 \in \sm{(R, \textsf{ind})}_\textit{At} \: / \: r_0 \textrm{ is a $k$-point} \} \\
& = & \bigcup_{(e, r) \textrm{is an injective atomic }k-\textrm{experiment of }(\LPSofPPS{R}, \textsf{ind})} \{ \rho \cdot r \: / \: \rho \textrm{ is a partial map from $A'$ to $A$} \} \: ,
\end{eqnarray*}
where $\rho \cdot r$ is defined by a straightforward generalization of Definition \ref{definition : actions on points}. In our proof we will only use Fact~\ref{fact:PSandLPS}, namely that for a PS $R=(\Phi,b)$, the restriction of $\sm{R}$ to the injective $k$-points which are $\sm{R}$-atomic is precisely the set of the results of the atomic injective $k$-experiments of $\Phi$ (up to the name of the atoms): 

\begin{fact}\label{fact:PSandLPS}
Let $k \in \mathbb{N}$ and let $(R, \textsf{ind}) \in \PSind$. We have $\{ r_0 \in \sm{(R, \textsf{ind})}_\textit{At} \: / \: r_0 \textrm{ is an injective $k$-point} \} = \bigcup_{(e, r) \textrm{is an injective atomic }k-\textrm{experiment of }(\LPSofPPS{R}, \textsf{ind})} \{ \rho \cdot r \: / \: \rho \in \pInj \textrm{ and } \codom{\rho} = A \}$.
\end{fact}

\begin{proof}
One of the two inclusions is easy to prove: given an injective atomic $k$-experiment $(e, r)$ of $(\LPSofPPS{R}, \textsf{ind})$ and given $\rho \in \pInj$ such that $\codom{\rho} = A$, there is an experiment $(e_{\rho},r_0)$ of $(R, \textsf{ind})$ such that $r_0=\rho\cdot r$. The experiment $(e_{\rho},r_0)$ of $(R, \textsf{ind})$ can be defined by induction on $(\LPSofPPS{R}, \textsf{ind})$ (see also Example~\ref{example:ExpPS-LPS}).

Conversely, let $r_{0}\in \sm{(R, \textsf{ind})}_\textit{At}$ be an injective $k$-point and let $(e_{0},r_{0})$ be an experiment of $(R, \textsf{ind})$. We prove that for every atomic injective $k$-experiment $(e,r)$ of $(\LPSofPPS{R}, \textsf{ind})$, there exists $\rho\in\pInj$ such that $\im{\rho}\subseteq\textit{At'}(r_{0})$ and $\rho\cdot r=r_{0}$: this immediately yields the missing inclusion. The proof is by induction on $\mes{\PortsofPPLPS{\LPSofPPS{R}}}$ (see Definition~\ref{def:measure}), the unique case deserving some details being the one where there is a unique terminal $!$-cell $v$ of $R$ and every other terminal cell is a $?$-cell having a unique auxiliary port which is an element of $\mapbangauxd{R}(v)$\footnote{In the standard terminology of linear logic proof-nets one would say that $R$ is an exponential box.}. The situation is represented in Figure~\ref{fig:PSandLPS}.
\begin{figure}
\begin{center}
\includegraphics[width=12cm,height=6cm]{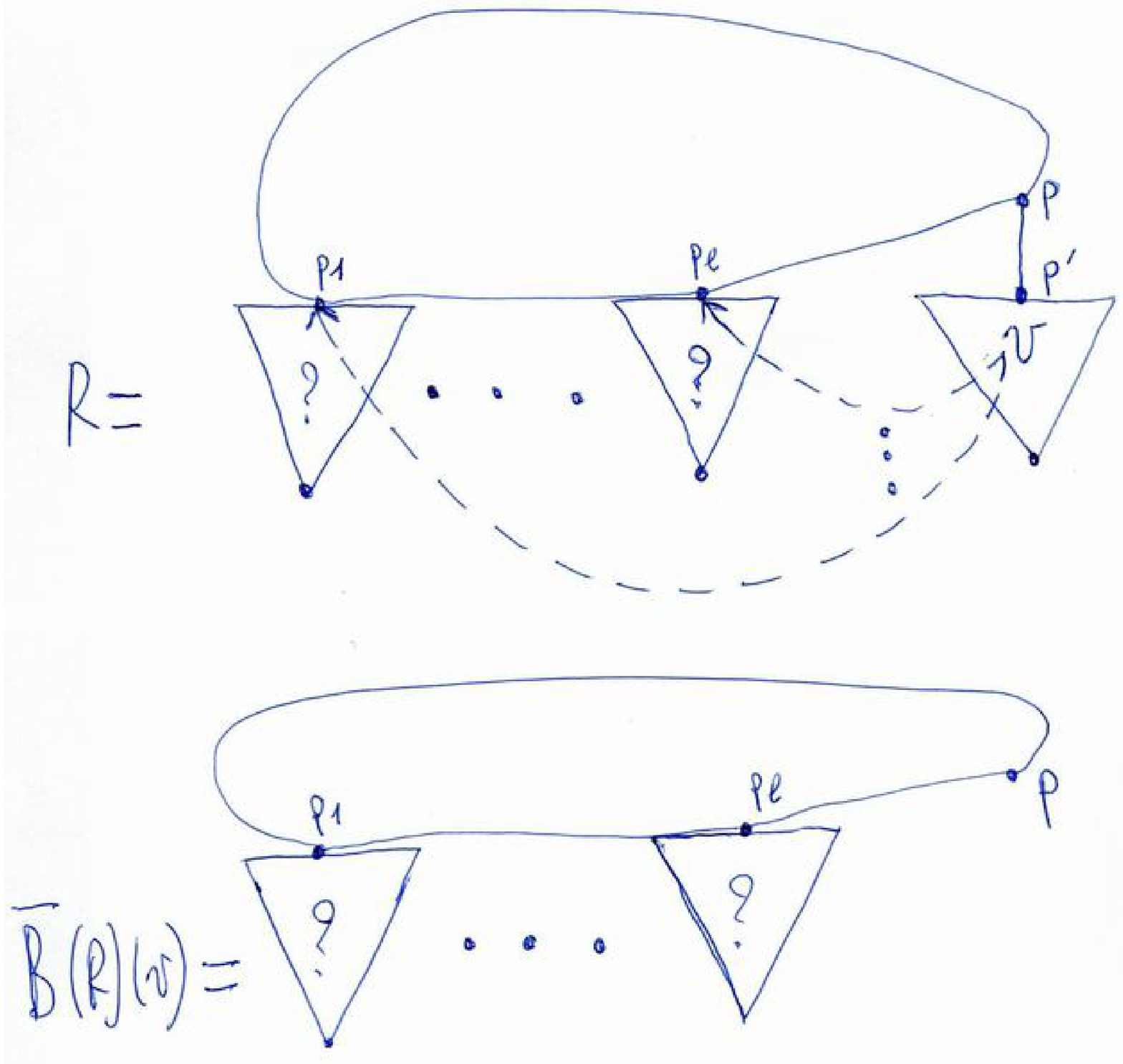}
\caption{\textbf{The critical case of Fact~\ref{fact:PSandLPS}.}\label{fig:PSandLPS} We have $p=p'$ if, and only if, $p' \in \bigcup \axioms{\LPSofPPS{R}}$.
}
\end{center}
\end{figure}
We set $\{p_{1},\ldots,p_{l}\}=\mapbangauxd{R}(v)$, we call $\overline{B}(R)(v)$ the box of $v$ (we still denote by  $\textsf{ind}$ the obvious bijection $\conclusions{\LPSofPPS{\overline{B}(R)(v)}} \simeq \integer{\textit{Card}(\conclusions{\LPSofPPS{\overline{B}(R)(v)}}}$) and we call $p$ the unique free port of $\overline{B}(R)(v)$ such that $\mappriports{\PortsofPPLPS{\LPSofPPS{R}}}(v) \leq_{\LPSofPPS{R}} p$.

In the sequel of the proof, it is important to distinguish between experiments of PS (Definition~\ref{def:expPS}) and $k$-experiments of LPS (Definition~\ref{def:experiment}): the experiments of PS have $0$ as index ($e_{0}$ and $f_{0}^{i}$), while all the others are $k$-experiments of LPS.\\ 
Let $e_{0}(v)=\multi{\multi{f_{0}^{1},\ldots,f_{0}^{1}}}$, where $(f_{0}^{i},r_{0}^{i})$ is an experiment of $(\overline{B}(R)(v), \textsf{ind})$. Clearly, $r_{0}^{i}\in \sm{(\overline{B}(R)(v), \textsf{ind})}_\textit{At}$ is an injective $k$-point. The restriction $(f,s)$ of $(e,r)$ to $\LPSofPPS{\overline{B}(R)(v)}$ is an atomic injective $k$-experiment of $(\LPSofPPS{\overline{B}(R)(v)}, \textsf{ind})$. We can then apply the induction hypothesis: for every $i\in\integer{k}$ there exists $\rho_{i}\in\pInj$ such that $\im{\rho_{i}}\subseteq\textit{At'}(r_{0}^{i})$ and $\rho_{i}\cdot s=r_{0}^{i}$\footnote{Notice that for every $i\in\integer{k}$ one has $\textit{At'}(s)\subseteq\dom{\rho_{i}}$.}.

Since $\im{\rho_{i}}\subseteq\textit{At'}(r_{0}^{i})$ and since $r_{0}$ is injective, one has $\textit{At'}(r_{0}^{i})\cap\textit{At'}(r_{0}^{j})=\emptyset$ when $i\neq j$ and thus $\im{\rho_{i}}\cap\im{\rho_{j}}=\emptyset$ when $i\neq j$. We can then define $\rho\in\pInj$ on the elements $\gamma\in \textit{At'}(r)$: since for every such $\gamma$ there exist a unique $i \in \integer{k}$ and a unique $\beta \in \textit{At'}(s)$ such that $\gamma = \dig(i)(\beta)$, we can set $\rho(\gamma) = \rho_i(\beta)$.

We now check that $\rho$ is indeed the function we look for. With the notations introduced we have:
\begin{itemize}
\item
$r_{0}=((-,\sum_{i=1}^k f_{0}^{i}(p_{1})),\ldots,(-,\sum_{i=1}^k f_{0}^{i}(p_{l})),(+,\sum_{i=1}^k f_{0}^{i}(p)))$
\item
$r_{0}^{i}=((-,f_{0}^{i}(p_{1})),\ldots,(-,f_{0}^{i}(p_{l})),\beta_{i})$, where $f_{0}^{i}(p)=\multi{\beta_{i}}$, for every $i\in\integer{k}$
\item
$s=((-,\multi{f(p_{1})}),\ldots,(-,\multi{f(p_{l})}),f(p))$
\item
$r=((-,\dig_1^k(\multi{f(p_{1})})),\ldots,(-,\dig_1^k(\multi{f(p_{l})})),(+,\dig_1^k(\multi{f(p)})))$.
\end{itemize}

Now notice that for every $j\in\integer{l}$ we have $\dig_1^k(\multi{f(p_{j})})= \sum_{i=1}^k \multi{\dig(i)\cdot f(p_{j})}$; and, since we have $\textit{At'}(f(p_{j}))\subseteq\textit{At'}(s)$, we can deduce for every $\beta\in\textit{At'}(f(p_{j}))$ and for every $i \in \integer{k}$ that $\dig(i)(\beta)\in\dom{\rho}$ and $\rho(\dig(i)(\beta))=\rho_{i}(\beta)$. This entails that for every $j\in\integer{l}$ one has $\rho\cdot \dig_1^k(\multi{f(p_{j})})= \sum_{i=1}^k \multi{\rho\cdot (\dig(i)\cdot f(p_{j}))}= \sum_{i=1}^k \multi{\rho_{i}\cdot f(p_{j})} = \sum_{i=1}^k \rho_i \cdot \multi{f(p_j)}$. In the same way, we have $\rho\cdot \dig_1^k(\multi{f(p)})= \sum_{i=1}^k \multi{\rho\cdot (\dig(i)\cdot f(p))}= \sum_{i=1}^k \multi{\rho_{i}\cdot f(p)}$. Then the following equalities hold:\\
$\rho\cdot r=((-,\rho\cdot\dig_1^k(\multi{f(p_{1})})),\ldots,(-,\rho\cdot\dig_1^k(\multi{f(p_{l})})),(+,\rho\cdot\dig_1^k(\multi{f(p)})))$\\
$=((-,\sum_{i=1}^k \rho_{i}\cdot\multi{f(p_{1})}),\ldots,(-,\sum_{i=1}^k \rho_{i}\cdot\multi{f(p_{l})}),(+, \sum_{i=1}^k \multi{\rho_i \cdot f(p)}))$\\
$=((-,\sum_{i=1}^k f_{0}^{i}(p_{1})),\ldots,(-,\sum_{i=1}^k f_{0}^{i}(p_{l})),(+,\sum_{i=1}^k f_{0}^{i}(p)))=r_{0}$.
\end{proof}

\begin{example}\label{example:ExpPS-LPS}
Consider the LPS $\Psi_2$ of Figure~ \ref{example : LPS}. If we take $\gamma_{1}\neq\gamma_{2}$, then the experiment $(e_{2},r_{2})$ considered in Example~\ref{example : experiment} is an injective atomic $3$-experiment of $(\Psi_2,\textsf{ind}_2)$. Let $\rho \in \pInj$ be such that for $j\in\integer{2}$ and $i\in\integer{3}$ one has $\rho(\gamma_j,i)=\gamma_{ji}$, where $\gamma_{ji}\in A$ (since $\rho \in \pInj$ the $\gamma_{ji}$s are pairwise different). Then for any\footnote{Corollary~\ref{cor:injectConnected} shows that in this particular case ($\Psi_2$ is a connected graph) there is actually a unique PS $R$ such that $\LPSofPPS{R}=\Psi_2$.} PS $R$ such that $\LPSofPPS{R}=\Psi_2$, there exists an experiment $e_0=(e_2)_{\rho}$ of $R$ with result $r_{0}=\rho\cdot r_{2}=((-,[\gamma_{11},\gamma_{12},\gamma_{13},\gamma_{21},\gamma_{22},\gamma_{23}]),(+,[(+,\gamma_{11},\gamma_{21}),$ $(+,\gamma_{12},\gamma_{22}),$ $(+,\gamma_{13},\gamma_{23})]))$. Indeed, if we call $v$ the unique $!$-cell of $R$, we can set $e_0(v)=[[f_1,f_2,f_3]]$, where $f_i$ is the experiment of $v$'s box obtained by setting $f_i(p_1)=[\gamma_{2i}]$ and $f_i(p_2)=[\gamma_{1i}]$ (which entirely determines $f_i$). One can easily check that $r_0$ is indeed $e_0$'s result.
\end{example}

\begin{thm}\label{theorem:injectLPS}
Let $(R, \textsf{ind}), (R', \textsf{ind'}) \in \PSind$. Let $k > \cosize{\PortsofPPLPS{\LPSofPPS{R}}},$ $\cosize{\PortsofPPLPS{\LPSofPPS{R'}}}$. If $\{ r_0 \in \sm{(R, \textsf{ind})}_\textit{At} \: / \: r_0 \textrm{ is an injective $k$-point} \} \cap \{ r_0 \in \sm{(R', \textsf{ind'})}_\textit{At} \: / \: r_0 \textrm{ is an injective $k$-point} \} \not= \emptyset$, then $(\LPSofPPS{R}, \textsf{ind}) \simeq (\LPSofPPS{R'}, \textsf{ind'})$.
\end{thm}

\begin{proof}
Let $r_0$ be an injective $\sm{(R, \textsf{ind})}$-atomic $k$-point of $\sm{(R, \textsf{ind})}$ which is also an injective $\sm{(R', \textsf{ind'})}$-atomic $k$-point of $\sm{(R', \textsf{ind'})}$. By Fact~\ref{fact:PSandLPS}, there exists an injective atomic $k$-experiment $(e, r)$ (resp.\ $(e', r')$) of $(\LPSofPPS{R}, \textsf{ind})$ (resp.\ $(\LPSofPPS{R'}, \textsf{ind'})$) and $\rho \in \pInj$ (resp.\ $\rho' \in \pInj$) such that $\rho \cdot r = r_0 = \rho' \cdot r'$. By Proposition~\ref{prop : KeyProposition} we thus have $(e, r) \simeq_{\textit{At}} (e', r')$ which implies $(\LPSofPPS{R}, \textsf{ind}) \simeq (\LPSofPPS{R'}, \textsf{ind'})$.
\end{proof}

\begin{remark}
Of course, as illustrated by Figure~\ref{figure: LPS versus PS}, there are different PS with the same LPS. The $k$-experiments of two PS\footnote{See Footnote~\ref{footnote : k-experiment of PS}.\newcounter{CD}\setcounter{CD}{\value{footnote}}} have the same results if, and only if, the PS have the same LPS, but we do not say anything about the results of the other experiments.

\begin{figure}
  \centering
    \includegraphics[width=13cm,height=3cm]{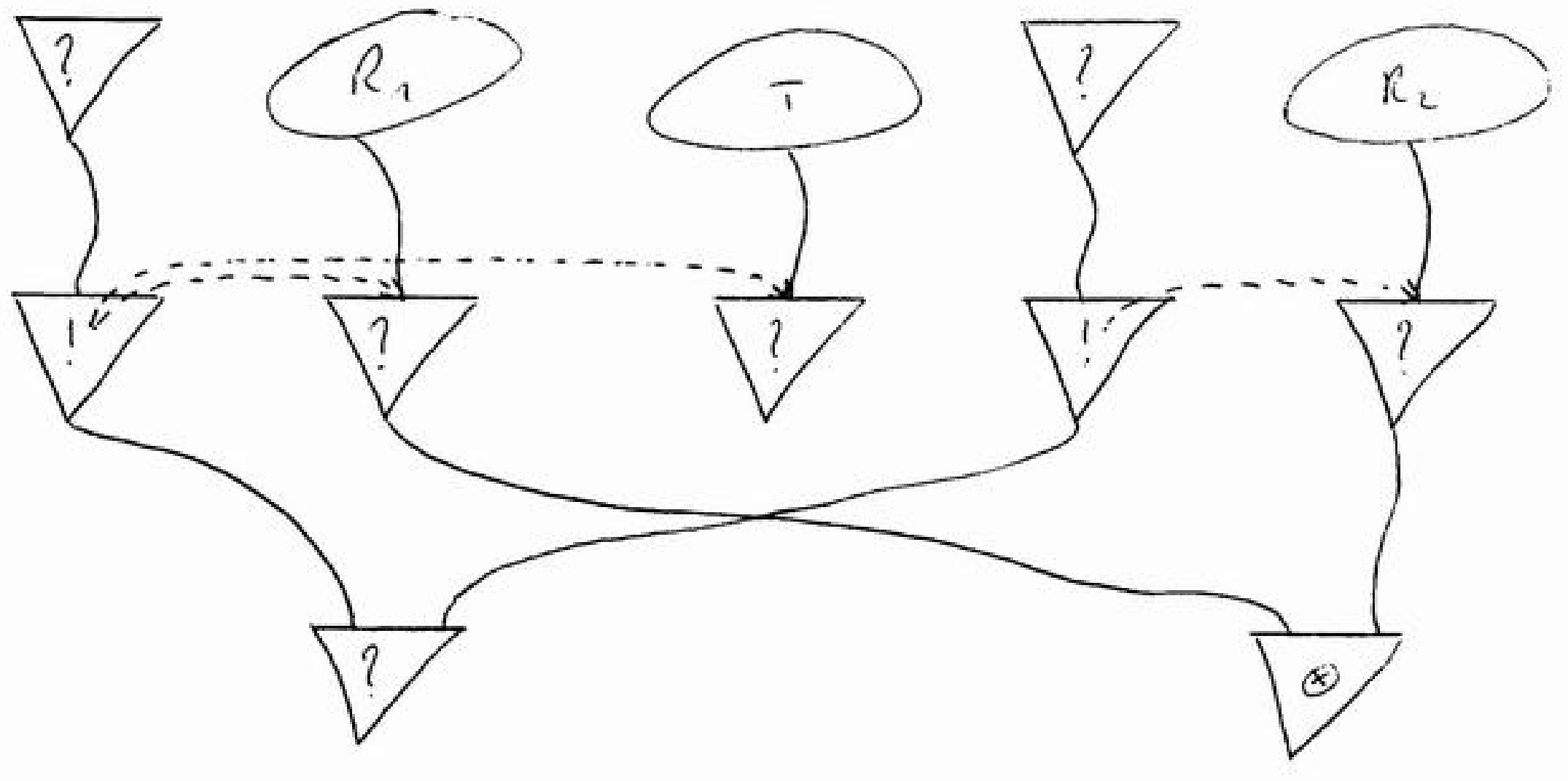}  
    \includegraphics[width=13cm,height=3cm]{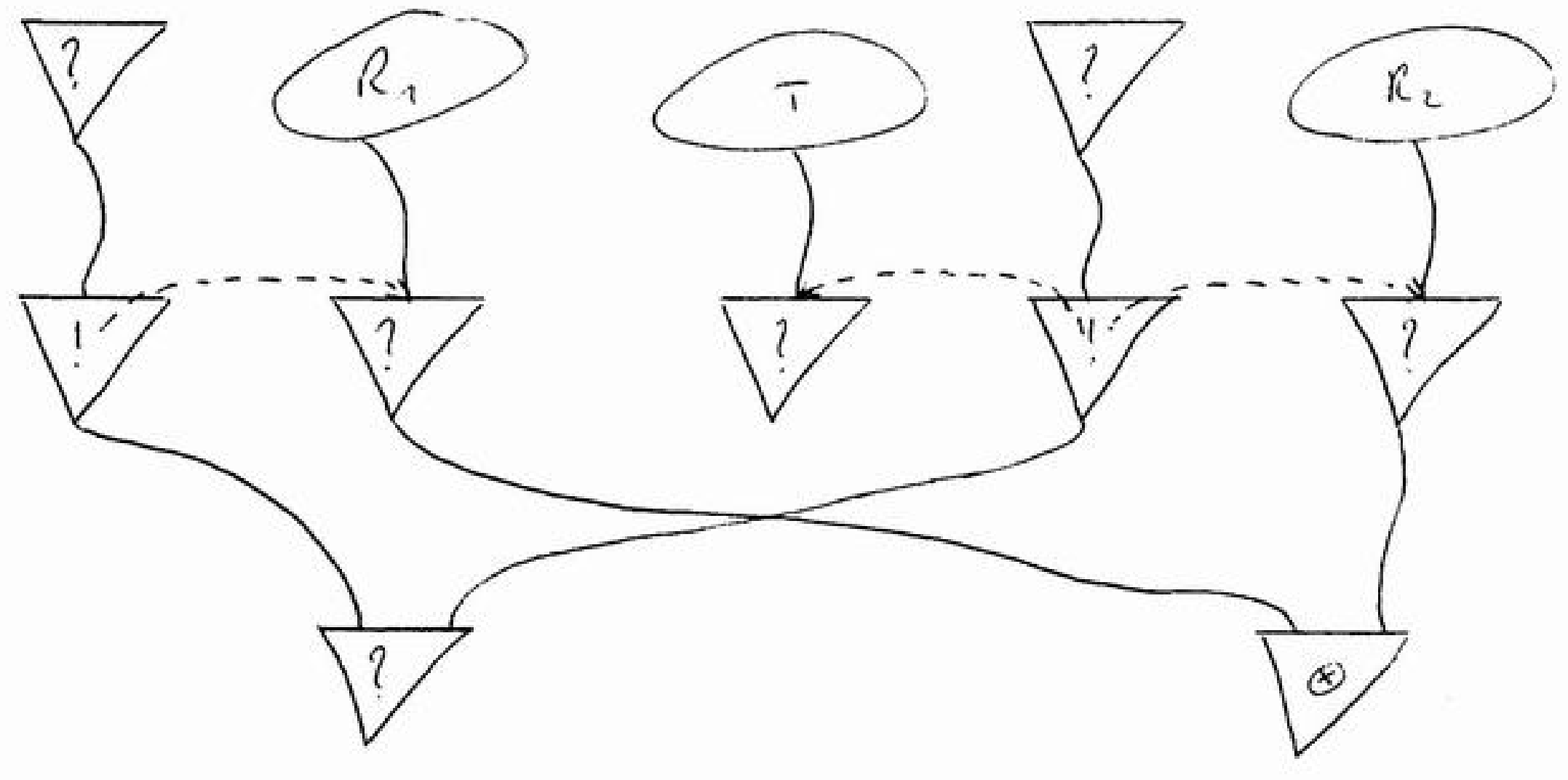}  
  \caption{\textbf{Two different PS with the same LPS.} The PS $R_1$, $R_2$ and $T$ are PS of depth $0$.\label{figure: LPS versus PS}}
  \label{Labelname}
\end{figure}
\end{remark}

\begin{cor}\label{cor:injectLPS}
Assume $A$ is infinite. Let $(R, \textsf{ind}), (R', \textsf{ind'}) \in \PSind$. If $\sm{(R, \textsf{ind})} = \sm{R', \textsf{ind'})}$, then $(\LPSofPPS{R}, \textsf{ind}) \simeq (\LPSofPPS{R'}, \textsf{ind'})$.
\end{cor}

\begin{proof}
Since $A$ is infinite, one has $\{ r_0 \in \sm{(R, \textsf{ind})}_\textit{At} \: / \: r_0 \textrm{ is an injective $k$-point} \} \cap \{ r_0 \in \sm{(R', \textsf{ind'})}_\textit{At} \: / \: r_0 \textrm{ is an injective $k$-point} \} \not= \emptyset$. Apply Theorem~\ref{theorem:injectLPS}.
\end{proof}

\begin{remark}
In the proof of Corollary~\ref{cor:injectLPS}, we use the fact that there always exists an $\sem{R}$-atomic injective $k$-point in the interpretation of any PS $R$ and thus there always exists an atomic injective $k$-experiment of $R$\footnotemark[\value{CD}]. It is worth noticing that such an atomic injective $k$-experiment\footnotemark[\value{CD}] is unique ``up to the names of the atoms''.

The reader acquainted with \emph{injective $k$-obsessional experiments} (see~\cite{phdtortora,injectcoh}) knows that, in the coherent model, not every PS has an injective $k$-obsessional experiment: this is precisely the reason why the proof of injectivity of the coherent model given in~\cite{phdtortora,injectcoh} for the $(?\wp)$LL fragment (already mentioned in the introduction) cannot be extended to $MELL$; and still for that reason injectivity of the coherent model fails for $MELL$ as shown in~\cite{phdtortora,injectcoh}.
\end{remark}

The following corollary is based on a simple and crucial remark, already used in~\cite{injectcoh} (for the same purpose): since in LPS the depth of every port is known, given two $!$-cells $v$ and $w$ with the same depth in a PS $(\Phi,b)$ and given an auxiliary port $p$ of some $?$-cell of $\Phi$, there might be an ambiguity on whether $p\in b(v)$ or $p\in b(w)$ (we would say in the standard terminology of linear logic proof-nets whether $p$ is an auxiliary door of $v$ or $w$'s box) only in case $\Phi$ is not a connected graph. Indeed (using again the standard terminology of linear logic proof-nets), in case $\Phi$ is connected, $p$ and $v$ are two ``doors of the same box'' iff there exists a path of $\Phi$ connecting $p$ and $v$ and crossing only cells with depth greater than the depth of $v$. More precisely:

\begin{cor}\label{cor:injectConnected}
Assume $A$ is infinite. Let $(R, \textsf{ind}), (R', \textsf{ind'}) \in \PSind$ such that $\LPSofPPS{R}$ is a connected graph. If $\sm{(R, \textsf{ind})} = \sm{(R', \textsf{ind'})}$, then $(R, \textsf{ind}) \simeq (R', \textsf{ind'})$.
\end{cor}

\begin{proof}
By Corollary~\ref{cor:injectLPS} $(\LPSofPPS{R}, \textsf{ind}) \simeq (\LPSofPPS{R'}, \textsf{ind'})$. Now notice that when $\LPSofPPS{R}$ is connected, there is a unique function $b$ such that $(\LPSofPPS{R},b)\in\PS$. Indeed, given $v\in\bangs{\PortsofPPLPS{\LPSofPPS{R}}}$, we have $p\in b(v)$ iff $\depth{\LPSofPPS{R}}{p} \leq \depth{\LPSofPPS{R}}{\mappriports{\PortsofPPLPS{\LPSofPPS{R}}}(v)}$ and there exists a path $d_{vp}$ of $\LPSofPPS{R}$ starting from $\mappriports{\PortsofPPLPS{\LPSofPPS{R}}}(v)$ and ending in $p$ such that for every port $q\not\in\{p, \mappriports{\PortsofPPLPS{\LPSofPPS{R}}}(v)\}$ crossed by $d_{vp}$ we have that $\depth{\LPSofPPS{R}}{q}>\depth{\LPSofPPS{R}}{
\mappriports{\PortsofPPLPS{\LPSofPPS{R}}}(v)}$.
\end{proof}

As already pointed out in the introduction, the theory of proof-nets is among the striking novelties introduced with Linear Logic. Right from the start (see~\cite{ll}), it appeared very natural to first introduce graphs (called like in this paper ``proof-structures'') not necessarily representing sequent calculus proofs, and then look for ``intrinsic'' (usually graph-theoretical) properties allowing to characterize, among proof-structures, precisely those corresponding to sequent calculus proofs (in this case the proof-structure is called \emph{proof-net}). Such a property is called \emph{correctness criterion}; the most used one is the Danos-Regnier criterion: a proof-structure $\pi$ of Multiplicative Linear Logic is a proof-net iff every correctness graph (every graph obtained from $\pi$ by erasing one of the two premisses of every $\parr$ link) is acyclic and connected.\\
As soon as one leaves the purely multiplicative fragment of Linear Logic, things become less simple; for Multiplicative and Exponential Linear Logic $MELL$, one often considers (like for example in~\cite{carvalhopaganitortora08}) a weaker correctness criterion: a proof-structure is a proof-net when every correctness graph is acyclic (and not necessarily connected); such a criterion corresponds to a particular version of Linear Logic sequent calculus (see for example~\cite{phdtortora}). But it is also well-known (see again for example~\cite{phdtortora}) that in the absence of weakening and $\bot$ links, the situation is much better, in the sense that one can strengthen the criterion so as to capture the standard Linear Logic sequent calculus (very much in the style of the purely multiplicative case): in this framework, an $MELL$ proof-structure is a proof-net iff every correctness graph is not only acyclic, but also connected. By $MELL$ net we mean in the following corollary the (indexed) untyped version (in the style of~\cite{carvalhopaganitortora08}) of this strong notion of proof-net:

\begin{cor}\label{cor:injectNoW}
Assume $A$ is infinite. Let $R$ and $R'$ be two $MELL$ nets without weakening nor $\bot$ links. If $\sm{R} = \sm{R'}$, then $R$ and $R'$ have the same (cut-free) normal form.
\end{cor}

\begin{proof}
Let $R_0$ (resp.\ $R_0'$) be a cut-free normal form of $R$ (resp.\ $R'$). Then $\sm{R}=\sm{R_0}=\sm{R'_0}=\sm{R'}$. Since we are in $MELL$ without weakening nor $\bot$, $\LPSofPPS{R_0}$ (so as $\LPSofPPS{R'_0}$) is a connected graph. Apply Corollary~\ref{cor:injectConnected}.
\end{proof}

\begin{remark}\label{rem:InjectStandardMELL}
Theorem~\ref{theorem:injectLPS}, Corollary~\ref{cor:injectLPS}, Corollary~\ref{cor:injectConnected} and Corollary~\ref{cor:injectNoW} hold for the standard typed $MELL$ proof-nets of~\cite{injectcoh}: in particular if every propositional variable of the logical language is interpreted by the infinite set $A$ and if $\pi$ and $\pi'$ are two cut-free typed proof-nets with atomic axioms, without weakenings nor $\bot$\footnote{We still refer here to the strong notion of proof-net corresponding to $MELL$ sequent calculus.}, and such that $\sm{\pi}=\sm{\pi'}$, then $\pi=\pi'$.
\end{remark}

\section{Proof of Proposition \ref{prop : KeyProposition}}\label{sect:InjectivityLPS}

In this last section, we use the tools previously introduced in order to prove the key-proposition (Proposition~\ref{prop : KeyProposition}) concerning only LPS (and not PS anymore). Since we need to consider isomorphisms between several kinds of objects (elements of $D'$, $t$-uples of elements of $D'$, finite multisets of $D'$, $t$-uples of finite multisets of $D'$,\ldots) we use the notion of groupoid (subsection~\ref{subsect:groupoid}). Subsections~\ref{subsect:contraction},~\ref{subsect:expunit} and~\ref{subsect:cbox} establish the main results that will be used in the different cases of the proof by induction of Proposition~\ref{prop : KeyProposition}, given in Subsection~\ref{subsect:key-proposition}.

\bigskip

Let $e$ be an atomic $k$-experiment of $\Phi \in \PLPS$ and suppose $e(p)=\alpha$ for $p\in\conclusions{\Phi}$. If $\alpha=(+,\alpha_{1},\alpha_{2})$, then since $e$ is atomic we can say that $p$ is not an axiom port, so that $p$ is necessarily the principal port of a cell of type $\otimes$. When $\alpha=(-,a)$ for some $a\in\setfmulti{D'}$, even if we know that $p$ is not an axiom port, there are several possibilities for the $?$-cell having $p$ as principal port. The following fact allows (in particular) to distinguish between $?$-cells having only auxiliary doors (remember Remark~\ref{rem:explainingPorts}) among their premises from the others.

\begin{fact}\label{fact : arity and paxnumber of contractions}
Let $\Phi \in \PLPS$. Let $l \in \contrlinks{\PortsofPPLPS{\Phi}}$. Let $k > \arity{\PortsofPPLPS{\Phi}}(l)$. Let $\mathcal{P}_0 \subseteq \mapauxports{\PortsofPPLPS{\Phi}}(l)$. Let $e$ be a $k$-experiment of $\Phi$. We set $a = \sum_{p \in \mathcal{P}_0} \dig_{\paxnumber{\PortsofPPLPS{\Phi}}(p)}^k (\multi{e(p)})$. Then $k$ divides $\textit{Card}(a)$ if, and only if, $(\forall p \in \mathcal{P}_0)$ $\paxnumber{\PortsofPPLPS{\Phi}}(p) \not= 0$.
\end{fact}

\begin{proof}
We have
\begin{eqnarray*}
\textit{Card}(a) & = & \sum_{p \in \mathcal{P}_0} k^{\paxnumber{\PortsofPPLPS{\Phi}}(p)} \\
& = & \textit{Card}(\{ p \in \mathcal{P}_0 \: / \: \paxnumber{\PortsofPPLPS{\Phi}}(p))=0 \}) + k \sum_{\begin{array}{c} p \in \mathcal{P}_0 \\ \paxnumber{\PortsofPPLPS{\Phi}}(p) \not= 0 \end{array}} k^{\paxnumber{\PortsofPPLPS{\Phi}}(p)-1}
\end{eqnarray*}
Hence $k$ divides $\textit{Card}(a)$ if, and only if, $k$ divides $\textit{Card}(\{ p \in \mathcal{P}_0 \: / \: \paxnumber{\PortsofPPLPS{\Phi}}(p)=0 \})$. Now 
\begin{eqnarray*}
\textit{Card}(\{ p \in \mathcal{P}_0 \: / \: \paxnumber{\PortsofPPLPS{\Phi}}(p)=0 \}) & \leq & \arity{\PortsofPPLPS{\Phi}}(l) \\
& < & k.
\end{eqnarray*}
So $k$ divides $\textit{Card}(\{ p \in \mathcal{P}_0 \: / \: \paxnumber{\PortsofPPLPS{\Phi}}(p)=0 \})$ if, and only if, $\textit{Card}(\{ p \in \mathcal{P}_0 \: / \: \paxnumber{\PortsofPPLPS{\Phi}}(p)=0 \}) = 0$ i.e. $(\forall p \in \mathcal{P}_0)$ $\paxnumber{\PortsofPPLPS{\Phi}}(p) \not= 0$.
\end{proof}

\subsection{Groupoids}\label{subsect:groupoid}

We recall that a groupoid is a category such that any morphism is an iso and that a morphism of groupoids is a functor between two groupoids. For any groupoid \textbf{G}, we will denote by $\textbf{G}_0$ the class of objects of the groupoid $\textbf{G}$. In the following, we sometimes think of a set as a groupoid such that the morphisms are identities on the elements of the set. We now define some useful groupoids:
\begin{minilist}
\item The groupoid $\groupoidD$: let $\groupoidD_0 = D'$ and $\rho : \alpha \rightarrow \alpha'$ in $\groupoidD$ if, and only if, $\rho \in \pInj \textrm{ such that } \rho \cdot \alpha = \alpha'$.
\item The groupoid $\groupoidsD$: let $\groupoidsD_0 = D'^{< \omega}$ and $\rho : (\alpha_1, \ldots, \alpha_n) \rightarrow (\alpha'_1, \ldots, \alpha'_{n'})$ in $\groupoidsD$ if, and only if, $n = n'$ and $(\forall i \in \integer{n}) \: \rho : \alpha_i \rightarrow \alpha'_i$ in $\groupoidD$.
\item The groupoid $\groupoidM$: let $\groupoidM_0 = \setfmulti{D'}$ and $\rho : a \rightarrow a'$ in $\groupoidM$ if, and only if, $\rho \cdot a = a'$.
\item The groupoid $\groupoidsDM$: let ${\groupoidsDM}_0 = (D'^{< \omega} \times \setfmulti{D'})$ and $\rho : (r, a) \rightarrow (r', a')$ in $\groupoidsDM$ if, and only if, $\rho : r \rightarrow r'$ in $\groupoidsD$ and $\rho : a \rightarrow a'$ in $\groupoidM$.
\item the groupoid $\groupoidpM$: let $\groupoidpM_0 = \setfparts{\setfmulti{D'}}$ and $\rho : \mathfrak{a} \rightarrow \mathfrak{a'}$ in $\groupoidpM$ if, and only if, for any $a' \in \setfmulti{D'}$, we have $a' \in \mathfrak{a'} \Leftrightarrow (\exists a \in \mathfrak{a}) \: \rho : a \rightarrow a'$ in $\groupoidM$.
\item The groupoid $\groupoidsM$: let $\groupoidsM_0 = {\setfmulti{{D'}^\textit{At}}}^{< \omega}$ and $\rho : (a_1, \ldots, a_n) \rightarrow (a'_1, \ldots, a'_n)$ in $\groupoidsM$ if, and only if, for any $i \in \integer{n}$, we have $\rho : a_i \rightarrow a'_i$ in $\groupoidM$.
\item the groupoid $\groupoidpsM$: let $\groupoidpsM_0 = \setfparts{\setfmulti{{D'}^\textit{At}}^{< \omega}}$ and $\rho : \mathfrak{r} \rightarrow \mathfrak{r'}$ in $\groupoidpsM$ if, and only if, for any $r' \in {\setfmulti{{D'}^\textit{At}}}^{< \omega}$, we have $r' \in \mathfrak{r'} \Leftrightarrow (\exists r \in \mathfrak{r}) \: \rho : r \rightarrow r'$ in $\groupoidsM$.
\item the groupoid $\groupoidppsM$: let $\groupoidppsM_0 = \setfparts{\setfparts{\setfmulti{{D'}^\textit{At}}^{< \omega}}}$ and $\rho : \mathcal{A} \rightarrow \mathcal{A'}$ in $\groupoidppsM$ if, and only if, for any $\mathfrak{a'} \in {\setfparts{\setfmulti{{D'}^\textit{At}}^{< \omega}}}$, we have $\mathfrak{a'} \in \mathcal{A'} \Leftrightarrow (\exists \mathfrak{a} \in \mathcal{A}) \: \rho : \mathfrak{a} \rightarrow \mathfrak{a'}$ in $\groupoidpsM$.
\item the groupoid $\groupoidBij$: objects are sets and morphisms are bijections.
\end{minilist}

In the sequel, we will write $\rho : r \rightarrow r'$ (referring to a given groupoid) in order to indicate that $\rho$ is an iso between $r$ and $r'$, while we will write $r\simeq r'$ meaning that there exists some iso $\rho : r \rightarrow r'$.

\begin{defin}
We denote by $\textit{Card}$ the morphism of groupoids $\textbf{M} \rightarrow \mathbb{N}$ defined by: 
$\textit{Card}(a) = \sum_{\alpha \in \textit{Supp}(a)} a(\alpha)$; 
and $\textit{Card}(\rho) = id_{\textit{Card}(a)}$ for any $\rho : a \rightarrow a'$.
\end{defin}

\subsection{The case of $\contrPLPS$}\label{subsect:contraction}

In the sequel, we split a multiset $a$ following an equivalence relation defined on $ \textit{Supp}(a)$:

\begin{defin}
Let $\mathcal{E}$ be a set and let $a \in \setfmulti{\mathcal{E}}$. Let $\mathcal{R}$ be an equivalence relation on $\mathcal{E}$. We set $a / \mathcal{R} = \{ a_0 \in \setfmulti{\mathcal{E}} \: / \: \textit{Supp}(a_0) \in \mathcal{E}/\mathcal{R} \textrm{ and } (\forall \alpha \in \textit{Supp}(a_0)) \: a_0(\alpha) = a(\alpha) \}$.
\end{defin}

Consider again the LPS $\Psi_{2}$ of Figure~\ref{example : LPS} and the $3$-experiment $(e_2,r_2)$ of $(\Psi_2, \textsf{ind}_2)$ already defined in Example~\ref{example : experiment}, where we suppose $\gamma_1\neq\gamma_2$. We have that $(r_{2},(\gamma_1, 1)),(r_{2},(\gamma_1, 2))\in\groupoidsD_{0}$ and if we define $\rho\in\pInj$ by setting $\rho(\gamma_1,1)=(\gamma_1,2)$, $\rho(\gamma_1,2)=(\gamma_1,1)$, $\rho(\gamma_2,1)=(\gamma_2,2)$ and $\rho(\gamma_2,2)=(\gamma_2,1)$, we have that $\rho:(r_{2},(\gamma_1, 1))\to(r_{2},(\gamma_1, 2))$ in $\groupoidsD$\footnote{Notice that we do not have, for example, $(r_{2},(\gamma_1, 1))\simeq (r_{2},(\gamma_2, 2))$ in $\groupoidsD$.}: the effect of the morphism $\rho$ of $\groupoidsD$ is to exchange two elements of $a_1 =$ $[(\gamma_1, 1),$ $(\gamma_2, 1),$ $(\gamma_1, 2),$ $(\gamma_2, 2),$ $(\gamma_1, 3),$ $(\gamma_2, 3)]$, without changing $r_{2}$. This allows to define an equivalence relation on any $a\in\setfmulti{D'}$ (w.r.t.\ a given $r\in\groupoidsD_{0}$):

\begin{defin}\label{def:Q}
For any $(r, a) \in {\groupoidsDM}_0$, we set $\textit{Q}(r, a) = a / \simeq$, where $\alpha_1 \simeq \alpha_2$ if, and only if, $(r, \alpha_1) \simeq (r, \alpha_2)$ in $\groupoidsD$. 
\end{defin}

\begin{fact}\label{fact:Qmorphism}
By extending the definition of $\textit{Q}$ to the morphisms of $\groupoidsDM$ in setting $\textit{Q}(\rho) = \rho$, we obtain a morphism of groupoids $\groupoidsDM \rightarrow \groupoidpM$.
\end{fact}

\begin{proof}
For any $\rho \in \pInj$, for any $(r, \alpha_1), (r, \alpha_2) \in {\groupoidsD}_0$, we have $(r, \alpha_1) \simeq (r, \alpha_2)$ in $\groupoidsD$ if, and only if, we have $(\rho \cdot r, \rho \cdot \alpha_1) \simeq (\rho \cdot r, \rho \cdot \alpha_2)$ in $\groupoidsD$.
\end{proof}

Suppose $(e,r)$ is an experiment  of $(\Phi,\textsf{ind}) \in \PLPSind$, suppose $e(\mappriports{\Phi}(l)) = (-, a)$ for some $l \in \contrlinks{\PortsofPPLPS{\Phi}}\cap \terminallinks{\Phi}$ and suppose that $e(p)=\alpha$ for $p\in\mapauxports{\PortsofPPLPS{\Phi}}(l)$ such that $\paxnumber{\PortsofPPLPS{\Phi}}(p)=d$. Then the idea is that (like we did in the example before Definition~\ref{def:Q}) one can exchange two ``copies'' of $\alpha$ in $a$ without changing $r$: the intuition is that for every $\alpha_1,\alpha_2 \in \textit{Supp}(\dig^k_d(\multi{\alpha}))$ one has $(r,\alpha_1)\simeq (r,\alpha_2)$ in $\groupoidsD$. More precisely, the following fact holds:

\begin{fact}\label{fact : morphism Q}
Let $k \in \mathbb{N}$. Let $(\Phi, \textsf{ind}) \in \PLPSind$. Let $l \in \contrlinks{\PortsofPPLPS{\Phi}} \cap \terminallinks{\Phi}$. Let $(e, r)$ be a $k$-experiment of $(\Phi, \textsf{ind})$. Let $a \in \setfmulti{D'}$ such that $e(\mappriports{\Phi}(l)) = (-, a)$. Let $a_0 \in \textsf{Q}(r, a)$. Then there exists $\mathcal{P}_0 \subseteq \mapauxports{\PortsofPPLPS{\Phi}}(l)$ such that $a_0 = \sum_{q \in \mathcal{P}_0} \dig_{\paxnumber{\PortsofPPLPS{\Phi}}(q)}^k(e(q))$.
\end{fact}

\begin{proof}
We prove, by induction on $d$ and using Fact \ref{fact : result invariant by permutations of dig}, that for any $d \in \mathbb{N}$, for any $\alpha \in D'$, for any $\alpha_1, \alpha_2 \in \textit{Supp}(\dig_{d}^k(\multi{\alpha}))$, we have $(r, \alpha_1) \simeq (r, \alpha_2)$ in $\groupoidsD$.
\end{proof}

\subsection{The case of $\contrunitPLPS$ and $\bangunitPLPS$}\label{subsect:expunit}

\begin{remark}\label{rem:toutestdig}
If $e$ is a $k$-experiment of $\Phi\in\PLPS$ and $l \in \contrlinks{\PortsofPPLPS{\Phi}}$, we know by Definition~\ref{def:experiment} that $e(\mappriports{\PortsofPPLPS{\Phi}}(l)) = (-,a)$, where $a=\sum_{p \in \mapauxports{\PortsofPPLPS{\Phi}}(l)} \dig_{\paxnumber{\PortsofPPLPS{\Phi}}(p)}^k(\multi{e(p)})$. When $l\in\contractionspax{\PortsofPPLPS{\Phi}}$ we have $\paxnumber{\PortsofPPLPS{\Phi}}(p)\geq 1$ for every $p\in\mapauxports{\PortsofPPLPS{\Phi}}(l)$, which implies that $a=\dig_{1}^{k}(b)$ for $b=\sum_{p \in \mapauxports{\PortsofPPLPS{\Phi}}(l)} \dig_{\paxnumber{\PortsofPPLPS{\Phi}}(p)-1}^k(\multi{e(p)}))$. It then follows that when $\Phi\in\contrunitPLPS$ there always exists $l\in\terminallinks{\Phi}$ such that $e(\mappriports{\PortsofPPLPS{\Phi}}(l)) = (-,\dig_{1}^{k}(b))$ for some $b\in\setfmulti{D'}$.
\end{remark}

The following fact will be used in the cases $\contrunitPLPS$ and $\bangunitPLPS$ of the proof of Proposition~\ref{prop : KeyProposition}: it intuitively states that given an (injective atomic) experiment $e$ (resp.\ $e'$) of $\Phi$ (resp.\ $\Phi'$) such that $e(\mappriports{\PortsofPPLPS{\Phi}}(l))\simeq e'(\mappriports{\PortsofPPLPS{\Phi'}}(l'))$ for some suitable teminal link $l$ (resp.\ $l'$), there exists $p\in\conclusions{\enleverunecoucheunitsweakenings{\Phi}{l}}$ such that for the ``corresponding'' $p'\in\conclusions{\enleverunecoucheunitsweakenings{\Phi'}{l'}}$ one has $\enleverunecoucheunitsweakenings{e}{l}(p)\simeq\enleverunecoucheunitsweakenings{e'}{l'}(p')$.

\begin{fact}\label{fact : enlever une couche aux unites : semantique}
Let $k \in \mathbb{N}$. Let $b, b' \in \setfmulti{D'}$. Let $\rho : $ $\dig_1^k (b) \to \dig_1^k (b')$ in $\groupoidM$. Then we have $\rho : $ $b^\ast +(\dig_1^k(b))^\textit{At}\to {b'}^\ast+ (\dig_1^k (b'))^\textit{At}$ in $\groupoidM$.
\end{fact}

\begin{proof}
We have $\dig_1^k (b^\ast) = {(\dig_1^k (b))}^\ast = {(\dig_1^k (b'))}^\ast = \dig_1^k ({b'}^\ast)$, hence $b^\ast = {b'}^\ast$. 
From $\rho : $ $\dig_1^k (b) \to \dig_1^k (b')$ one deduces that $\rho : $ $(\dig_1^k (b))^\textit{At} \to (\dig_1^k (b'))^\textit{At}$, and since for $\rho\in\pInj$ we already noticed that $\rho (b^\ast) = b^\ast$, we can conclude that $\rho : $ $b^\ast + (\dig_1^k (b))^\textit{At} \to b^\ast+(\dig_1^k(b'))^\textit{At}={b'}^\ast+(\dig_1^k (b'))^\textit{At}$.
\end{proof}

\begin{fact}\label{fact : bangunit : [beta]^ast = [beta]}
Let $k \in \mathbb{N}$. Let $\beta \in D'$ such that $(\dig_1^k(\multi{\beta}))^\ast \not= \multi{}$. Then $(\multi{\beta})^\ast = \multi{\beta}$.
\end{fact}

\begin{proof}
From $(\dig_1^k(\multi{\beta}))^\ast \not= \multi{}$, we deduce that $\textit{At'}(\beta) = \emptyset$.
\end{proof}

\subsection{The case of $\cboxPLPS$}\label{subsect:cbox}

We denote by $\textit{U}$ the forgetful functor $\groupoidppsM \rightarrow \groupoidBij$.

\bigskip

In the following informal discussion, we fix an LPS $\Phi$ and an atomic $k$-experiment $(e,r)$ of $(\Phi, \textsf{ind})$. Suppose $\Phi$ consists of 2 cells: a $!$-cell and a $?$-cell with a unique auxiliary port $p$ such that $\paxnumber{\PortsofPPLPS{\Phi}}(p) = 1$, and suppose that the two auxiliary ports of the two cells are connected by an axiom (in the language of the usual theory of linear logic proof-nets, $\Phi$ would correspond to an axiom link inside an exponential box). In this case $r=((-,\dig_1^k (\multi{\delta})),(+,\dig_1^k (\multi{\delta})))\in {D'}^\textit{At}\times{D'}^\textit{At}$ for some $\delta\in A$. If $\alpha,\alpha'\in \textit{Supp}(\dig_1^k (\multi{\delta}))$, then $\textit{At'}(\alpha) \cap \textit{At'}(\alpha')= \emptyset$: two elements of the multiset associated with the principal port of the $?$-cell have no atom in common, since they ``come from'' two different copies of the content of the box.\\ 
Suppose now that, more generally, $\Phi\in \cboxPLPS$ has two conclusions, one is the principal port of a $!$-cell and the other one is the principal port of a $?$-cell, but now this last cell has several auxiliary ports and for every such port $p$ one has $\paxnumber{\Phi}(p)\geq 1$; suppose also that the graph obtained by removing this $?$-cell is connected (in the language of the usual theory of linear logic proof-nets, $\Phi$ would now correspond to a connected proof-net inside an exponential box, where the $?$-conclusions of the box are contracted): an example of such an $\LPS$ is in Figure~\ref {example : LPS} (see also the following Example~\ref{example:bridges}). The previous remark can be generalized to such an LPS: let $a$ (resp.\ $b$) be the multiset associated by $e$ with the principal port of the $?$-cell (resp.\ $!$-cell) conclusion of $\Phi$; we have that $\alpha,\alpha'\in \textit{Supp}(a)$ ``come from'' the same copy of the content of the box if and only if there is a ``bridge'' between $\alpha$ and $\alpha'$\footnote{Notice that by Definition~\ref{def:ListPLPS} $\Phi \notin\contrunitPLPS\cup\bangunitPLPS$, so that $\alpha,\alpha'\in{D'}^\textit{At}$.}, meaning that there is a sequence $\alpha_0, \ldots, \alpha_n$ such that $\alpha_{i}\in \textit{Supp}(a+b)$ and $\alpha_0 = \alpha$, $\alpha_n = \alpha'$ and for any $i \in \integer{n}$, we have $\textit{At'}(\alpha_{i-1}) \cap \textit{At'}(\alpha_i) \not= \emptyset$. This means that one can split the multiset $a$ into equivalence classes given by the relation ``being connected by a bridge'', and every equivalence class will identify a copy of the box.\\ 
For general $\Phi\in \cboxPLPS$, the situation is more complex: it might be the case that the elements $\alpha$ and $\alpha'$ above come from the same copy of a box even though they are not connected by a bridge. On the other hand, the converse still holds: when there is a bridge between $\alpha$ and $\alpha'$ they do come from the same copy of the box. We thus define a function $sB$, that splits the result $r$ of the experiment $e$ into equivalence classes of this relation.

\begin{defin}\label{def:ponts}
For any $D_0 \subseteq {D'}^\textit{At}$, we define the equivalence relation $\simeq_{D_0}$ on $D_0$ as follows: $\alpha \simeq_{D_0} \alpha'$ if, and only if, there exist $\alpha_0, \ldots, \alpha_n \in D_0$ such that $\alpha_0 = \alpha$, $\alpha_n = \alpha'$ and for any $i \in \integer{n}$, we have $\textit{At'}(\alpha_{i-1}) \cap \textit{At'}(\alpha_i) \not= \emptyset$.
\end{defin}

\begin{defin}
We denote by $\bridge$ the function $\setfparts{{D'}^\textit{At}} \rightarrow \setfparts{\setfparts{{D'}^\textit{At}}}$ defined by $\bridge(D_0) = D_0 / \simeq_{D_0}$.
\end{defin}

The function $\textit{sB}$ that we are going to define ``splits'' a $t$-uple of multisets, following the equivalence classes of the ``bridge'' equivalence relation:

\begin{defin}\label{def:sB}
We denote by $\textit{sB}$ the morphism of groupoids $\groupoidsM \rightarrow \groupoidpsM$ defined by: $\textit{sB}(a_1, \ldots, a_n) = \{ (\restriction{a_1}{\textit{Supp}(a_1) \cap \mathfrak{a}},$ $ \ldots,$ $ \restriction{a_n}{\textit{Supp}(a_n) \cap \mathfrak{a}}) \: / \: $ $\mbox{$\mathfrak{a} \in \bridge(\textit{Supp}(\sum_{i=1}^n a_i))$} \}$; and $\textit{sB}(\rho) = \rho$.
\end{defin}

\begin{example}\label{example:bridges}
Let $a_1$ and $a_2$ be as in Example \ref{example : experiment}. Assume that $\gamma_1 \neq \gamma_2$. 
Then we have $\textit{B}(\textit{Supp}(a_1 + a_2)) = $ $\{ c_1, c_2, c_3 \}$, where $c_z = $ $\{ (\gamma_1, z), (\gamma_2, z), (+, (\gamma_1, z), (\gamma_2, z)) \},$ and $\textit{sB}(a_1, a_2) =$ $\{ r_1, r_2, r_3 \}$, where $r_z = $ $ ([(\gamma_1, z), (\gamma_2, z)],$ $[(+, ({\gamma_1}, z), ({\gamma_2}, z))])$. 
Notice that every element of $\textit{sB}(a_1, a_2)$ corresponds to a copy of the box.
\end{example}

Given $r=(a_{1},\ldots,a_{n}) \in \groupoidsM_0$ and two different equivalence classes $\mathfrak{a},\mathfrak{b}\in \bridge(\textit{Supp}(\sum_{i=1}^n a_i))$, we clearly have that $\textit{At'}(\mathfrak{a}) \cap \textit{At'}(\mathfrak{b})= \emptyset$. This implies that any element of the restriction of $r$ to the elements of  $\mathfrak{a}$ has no atom in common with any element of the restriction of $r$ to the elements of $\mathfrak{b}$, as the following fact precisely states. A consequence that will be used in Lemma~\ref{lemma : Key-Proposition : enleverunecouche a r} is that if for some $r,r' \in \groupoidsM_0$ one has $\rho:\textit{sB}(r)\to\textit{sB}(r')$ in $\groupoidpsM$, then $\rho:r\to r'$ in $\groupoidsM$.

\begin{fact}\label{fact : no common atoms in sB(r)}
Let $r \in \groupoidsM_0$. For any $r_1, r_2 \in \textit{sB}(r)$, we have $\textit{At'}(r_1) \cap \textit{At'}(r_2) \not= \emptyset \Rightarrow r_1 = r_2$.
\end{fact}

\begin{proof}
Suppose $r=(a_{1},\ldots,a_{n})$, $r_{1}=(c_{1},\ldots,c_{n})$ and $r_{2}=(d_{1},\ldots,d_{n})$. By Definition~\ref{def:sB}, for every $i\in\{1,\ldots,n\}$ we have that $c_{i}=\restriction{a_{i}}{\textit{Supp}(a_{i})\cap\mathfrak{a}}$ and $d_{i}=\restriction{a_{i}}{\textit{Supp}(a_{i})\cap\mathfrak{b}}$ for some $\mathfrak{a},\mathfrak{b}\in\bridge(\textit{Supp}(\sum_{i=1}^n a_i))$.

If $\textit{At'}(r_1) \cap \textit{At'}(r_2) \not= \emptyset$, then since $\textit{At'}(r_1)\subseteq\textit{At'}(\mathfrak{a})$ and $\textit{At'}(r_2)\subseteq\textit{At'}(\mathfrak{b})$, we have $\textit{At'}(\mathfrak{a}) \cap \textit{At'}(\mathfrak{b}) \not= \emptyset$, which means that $\textit{At'}(\xi) \cap \textit{At'}(\eta) \not= \emptyset$ for some $\xi\in\mathfrak{a}$ and $\eta\in\mathfrak{b}$: this implies by Definition~\ref{def:ponts} that $\xi \simeq_{\textit{Supp}(\sum_{i=1}^n a_i)} \eta$ and thus $\mathfrak{a}=\mathfrak{b}$ and $r_{1}=r_{2}$.
\end{proof}


In the language of the usual theory of linear logic proof-nets, given a proof-net one can ``box it''; we have generalized this boxing operation in the framework of $LPS$: 
for $\Phi\in\cboxPLPS$ this corresponds to the passage from $\enleverunecouche{\Phi}$ to $\Phi$. 
From an experiment $(e_{1},r_{1})$ of $\enleverunecouche{(\Phi,\textsf{ind})}$, one can naturally obtain an experiment $(e,r)$ of $(\Phi,\textsf{ind})$. 
The following lemma (intuitively) relates the effect of applying the splitting function $\textit{sB}$ after boxing to the effect of applying the splitting function $\textit{sB}$ before boxing.

\begin{lem}\label{lemma : sB}
Let $k, n \in \mathbb{N}$ such that $k>0$. Let $b_1, \ldots, b_n \in \setfmulti{{D'}^\textit{At}}$. We have $\textit{sB}(\dig_1^k(b_{1}),\ldots,\dig_1^k(b_{n}))= \{ (\dig(j_0) \cdot f_1, \ldots, \dig(j_0) \cdot f_n) \: / \: j_0 \in \integer{k} \textrm{ and } (f_1, \ldots, f_n) \in \textit{sB}(b_1, \ldots, b_n) \}$.
\end{lem}

\begin{proof}
For any $b \in \setfmulti{{D'}^\textit{At}}$, we have $\bridge(\textit{Supp}(\dig_1^k(b)))=\bridge(\textit{Supp}(\sum_{j=1}^k \dig(j) \cdot b)) = \{ \{ \dig(j_0) \cdot \beta \: / \: \beta \in \mathfrak{b} \} \: / \: j_0 \in \integer{k} \textrm{ and }\mathfrak{b} \in \bridge(\textit{Supp}(b)) \}$. 
Now notice that $\dig_1^k(\sum_{i=1}^n b_i)=\sum_{i=1}^n\dig_1^k(b_i)$; hence
$\bridge(\textit{Supp}(\sum_{i=1}^n\dig_1^k(b_i)))=\bridge(\textit{Supp}(\dig_1^k(\sum_{i=1}^n b_i)))= \{ \dig(j_0) \cdot \beta \: / \: \beta \in \mathfrak{b} \} \: / \: j_0 \in \integer{k} \textrm{ and }\mathfrak{b} \in \bridge(\textit{Supp}(\sum_{i=1}^n b_i)) \}$.
Thus
\begin{eqnarray*}
& & \textit{sB}(\dig_1^k(b_{1}),\ldots,\dig_1^k(b_{n}))\\
& = & \{ (\restriction{(\dig_1^k(b_1))}{\textit{Supp}(\dig_1^k(b_1)) \cap \mathfrak{a}} , \dots ,\restriction{(\dig_1^k(b_n))}{\textit{Supp}(\dig_1^k(b_n)) \cap \mathfrak{a}} ) \: / \: \\
& & \: \: \: \mathfrak{a} \in \bridge(\textit{Supp}(\sum_{i=1}^n \dig_1^k(b_i))) \} \allowdisplaybreaks\\
& = & \{ (\restriction{(\dig_1^k(b_1))}{\textit{Supp}(\dig_1^k(b_1)) \cap \{ \dig(j_0) \cdot \beta \: / \: \beta \in \mathfrak{b}  \}
} , \dots ,\restriction{(\dig_1^k(b_n))}{\textit{Supp}(\dig_1^k(b_n)) \cap \{ \dig(j_0) \cdot \beta \: / \: \beta \in \mathfrak{b} \}} ) \: / \\ 
& & \: \: \: j_0 \in \integer{k} \textrm{ and } \mathfrak{b} \in  \bridge(\textit{Supp}(\sum_{i=1}^n b_i))\} \allowdisplaybreaks\\
& = & \{ (\restriction{(\dig_1^k(b_1))}{\{ \dig(j_0) \cdot \beta \: / \: \beta \in \textit{Supp}(b_1) \cap \mathfrak{b}  \}} , \dots ,\restriction{(\dig_1^k(b_n))}{\{ \dig(j_0) \cdot \beta \: / \: \beta \in \textit{Supp}(b_n) \cap \mathfrak{b} \}} ) \: / \\ 
& & \: \: \: j_0 \in \integer{k} \textrm{ and } \mathfrak{b} \in  \bridge(\textit{Supp}(\sum_{i=1}^n b_i))\} \allowdisplaybreaks\\
& = & \{ (\restriction{(\dig(j_0) \cdot b_1)}{\{ \dig(j_0) \cdot \beta \: / \: \beta \in \textit{Supp}(b_1) \cap \mathfrak{b}  \}} , \dots ,\restriction{(\dig(j_0) \cdot b_n)}{\{ \dig(j_0) \cdot \beta \: / \: \beta \in \textit{Supp}(b_n) \cap \mathfrak{b} \}} ) \: / \\ 
& & \: \: \: j_0 \in \integer{k} \textrm{ and } \mathfrak{b} \in  \bridge(\textit{Supp}(\sum_{i=1}^n b_i))\} \allowdisplaybreaks\\
& = & \{ (\dig(j_0) \cdot \restriction{b_1}{\textit{Supp}(b_1) \cap \mathfrak{b}}, \ldots, \dig(j_0) \cdot \restriction{b_n}{\textit{Supp}(b_n) \cap \mathfrak{b}}) \: / \: j_0 \in \integer{k} \textrm{ and } \mathfrak{b} \in  \bridge(\textit{Supp}(\sum_{i=1}^n b_i))\} \\
& = & \{ (\dig(j_0) \cdot f_1, \ldots, \dig(j_0) \cdot f_n) \: / \: j_0 \in \integer{k} \textrm{ and } (f_1, \ldots, f_n) \in \textit{sB}(b_1, \ldots, b_n) \}.
\end{eqnarray*}
\end{proof}

Our aim is now to prove Lemma~\ref{lemma : Key-Proposition : enleverunecouche a r}: both the following Definition~\ref{def:R} and Fact~\ref{fact : iso in psM} are just tools to prove this result (in order to get some intuition, see Example~\ref{example:psM}).

\begin{defin}\label{def:R}
We denote by $\textit{R}$ the morphism of groupoids $\groupoidpsM \rightarrow \groupoidppsM$ defined by: $\textit{R}(\mathfrak{a}) = \mathfrak{a} / \simeq_{\groupoidsM}$, where $r \simeq_{\groupoidsM} r'$ if, and only if, $r \simeq r'$ in $\groupoidsM$; and $\textit{R}(\rho) = \rho$.
\end{defin}

\begin{fact}\label{fact : iso in psM}
Let $k \in \mathbb{N} \setminus \{ 0 \}$. Let $r, r' \in \groupoidsM_0$. Let $\mathfrak{b} \in \textit{R}(\textit{sB}(r)), \mathfrak{b'} \in \textit{R}(\textit{sB}(r'))$ such that $\{ \dig(j_0) \cdot r_0 \: / \: j_0 \in \integer{k} \textrm{ and } r_0 \in \mathfrak{b} \} \simeq \{ \dig(j_0) \cdot r_0' \: / \: j_0 \in \integer{k} \textrm{ and } r_0' \in \mathfrak{b'} \}$ in $\groupoidpsM$. Then we have $\mathfrak{b} \simeq \mathfrak{b'}$ in $\groupoidpsM$.
\end{fact}

\begin{proof}
Let $\rho : \{ \dig(j_0) \cdot r_0 \: / \: j_0 \in \integer{k} \textrm{ and } r_0 \in \mathfrak{b} \} \rightarrow \{ \dig(j_0) \cdot r_0' \: / \: j_0 \in \integer{k} \textrm{ and }r_0' \in \mathfrak{b'} \}$ in $\groupoidpsM$. Let $r_0 \in \mathfrak{b}$. Let $r'_0 \in \mathfrak{b'}$ and $j_0 \in \integer{k}$ such that $\rho : \dig(1) \cdot r_0 \rightarrow \dig(j_0) \cdot r'_0$ in $\groupoidsM$; then we have $r_0 \simeq r'_0$ in $\groupoidsM$. Thus the following holds:
\begin{minilist}
\item there exists $r_0 \in \mathfrak{b}$, $r'_0 \in \mathfrak{b'}$ such that $r_0 \simeq r'_0$ in $\groupoidsM$;
\item for any $r_1, r_2 \in \mathfrak{b}$, we have $r_1 \simeq r_2$ in $\groupoidsM$ and for any $r'_1, r'_2 \in \mathfrak{b'}$, $r'_1 \simeq r'_2$ in $\groupoidsM$;
\item for any $r_1, r_2 \in \mathfrak{b}$, we have $\textit{At'}(r_1) \cap \textit{At'}(r_2) \not= \emptyset \Rightarrow r_1 = r_2$ and for any $r'_1, r'_2 \in \mathfrak{b'}$, we have $\textit{At'}(r'_1) \cap \textit{At'}(r'_2) \not= \emptyset \Rightarrow r'_1 = r'_2$ (by Fact \ref{fact : no common atoms in sB(r)});
\item $\textit{Card}(\mathfrak{b}) = \textit{Card}(\mathfrak{b'})$.
\end{minilist}
Hence $\mathfrak{b} \simeq \mathfrak{b'}$ in $\groupoidpsM$. Indeed: let $\tau : r_0 \to r'_0$ in $\groupoidsM$ and let $\varphi : \mathfrak{b} \to \mathfrak{b'}$ in $\groupoidBij$; for any $r_1 \in \mathfrak{b}$, let $\tau_{r_1} : r_1 \to r_0$ in $\groupoidsM$; for any $r'_1 \in \mathfrak{b'}$, let $\tau'_{r'_1} : r'_0 \to r'_1$ in $\groupoidsM$; for any $r_1 \in \mathfrak{b}$, we set $\rho_{r_1} = \tau'_{\varphi(r_1)} \circ \tau \circ \tau_{r_1}$; we define $\rho' : \mathfrak{b} \to \mathfrak{b'}$ in $\groupoidpsM$ by setting $\rho'(\delta) = \rho_{r_1}(\delta)$ if $\delta \in \textit{At'}(r_1)$.
\end{proof}

\begin{example}\label{example:psM}
In order to help the reader to get some intuition of what we want to do here, let us consider the following LPS $\Phi$: the contraction of two auxiliary doors $p_{1}$ and $p_{2}$ such that $\paxnumber{\PortsofPPLPS{\Phi}}(p_{1})=\paxnumber{\PortsofPPLPS{\Phi}}(p_{2})=1$; above each auxiliary door, a $\parr$; above each $\parr$, an axiom. 
Let $e = e'$ be the injective atomic $k$-experiment of $\Phi$ such that the label associated by $e$ with every auxiliary port of the $?$-cell is $(-, \gamma_z, \gamma_z)$, where $\gamma_z \in A$, $z\in\integer{2}$ and $\gamma_{1}\neq\gamma_{2}$. The result $r = r'$ is $(-, \sum_{1 \leq j \leq k, 1 \leq z \leq 2} \multi{(-, (\gamma_z, j), (\gamma_z, j))})$. We have $\rho : \mathfrak{a} \rightarrow \mathfrak{a'}$ in $\groupoidpsM$, where $\mathfrak{a} = \bigcup_{1 \leq j \leq k, 1 \leq z \leq 2} \{ ( \multi{(-, (\gamma_z, j), (\gamma_z, j))}) \} = \mathfrak{a'}$, with $\rho$ that can send any $(\gamma_z, j)$ to any $(\gamma_{z'}, j')$. Fact \ref{fact : iso in psM} will be useful to deduce very generally that in situations of this kind, we have $\mathfrak{b} \simeq \mathfrak{b'}$ in $\groupoidpsM$, where here $\mathfrak{b} = \{ ([(-, \gamma_1, \gamma_1)]), ([(-, \gamma_2, \gamma_2)]) \} = \mathfrak{b'}$.
\end{example}

The following lemma is the crucial step allowing to apply the induction hypothesis in the proof of the key-Proposition~\ref{prop : KeyProposition} in the $\cboxPLPS$ case: it intuitively states that if there is an isomorphism between the results of two experiments of $\Phi_1,\Phi_2\in\cboxPLPS$, then there exists also an isomorphism between the results of two experiments of $\enleverunecouche{\Phi_1}$ and $\enleverunecouche{\Phi_2}$:

\begin{lem}\label{lemma : Key-Proposition : enleverunecouche a r}
Let $k, n \in \mathbb{N}$ such that $k>0$. Let $b_1, \ldots, b_n, b'_1, \ldots, b'_n \in \setfmulti{{D'}^\textit{At}}$ such that $(\dig_1^k(b_1),\ldots,\dig_1^k(b_n)) \simeq (\dig_1^k(b'_1),\ldots,\dig_1^k(b'_n))$ in $\groupoidsM$. Then we have $(b_1, \ldots, b_n) \simeq (b'_1, \ldots, b'_n)$ in $\groupoidsM$.
\end{lem}

\begin{proof}
We set $\mathfrak{a} = \{ \dig(j_0) \cdot (f_1, \ldots, f_n) \: / \: j_0 \in \integer{k} \textrm{ and }(f_1, \ldots, f_n) \in \textit{sB}(b_1, \ldots, b_n) \}$ and $\mathfrak{a'} = \{ \dig(j_0) \cdot (f'_1, \ldots, f'_n) \: / \: j_0 \in \integer{k} \textrm{ and } (f'_1, \ldots, f'_n) \in \textit{sB}(b'_1, \ldots, b'_n) \}$. Since $sB$ is a morphism of groupoids, by Lemma \ref{lemma : sB}, there exists $\rho : \mathfrak{a} \to \mathfrak{a'}$ in $\groupoidpsM$. 

Since for any $r,r' \in {\groupoidsM}_0$, for any $j_1, j_2 \in \integer{k}$, we have $\dig(j_1) \cdot r \simeq \dig(j_2) \cdot r'$ in $\groupoidsM$ if, and only if, $r \simeq r'$ in $\groupoidsM$, we can define 
$\varphi : \textit{U}(\textit{R}(\textit{sB}(b_1, \ldots, b_n))) \rightarrow \textit{U}(\textit{R}(\mathfrak{a}))$ in $\groupoidBij$ by setting $\varphi(\{ (f_1^1, \ldots, f_n^1), \ldots, (f_1^q, \ldots, f_n^q)\}
) = \{ \dig(j) \cdot (f_1^z, \ldots, f_n^z) \: / \: j \in \integer{k} \textrm{ and } z \in \integer{q} \}$ and $\varphi' : \textit{U}(\textit{R}(\textit{sB}(b'_1, \ldots, b'_n))) \rightarrow \textit{U}(\textit{R}(\mathfrak{a'}))$ in $\groupoidBij$ by setting $\varphi'(\{ ({f'}_1^1, \ldots, {f'}_n^1), \ldots, ({f'}_1^q, \ldots, {f'}_n^q)\}) = \{ \dig(j) \cdot ({f'}_1^z, \ldots, {f'}_n^z) \: / \: j \in \integer{k} \textrm{ and } z \in \integer{q} \}$. 
We have ${\varphi'}^{-1} \circ \textit{U}(\textit{R}(\rho)) \circ \varphi : \textit{U}(\textit{R}(\textit{sB}(b_1, \ldots, b_n))) \to \textit{U}(\textit{R}(\textit{sB}(b'_1, \ldots, b'_n)))$ in $\groupoidBij$. 

For any $\mathfrak{b} \in \textit{U}(\textit{R}(\textit{sB}(b_1, \ldots, b_n)))$, we have $\rho : \varphi(\mathfrak{b})=\{\dig(j_{0}) \cdot r_{0} \: / \: j_0 \in \integer{k} \textrm{ and } r_{0}\in\mathfrak{b}\}\rightarrow \{\dig(j_{0}) \cdot r'_{0} \: / \: j_0 \in \integer{k} \textrm{ and } r'_{0}\in({\varphi'}^{-1} \circ \textit{U}(\textit{R}(\rho)) \circ \varphi)(\mathfrak{b})\}=(U(R(\rho)) \circ \varphi)(\mathfrak{b})$ in $\groupoidpsM$. Hence by Fact \ref{fact : iso in psM}, for any $\mathfrak{b} \in \textit{U}(\textit{R}(\textit{sB}(b_1, \ldots, b_n)))$ there exists $\tau_\mathfrak{b} : \mathfrak{b} \rightarrow ({\varphi'}^{-1} \circ \textit{U}(\textit{R}(\rho)) \circ \varphi)(\mathfrak{b})$ in $\groupoidpsM$.

Now, by applying a first time Fact~\ref{fact : no common atoms in sB(r)}, we can define an application $\tau : \bigcup_{r \in \textit{sB}(b_1, \ldots, b_n)} \textit{At'}(r) \rightarrow \bigcup_{r' \in \textit{sB}(b'_1, \ldots, b'_n)} \textit{At'}(r')$ by setting $\tau(\delta) = \tau_\mathfrak{b}(\delta)$ for $\delta \in \textit{At'}(r)$, $r \in \mathfrak{b}$ and $\mathfrak{b}\in\textit{R}(\textit{sB}(b_1, \ldots, b_n))$. 

We thus obtain $\tau : \textit{R}(\textit{sB}(b_1, \ldots, b_n)) \rightarrow \textit{R}(\textit{sB}(b'_1, \ldots, b'_n))$ in $\groupoidppsM$. 
By applying a second time Fact~\ref{fact : no common atoms in sB(r)}, we obtain $\tau : \bigcup R(\textit{sB}(b_1, \ldots, b_n)) = \textit{sB}(b_1, \ldots, b_n) \to \textit{sB}(b'_1, \ldots, b'_n) = \bigcup R(\textit{sB}(b'_1, \ldots, b'_n))$ in $\groupoidpsM$. Lastly, by applying a third time Fact~\ref{fact : no common atoms in sB(r)}, we obtain $\tau : (b_1, \ldots, b_n) \to (b'_1, \ldots, b'_n)$ in $\groupoidsM$.
\end{proof}

\subsection{Key-Proposition}\label{subsect:key-proposition}

When (for some $\Phi\in\PLPS$) ``above'' an auxiliary port $p$ of $l\in\contrlinks{\PortsofPPLPS{\Phi}}\cup\bangs{\PortsofPPLPS{\Phi}}$\footnote{In case $l\in\bangs{\PortsofPPLPS{\Phi}}$ such a premise is the unique premise of $l$.} there are no axiom ports, it is obvious that whatever $k$-experiment $e$ of $\Phi$ one considers, the label $\alpha=e(p)$ of $p$ contains no atom. And the converse holds too when $e$ is atomic: if $\textit{At'}(e(p))=\emptyset$, there are no axiom ports ``above'' $p$. This implies that $e(\mappriports{\PortsofPPLPS{\Phi}}(l)) = (\iota,b)$ for some $b\in\setfmulti{D'}$ and $b^{\ast}\neq\multi{}$ iff ``above'' one of the auxiliary ports of $l$ there are no axiom ports, as the following fact shows. 

\begin{fact}\label{fact:axiomes-etoiles}
Let $k \in \mathbb{N}$, let $\Phi\in \PLPS$ and let $e$ be an atomic $k$-experiment of $\Phi$. Suppose that $l\in\links{\PortsofPPLPS{\Phi}}$ and $e(\mappriports{\PortsofPPLPS{\Phi}}(l)) = (\iota,b)$ for some $b\in\setfmulti{D'}$.

We have that $b^{\ast}\neq\multi{}$ iff there exists $p \in \mapauxports{\PortsofPPLPS{\Phi}}(l)$ such that  for every $q \geq_{\Phi} p$ one has  $q \notin \bigcup{\axioms{\Phi}}$.
\end{fact}

\begin{proof}
Since $e$ is atomic\footnote{In case $e$ is not atomic, one might have for example $e(q)=(+,\ast)$ for some $q\in\bigcup \axioms{\Phi}$.}, we have $\textit{At'}(e(q)) \not= \emptyset$ for any $q \in \bigcup \axioms{\Phi}$, hence one can easily prove, 
by induction on the number of ports ``above'' the port $p$ of $\Phi$ (that is on $Card(\{q\in\ports{\PortsofPPLPS{\Phi}}\:/\: q \geq_{\Phi} p\})$), that there exists $q \geq_{\Phi} p$ such that $q \in \bigcup{\axioms{\Phi}}$ iff $\textit{At'}(e(p))\neq\emptyset$. This immediately yields the conclusion: for every $p \in \mapauxports{\PortsofPPLPS{\Phi}}(l)$ there exists  $q \geq_{\Phi} p$ such that $q \in \bigcup{\axioms{\Phi}}$ iff $\textit{At'}(\alpha)\neq\emptyset$ for every $\alpha\in b$ iff $b^{\ast}=\multi{}$.
\end{proof}

\addtocounter{prop}{-1}

\begin{prop}
Let $(\Phi, \textsf{ind}), (\Phi', \textsf{ind'}) \in \LPSind$, let $k > \cosize{\PortsofPPLPS{\Phi}}, \cosize{\PortsofPPLPS{\Phi'}}$, let $(e, r)$ (resp.\ $(e', r')$) be an atomic injective $k$-experiment of $(\Phi, \textsf{ind})$ (resp.\ $(\Phi', \textsf{ind'})$). If $r \simeq r'$ in \textbf{sD}, then $(e, r) \simeq_{\textit{At}} (e', r')$.
\end{prop}

\begin{proof}
The proof is by induction on $\mes{\PortsofPPLPS{\Phi}}$. We have $\mes{\PortsofPPLPS{\Phi}} = (0, 0)$ if, and only if, $\Phi \in \emptyPLPS$; in this case, it is obvious that we have $(e, r) \simeq (e', r')$. If $\mes{\PortsofPPLPS{\Phi}} > (0, 0)$, then let $\rho : r \rightarrow r'$ in $\groupoidsD$, we set $n = \textit{Card}(\conclusions{\Phi})$ and we distinguish between the several cases.

\begin{minilist}
\item In the case where $\Phi \in \axPLPS$, let $w = \{ p_0, q_0 \} \in \isolatedaxioms{\Phi}$ and let $i_0, j_0 \in \integer{n}$ such that $\textsf{ind}(p_0) = i_0$ and $\textsf{ind}(q_0) = j_0$. Let $p_0', q_0' \in \conclusions{\Phi'}$ such that $\textsf{ind'}(p_0') = i_0$ and $\textsf{ind'}(q_0') = j_0$. As $e$ is atomic and $e'$ is injective, we have $w' = \{ p_0', q_0' \} \in \isolatedaxioms{\Phi'}$.

Let $(\Phi_1, \textsf{ind}_1) \in \PLPSind$ (resp. $(\Phi'_1, \textsf{ind'}_1) \in \PLPSind$) obtained from $(\Phi, \textsf{ind})$ (resp. $(\Phi', \textsf{ind'})$) by removing $w$ (resp. $w'$).\footnote{See the appendix for a formal definition of $(\Phi_1, \textsf{ind}_1)$ and $(\Phi'_1, \textsf{ind'}_1)$.} Since $\Phi, \Phi' \in \LPS$, we have $\Phi_1, \Phi'_1 \in \LPS$.
We set $e_1 = \restriction{e}{\ports{\PortsofPPLPS{\Phi_1}}}$ and $e'_1 = \restriction{e'}{\ports{\PortsofPPLPS{\Phi'_1}}}$. We set $r_1 = e \circ {\textsf{ind}_1}^{-1}$ and $r'_1 = e' \circ {\textsf{ind'}_1}^{-1}$: it is immediate that $(e_1, r_1)$ is an injective atomic experiment of $(\Phi_1, \textsf{ind}_1)$ and that $(e'_1, r'_1)$ is an injective atomic experiment of $(\Phi'_1, \textsf{ind'}_1)$ and that from $\rho : r\to r'$ one deduces $\rho:r_{1}\to r'_{1}$. Notice that $\mes{\PortsofPPLPS{\Phi_1}} < \mes{\PortsofPPLPS{\Phi}}$: by induction hypothesis we have $(e_1, r_{1}) \simeq (e'_1, r'_{1})$, which, since $e$ is atomic and injective, yields $(e,r) \simeq (e',r')$. 

\item In the case where $\Phi \in \contrPLPS$, let $l_0 \in \contractions{\PortsofPPLPS{\Phi}} \cap \terminallinks{\Phi}$ and let $i_0 \in \integer{n}$ such that $\textsf{ind}(\mappriports{\PortsofPPLPS{\Phi}}(l_0)) = i_0$.
As $e'$ is atomic, there exists $l_0' \in \contrlinks{\PortsofPPLPS{\Phi'}} \cap \terminallinks{\Phi}$ such that $\mappriports{\PortsofPPLPS{\Phi'}}(l_0') = \textsf{ind'}^{-1}(i_0)$. 
Let $a \in \setfmulti{D'}$ such that $e(\mappriports{\PortsofPPLPS{\Phi}}(l_0)) = (-, a)$. Let $a' \in \setfmulti{D'}$ such that $\rho \cdot (-, a) = (-, a')$. Let $p \in \mapauxports{\PortsofPPLPS{\Phi}}(l_0)$ such that $\paxnumber{\PortsofPPLPS{\Phi}}(p) = 0$. We set $\beta = e(p)$. We have $\beta \in \textit{Supp}(a)$, hence there exists $a_0 \in \textit{Q}(r, a)$ such that $\beta \in \textit{Supp}(a_0)$. 
By Fact \ref{fact : morphism Q}, there exists $\mathcal{P}_0 \subseteq \mapauxports{\PortsofPPLPS{\Phi}}(l_0)$ such that $a_0 = \sum_{q \in \mathcal{P}_0} \dig_{\paxnumber{\PortsofPPLPS{\Phi}}(q)}^k(e(q))$. We have $p \in \mathcal{P}_0$ (otherwise, we would have $a(\beta) > a_0(\beta)$). Hence, by Fact \ref{fact : arity and paxnumber of contractions}, $k$ does not divide $\textit{Card}(a_0) = \textit{Card}(\rho \cdot a_0)$. As we have $\rho : (r, a) \rightarrow (r', a')$ in $\groupoidsDM$ and by Fact~\ref{fact:Qmorphism} $\textit{Q}$ is a morphism of groupoids, we have $\rho \cdot a_0 \in \textit{Q}(r', a')$. Hence, by Fact \ref{fact : morphism Q}, there exists $\mathcal{P}_0' \subseteq \mapauxports{\PortsofPPLPS{\Phi'}}(l_0')$ such that $\rho \cdot a_0 = \sum_{q \in \mathcal{P}_0'} \dig_{\paxnumber{\PortsofPPLPS{\Phi'}}(q)}^k(e'(q))$. By Fact \ref{fact : arity and paxnumber of contractions}, there exists $p' \in \mathcal{P}_0'$ such that $\paxnumber{\PortsofPPLPS{\Phi'}}(p') = 0$. Let $\beta' = e'(p')$; we have $(r', \rho \cdot \beta) \simeq (r', \beta')$ and $(r, \beta) \simeq (r', \rho \cdot \beta)$ in $\groupoidsD$, hence $(r, \beta) \simeq (r', \beta')$ in $\groupoidsD$. 

Let $\Phi_1 \in \PLPS$ (resp. $\Phi'_1 \in \PLPS$) obtained from $\Phi$ (resp. $\Phi'$) by removing $p$ (resp. $p'$) from the auxiliary ports of $l_0$ (resp. $l_0'$).\footnote{See the appendix for a formal definition of $(\Phi_1, \textsf{ind}_1)$ and $(\Phi'_1, \textsf{ind'}_1)$.} Notice that $\mes{\PortsofPPLPS{\Phi_1}} < \mes{\PortsofPPLPS{\Phi}}$. Both $\Phi_1$ and $\Phi'_1$ have $n+1$ free ports: for $\Phi_1$, those of $\Phi$ and a new free port $p_0$; for $\Phi'_1$, those of $\Phi'$ and a new free port $p'_0$. 
We set 

$\textsf{ind}_1(q) = \left\lbrace \begin{array}{ll} \textsf{ind}(q) & \textrm{if $q \not= p_0$;} \\ n+1 & \textrm{if $q = p_0$;} \end{array} \right.$ and $\textsf{ind'}_1(q) = \left\lbrace \begin{array}{ll} \textsf{ind'}(q) & \textrm{if $q \not= p'_0$;} \\ n+1 & \textrm{if $q = p'_0$.} \end{array} \right.$ 

We have $(\Phi_1, \textsf{ind}_1), (\Phi'_1, \textsf{ind'}_1) \in \LPSind$. For any $q \in \ports{\PortsofPPLPS{\Phi_1}} \setminus \{ \mappriports{\PortsofPPLPS{\Phi_1}}(l_0) \}$, we set $e_1(q) = e(q)$. Let $b \in \setfmulti{D'}$ such that $a = b + \multi{\beta} $; we set $e_1(\mappriports{\PortsofPPLPS{\Phi_1}}(l_0)) = (-, b)$. For any $q \in \ports{\PortsofPPLPS{\Phi'_1}} \setminus \{ \mappriports{\PortsofPPLPS{\Phi'_1}}(l') \}$, we set $e'_1(q) = e'(q)$. Let $b' \in \setfmulti{D'}$ such that $a' = b' + \multi{\beta'}$; we set $e'_1(\mappriports{\PortsofPPLPS{\Phi'_1}}(l_0')) = (-, b')$. \\
We set $r_1(i) = \left\lbrace \begin{array}{l} r(i) \textrm{ if $i \notin \{ i_0, n + 1 \}$;} \\ (-, b) \textrm{ if $i = i_0$;} \\ \beta \textrm{ if $i = n + 1$.} \end{array} \right.$ \\
We set $r_1'(i) = \left\lbrace \begin{array}{l} r'(i) \textrm{ if $i \notin \{ i_0, n + 1 \}$;} \\ (-, b') \textrm{ if $i = i_0$;} \\ \beta' \textrm{ if $i = n + 1$.} \end{array} \right.$ 

Since $(e, r)$ (resp. $(e', r')$) is an atomic injective $k$-experiment of $(\Phi, \textsf{ind})$ (resp. $(\Phi', \textsf{ind'})$), $(e_1, r_1)$ (resp. $(e'_1, r'_1)$) is an atomic injective $k$-experiment of $(\Phi_1, \textsf{ind}_1)$ (resp. $(\Phi'_1, \textsf{ind'}_1)$) and since $(r, \beta) \simeq (r', \beta')$ in $\groupoidsD$ we have $r_1 \simeq r_1'$ in $\groupoidsD$. By induction hypothesis we deduce that $(e_1, r_{1}) \simeq (e'_1, r'_{1})$, from which the conclusion $(e,r) \simeq (e',r')$ immediately follows.

\item In the case where $\Phi \in\bangunitPLPS$, by Fact~\ref{fact:axiomes-etoiles}, there exists $l_0 \in \bangs{\PortsofPPLPS{\Phi}} \cap \terminallinks{\Phi}$ and $\beta \in D'$ such that $e(\mappriports{\PortsofPPLPS{\Phi}}(l_0)) = (+, \dig_1^k (\multi{\beta}))$ and ${(\dig_1^k (\multi{\beta}))}^\ast \not= \multi{}$. As $e'$ is atomic, there exists $l_0' \in \bangs{\PortsofPPLPS{\Phi'}} \cap \terminallinks{\Phi'}$ such that $\mappriports{\PortsofPPLPS{\Phi'}}(l_0') = \textsf{ind'}^{-1}(i_0)$.
Since $\rho : r\to r'$ one has $\rho : e(\mappriports{\PortsofPPLPS{\Phi}}(l_0))\to e'(\mappriports{\PortsofPPLPS{\Phi'}}(l'_0))$, so that there exists $\beta'\in D'$ such that $e'(\mappriports{\PortsofPPLPS{\Phi'}}(l'_0)) = (+, \dig_1^k (\multi{\beta'}))$ and $\rho : \dig_1^k (\multi{\beta}) \to \dig_1^k (\multi{\beta'})$. 
Hence $(\dig_1^k(\multi{\beta'}))^\ast \not= \multi{}$ and, by Fact~\ref{fact : enlever une couche aux unites : semantique}, $\rho : $ $(\multi{\beta})^\ast + (\dig_1^k (\multi{\beta}))^\textit{At}\to (\multi{\beta'})^\ast + (\dig_1^k((\multi{\beta'}))^\textit{At}$ in $\groupoidM$: by Fact~\ref{fact : bangunit : [beta]^ast = [beta]}, we obtain $\rho : \beta\to\beta'$ and thus $\rho:\enleverunecoucheunitsweakenings{r}{l_0}\to\enleverunecoucheunitsweakenings{r'}{l'_0}$, where $\enleverunecoucheunitsweakenings{r}{l_0}$ and $\enleverunecoucheunitsweakenings{r'}{l'_0}$ have been defined in Fact~\ref{fact : bangunit : syntaxe}. By this fact and by Fact~\ref{fact : Phi LPS => Phi[l] LPS}, we can apply the induction hypothesis and deduce that $(\enleverunecoucheunitsweakenings{e}{l_0}, \enleverunecoucheunitsweakenings{r}{l_0})\simeq_\textit{At} (\enleverunecoucheunitsweakenings{e'}{l'_0}, \enleverunecoucheunitsweakenings{r'}{l'_0})$, and by Fact~\ref{fact : bangunit : e[l]=e'[l] => e=e'} we conclude $(e, r) \simeq_{\textit{At}} (e', r')$.

\item In the case where $\Phi \in\contrunitPLPS$, by Remark~\ref{rem:toutestdig} and Fact~\ref{fact:axiomes-etoiles}, there exists $l_0 \in \contrlinks{\PortsofPPLPS{\Phi}} \cap \terminallinks{\Phi}$ and $b \in \setfmulti{D'}$ such that $e(\mappriports{\PortsofPPLPS{\Phi}}(l_0)) = (-, \dig_1^k (b))$ and ${(\dig_1^k (b))}^\ast \not= \multi{}$. As $e'$ is atomic, there exists $l_0' \in \contrlinks{\PortsofPPLPS{\Phi'}} \cap \terminallinks{\Phi}$ such that $\mappriports{\PortsofPPLPS{\Phi'}}(l_0') = \textsf{ind'}^{-1}(i_0)$. We have $\Phi\not\in\contrPLPS$, so that by Fact~\ref{fact : arity and paxnumber of contractions}, $k$ divides $\textit{Card}(a)$ for every $a\in \setfmulti{D'}$ such that $r(i) = (-, a)$ (where $i\in\integer{n}$). Still by Fact~\ref{fact : arity and paxnumber of contractions}, we obtain that $\Phi'\not\in\contrPLPS$. From $\rho : r \to r'$, we can deduce (using again Remark~\ref{rem:toutestdig}) that $\rho : \dig_1^k (b) \to \dig_1^k (b')$ in $\groupoidM$, hence, by Fact~\ref{fact : enlever une couche aux unites : semantique}, we get $\rho : $ $b^\ast + (\dig_1^k(b))^\textit{At}\to {b'}^\ast + \dig_1^k({b'})^\textit{At}$ in $\groupoidM$  and thus $\rho:\enleverunecoucheunitsweakenings{r}{l_0}\to\enleverunecoucheunitsweakenings{r'}{l'_0}$, where $\enleverunecoucheunitsweakenings{r}{l_0}$ and $\enleverunecoucheunitsweakenings{r'}{l'_0}$ have been defined in Fact~\ref{fact : contrunit : syntaxe}. By this fact and by Fact~\ref{fact : Phi LPS => Phi[l] LPS}, we can apply the induction hypothesis and deduce that $(\enleverunecoucheunitsweakenings{e}{l_0}, \enleverunecoucheunitsweakenings{r}{l_0})\simeq_\textit{At} (\enleverunecoucheunitsweakenings{e'}{l'_0}, \enleverunecoucheunitsweakenings{r'}{l'_0})$, and by Fact~\ref{fact : contrunit : e[l]=e'[l] => e=e'} we conclude $(e, r) \simeq_{\textit{At}} (e', r')$.
\item In the case where $\Phi \in \cboxPLPS$, for every $i\in\integer{n}$ we have that $r(i)=(\iota_i,b_i)$ for some $b_i\in\setfmulti{D'}$ and, by Fact~\ref{fact : arity and paxnumber of contractions}, $k$ divides $\textit{Card}(b_i)$. From $r\simeq r'$, we deduce that $r'(i)=(\iota_i,b'_i)$ for some $b'_i\in\setfmulti{D'}$ with $\textit{Card}(b_i)=\textit{Card}(b'_i)$. Since $e'$ is atomic, by applying again Fact~\ref{fact : arity and paxnumber of contractions}, we can conclude that $\Phi' \in \cboxPLPS$.
We can thus now apply Fact~\ref{fact : enleverunecouche a un resultat} twice:
\begin{enumerate}
\item
 there exists a unique atomic and injective $k$-experiment $(\enleverunecouche{e}, \enleverunecouche{r})$ of $\enleverunecouche{(\Phi, \textsf{ind})} = (\enleverunecouche{\Phi}, \enleverunecouche{\textsf{ind}})\in\LPSind$ such that
\begin{minilist}
\item for any $p \in (\ports{\PortsofPPLPS{\Phi}} \setminus \conclusions{\Phi}) \cap \ports{\PortsofPPLPS{\enleverunecouche{\Phi}}}$, we have $\enleverunecouche{e}(p) = e(p)$;
\item if $r(i) = (+, \dig_1^k(\multi{\alpha_{i}}))$ for some $\alpha_{i} \in D'$, then $\enleverunecouche{r}(i) = \alpha_{i}$ and if $r(i) = (-, \dig_1^k(c_{i}))$ then $\enleverunecouche{r}(i) = (-, c_{i})$.
\end{minilist}
\item
there exists a unique atomic and injective $k$-experiment $(\enleverunecouche{e'}, \enleverunecouche{r'})$ of $\enleverunecouche{(\Phi', \textsf{ind'})} = (\enleverunecouche{\Phi'}, \enleverunecouche{\textsf{ind'}})\in\LPSind$ such that
\begin{minilist}
\item for any $p \in (\ports{\PortsofPPLPS{\Phi'}} \setminus \conclusions{\Phi'}) \cap \ports{\PortsofPPLPS{\enleverunecouche{\Phi'}}}$, we have $\enleverunecouche{e'}(p) = e'(p)$;
\item if $r'(i) = (+, \dig_1^k(\multi{\alpha'_{i}}))$ for some $\alpha'_{i} \in D'$, then $\enleverunecouche{r'}(i) = \alpha'_{i}$ and if $r'(i) = (-, \dig_1^k(c'_{i}))$ then $\enleverunecouche{r'}(i) = (-, c'_{i})$.
\end{minilist}
\end{enumerate}
If we set $b_i=c_i$ (resp.\ $b_{i}=\multi{\alpha_{i}}$) if $\enleverunecouche{r}(i) =c_{i}$ (resp.\ $\enleverunecouche{r}(i) =\alpha_{i}$), and 
$b'_i=c'_i$ (resp.\ $b'_{i}=\multi{\alpha'_{i}}$) if $\enleverunecouche{r'}(i) =c'_{i}$ (resp.\ $\enleverunecouche{r'}(i) =\alpha'_{i}$), then  $r\simeq r'$ is equivalent to $(\dig_1^k(b_1),\ldots,\dig_1^k(b_n)) \simeq (\dig_1^k(b'_1),\ldots,\dig_1^k(b'_n))$. By Lemma~\ref{lemma : Key-Proposition : enleverunecouche a r} we can then conclude that $(b_1, \ldots, b_n) \simeq (b'_1, \ldots, b'_n)$, which immediately yields $\enleverunecouche{r}\simeq\enleverunecouche{r'}$. Since $\mes{\PortsofPPLPS{\enleverunecouche{\Phi}}} < \mes{\PortsofPPLPS{\Phi}}$, by induction hypothesis we deduce that $(\enleverunecouche{e},\enleverunecouche{\Phi})\simeq (\enleverunecouche{e'},\enleverunecouche{\Phi'})$, and by Fact~\ref{fact : cbox : enleverunecouche de e = enleverunecouche de e' => e = e'} we have $(e,r)\simeq (e',r')$.
\item the other cases are easier and left to the reader.
\end{minilist}
\end{proof}

\begin{remark}
A crucial point in the case $\Phi \in \contrPLPS$ of the proof is that we have $\rho \cdot \beta \simeq \beta'$, but we do not necessarily have $\rho \cdot \beta = \beta'$ and this corresponds to the fact that, as illustrated in the introduction by an example using the PS of Figure \ref{fig:NoUniqueExp}, there are different atomic $k$-experiments of PS\footnote{See Footnote~\ref{footnote : k-experiment of PS}.} having the same injective result. 
Consider again this figure and let $\Phi$ be the LPS of this PS. 
Let $e = e'$ be a $3$-experiment of $\Phi$ such that $e(p_z) = (-, \lambda_z, \lambda_z)$ with $\lambda_z \in A$ and $z\in\integer{2}$. 
We have $e(c_1) = (-, a)$ with $a = \multi{(-, \lambda_1, {\lambda_1})} + \sum_{j=1}^3 \multi{(-, (\lambda_2, j), (\lambda_2, j))}$. Let $r=r'$ be the result of $e=e'$. 
We have $\textit{Q}(r, a) = \{ a \} $, hence we can consider, for example, $\rho : (-, a) \rightarrow (-, a)$ in $\groupoidsD$ such that $\rho(\lambda_1) = (\lambda_2, 1)$. We have $\beta = (-, \lambda_1, {\lambda_1}) = \beta' \not= \rho \cdot \beta$.
\end{remark}

\bibliographystyle{unsrt}
\bibliography{ll}

\newpage

\section*{Technical appendix}

\section{Syntax}

\subsection{Pre-Linear Proof-Structures (PLPS)}

\begin{defin}\label{definition : <=}
For any $\Phi \in \PPLPS$, we define the binary relation $\below{\Phi}$ on $\ports{\PortsofPPLPS{\Phi}}$ as follows: $p \below{\Phi} p'$ if, and only if, one of the following conditions holds:
\begin{minilist}
\item 
there exists a cell $l$ of $\Phi$  such that $p$ (resp.\ $p'$) is the principal (resp.\ an auxiliary) port of $l$
\item
$p'$ (resp.\ $p$) is the principal (resp.\ an auxiliary) port of some cell $l'$ (resp.\ $l$) of $\Phi$ and $\{p,p'\}$ is a wire of $\Phi$.
\end{minilist}
The binary relation $\leq_\Phi$ (or simply $\leq$) on $\ports{\PortsofPPLPS{\Phi}}$ is the transitive reflexive closure of $\below{\Phi}$.
\end{defin}

We introduce a weaker notion than the one of PPLPS: \emph{$\omega$PPLPS}. An $\omega$PPLPS\footnote{$\omega$ is reminiscent of the definition of \emph{$\omega$-reduction} in \cite{mazzaevent}} is a PPLPS, except that Condition \ref{item:3} of Definition \ref{definition : PPLPS} is not required.

\begin{defin}
Let $\omega\PPLPS$ be the set of pairs $\Phi=(\mathbb{P}, \mathcal{W})$ such that
\begin{minilist}
\item $\mathbb{P} \in \Ports$; the ports of $\Phi$ are the ports of $\mathbb{P}$ 
\item $\mathcal{W} \subseteq \setofpairs{\ports{\mathbb{P}}}$ such that
\begin{enumerate}
\item for any $w, w' \in \mathcal{W}$, we have $(w \cap w' \not= \emptyset \Rightarrow w=w')$;
\item for any $p \in \ports{\mathbb{P}} \setminus \priports{\mathbb{P}}$, there exists $q \in \ports{\mathbb{P}}$ such that $\{ p, q \} \in \mathcal{W}$;
\item for any $w \in \mathcal{W}$, there exists $p \in w$ such that $p \notin \priports{\mathbb{P}}$.
\end{enumerate}
\end{minilist}
We set $\PortsofPPLPS{\Phi} = \mathbb{P}$ and $\edges{\Phi} = \mathcal{W}$.
\end{defin}

With every $\omega$PPLPS $\Phi$, we associate a unique PPLPS $ \omega(\Phi)$:

\begin{defin}
Let $\omega$ be the function $\omega\PPLPS \rightarrow \PPLPS$ defined as follows: $\omega(\Phi) = \Phi'$ is defined as follows:
\begin{itemize}
\item $\LinksofPorts{\PortsofPPLPS{\Phi'}} = \LinksofPorts{\PortsofPPLPS{\Phi}}$;
\item $\ports{\PortsofPPLPS{\Phi'}} = \ports{\PortsofPPLPS{\Phi}} \setminus \{ p \in \conclusions{\Phi} \: / \: (\exists q \in \priports{\PortsofPPLPS{\Phi}}) \: \{ p , q \} \in \edges{\Phi} \}$;
\item $\portsoflink{\PortsofPPLPS{\Phi'}} = \portsoflink{\PortsofPPLPS{\Phi}}$, $\mappriports{\PortsofPPLPS{\Phi'}} = \mappriports{\PortsofPPLPS{\Phi}}$, $\mapleftports{\PortsofPPLPS{\Phi'}} = \mapleftports{\PortsofPPLPS{\Phi}}$ and $\paxnumber{\PortsofPPLPS{\Phi'}} = \paxnumber{\PortsofPPLPS{\Phi}}$;
\item $\edges{\Phi'} = \{ w \in \edges{\Phi} \: / \: w \subseteq \ports{\PortsofPPLPS{\Phi'}} \}$.
\end{itemize}
\end{defin}

We give here the formal definition of \emph{the PLPS $\Psi$ obtained from $\Phi$ by removing $\mathcal{C}_0$}, where $\mathcal{C}_0 \subseteq \terminallinks{\Phi}$ is such that $(\mathcal{C}_0 = \{ l \} \textrm{ and }l \in \multlinks{\PortsofPPLPS{\Phi}} \cup \derelictions{\PortsofPPLPS{\Phi}})$ or $\mathcal{C}_0 \subseteq \bangs{\PortsofPPLPS{\Phi}}$:

\begin{defin}\label{def:eliminate-l}
Let $\Phi\in\PLPS$ and let $\mathcal{C}_0 \subseteq \terminallinks{\Phi}$ such that $(\mathcal{C}_0 = \{ l \} \textrm{ and }l \in \multlinks{\PortsofPPLPS{\Phi}} \cup \derelictions{\PortsofPPLPS{\Phi}})$ or $\mathcal{C}_0 \subseteq \bangs{\PortsofPPLPS{\Phi}}$. 
\emph{The PLPS $\Psi$ obtained from $\Phi$ by removing $\mathcal{C}_0$} is $\omega(\Phi')$, where $\Phi'$ is the $\omega$PPLPS defined as follows:
\begin{itemize}
\item $\links{\PortsofPPLPS{\Phi'}}=\links{\PortsofPPLPS{\Phi}}\setminus \mathcal{C}_0$;
\item $\ports{\PortsofPPLPS{\Phi'}}=\ports{\PortsofPPLPS{\Phi}}\setminus \bigcup_{l \in \mathcal{C}_0} \{\textsf{P}^\textsf{pri}(l)\}$;
\item $\portsoflink{\PortsofPPLPS{\Phi'}}$ (resp.\ $\mappriports{\PortsofPPLPS{\Phi'}}$, $\mapleftports{\PortsofPPLPS{\Phi'}}$) is the restriction of $\portsoflink{\PortsofPPLPS{\Phi}}$ (resp.\ $\mappriports{\PortsofPPLPS{\Phi}}$, $\mapleftports{\PortsofPPLPS{\Phi}}$) to $\ports{\PortsofPPLPS{\Phi'}}$;
\item $\paxnumber{\PortsofPPLPS{\Phi'}}=\paxnumber{\PortsofPPLPS{\Phi}}$;
\item $\edges{\Phi'}= \{ w \in \edges{\Phi} \: / \: w \subseteq \ports{\PortsofPPLPS{\Phi'}} \}$.
\end{itemize}
\end{defin}

\subsection{Linear Proof-Structures (LPS)}


We give the formal definition of $\enleverunecouche{\Phi}$ for $\Phi \in \cboxPLPS \cap \LPS$:

\begin{defin}\label{definition : enleverunecouche}
With $\Phi\in\cboxPLPS\cap\LPS$ one can associate the PLPS $\Phi_{-1}$ obtained from $\Phi$ by modifying the function $\#$ (all the rest is unchanged): $\contrlinks{\mathbb{P}(\Phi_{-1})}\cap\terminallinks{\Phi_{-1}} = \contractionspax{\mathbb{P}(\Phi)}\cap\terminallinks{\Phi}$ and for every cell $l \in \contractionspax{\mathbb{P}(\Phi)}\cap\terminallinks{\Phi}$, the auxiliary ports of $l$ in $\Phi$ are exactly those of $l$ in $\Phi_{-1}$; we can thus set $\paxnumber{\PortsofPPLPS{\Phi_{-1}}}(p)=\paxnumber{\PortsofPPLPS{\Phi}}(p)-1$ for such an auxiliary port $p$\footnote{We use here the crucial hypothesis that $l\in\contractionspax{\mathbb{P}}$ which means that $\paxnumber{\PortsofPPLPS{\Phi}}(p)>0$.}. For every $l\in\contrlinks{\mathbb{P}(\Phi_{-1})}\setminus(\contractionspax{\mathbb{P}(\Phi_{-1})}\cap\terminallinks{\Phi_{-1}})$ and for every auxiliary port $p$ of $l$, we set $\paxnumber{\PortsofPPLPS{\Phi_{-1}}}(p)=\paxnumber{\PortsofPPLPS{\Phi}}(p)$.

The PLPS $\overline{\Phi}$ is then obtained from $\Phi_{-1}$ by removing $\bangs{\mathbb{P}(\Phi_{-1})}\cap\terminallinks{\Phi_{-1}}$\footnote{following Definition~\ref{def:eliminate-l}}. 
\end{defin}

\subsection{Proof-Structures (PS)}

In the same way that we introduced indexed PPLPS, indexed PLPS, indexed LPS and indexed PS, we introduce the notion of indexed $\omega$PPLPS. Now, to every $(\Phi, \textsf{ind}) \in \omegaPPLPSind$, we associate the indexed PPLPS $\omega(\Phi) = (\omega(\Phi), \textsf{ind}_1)$ defined as follows: $\textsf{ind}_1(p) = \textsf{ind}(\conclusionunder{\Phi}{p})$.

\section{Experiments}

\begin{defin}\label{def:ortogonal}
We call \emph{depth of an element $\alpha \in D$} the least number $n\in\Nat$ such that $\alpha \in D_n$.\footnote{The definition of $D_n$ has been given in Definition \ref{definition : D}.} 

Let $+^\perp=-$ and $-^\perp=+$. We define $\alpha^\perp$ for any $\alpha \in D$, by induction on the depth of $\alpha$:
\begin{minilist}
\item for $\gamma \in A$, $\gamma^\perp = \gamma$ and for $\gamma=\ast$, $(\iota, \gamma)^\perp = (\iota^\perp, \gamma)$;
\item else, $(\iota, \alpha, \beta)^\perp = (\iota^\perp, \alpha^\perp, \beta^\perp)$, and $(\iota, [\alpha_1,\dots, \alpha_n])^\perp = (\iota^\perp, [\alpha_1^\perp,\dots, \alpha_n^\perp])$.
\end{minilist}
\end{defin}

\begin{defin}\label{definition : appears}
For any $\alpha \in D$, we define, by induction on the depth of $\alpha$, $\textit{Sub}(\alpha) \in \setfmulti{D}$ as follows:
\begin{minilist}
\item $\textit{Sub}(\delta) = \multi{\delta}$ if $\delta \in A\cup(\{ +, - \} \times\{ \ast \})$;
\item $\textit{Sub}(\iota, \alpha, \beta) = \multi{(\iota, \alpha, \beta)} + \textit{Sub}(\alpha) + \textit{Sub}(\beta)$;
\item $\textit{Sub}(\iota, \multi{\alpha_1, \ldots, \alpha_m}) = \multi{(\iota, \multi{\alpha_1, \ldots, \alpha_m})} + \sum_{j=1}^m \textit{Sub}(\alpha_j)$.
\end{minilist}
For any $(\alpha_1, \ldots, \alpha_n) \in D^{< \omega}$, we set $\textit{Sub}(\alpha_1, \ldots, \alpha_n) = \sum_{i=1}^n \textit{Sub}(\alpha_i)$.


For any $\beta \in D$, for any $r \in D^{< \omega}$, we say that \emph{$\beta$ occurs in $r$} if $\beta \in \textit{Supp}(\textit{Sub}(r))$.

For any $\gamma \in A$, for any $r \in D^{< \omega}$, for any $m \in \mathbb{N}$, we say that \emph{there are exactly $m$ occurrences of $\gamma$ in $r$} if $\textit{Sub}(r)(\gamma) = m$.
\end{defin}

The following precise definition of substitution clearly entails that for every $\alpha\in D$ and for every substitution $\sigma: D\rightarrow D$, one has $\sigma(\alpha^\perp)=\sigma(\alpha)^\perp$:

\begin{defin}\label{def:substitution}
A \emph{substitution} is a function $\sigma: D\rightarrow D$ induced by a function $\sigma^{A}:A\rightarrow D$ and defined by induction on the depth of elements of $D$, as follows (as usual $\iota \in \{ +, - \}$ and $\gamma \in A$):
\begin{minilist}
\item $\sigma(\gamma) = \sigma^{A}(\gamma)$ and $\sigma(\iota,\ast) = (\iota,\ast)$;
\item $\sigma(\iota, \alpha, \beta) \mathrel{:=} (\iota,\sigma(\alpha),\sigma(\beta))$
\item $\sigma(\iota, [\alpha_1,\dots, \alpha_n]) = (\iota, [\sigma(\alpha_1),\dots,\sigma(\alpha_n)])$.
\end{minilist}
\end{defin}

\section{Proof of Proposition \ref{prop : KeyProposition}}

\subsection{The case of $\axPLPS$}

We give here the formal definition of $(\Phi_1, \textsf{ind}_1)$ and $(\Phi'_1, \textsf{ind'}_1)$ of the proof of Proposition~\ref{prop : KeyProposition} (case $\Phi \in \axPLPS$).

We set $m_0 = \min \{ i_0, j_0 \}$ and $M_0 = \max \{ i_0, j_0 \}$. We define $(\Phi_1, \textsf{ind}_1), (\Phi'_1, \textsf{ind'}_1) \in \PLPSind$ as follows:
\begin{minilist}
\item $\LinksofPorts{\PortsofPPLPS{\Phi_1}} = \LinksofPorts{\PortsofPPLPS{\Phi}}$ and $\LinksofPorts{\PortsofPPLPS{\Phi'_1}} = \LinksofPorts{\PortsofPPLPS{\Phi'_1}}$;
\item $\ports{\PortsofPPLPS{\Phi_1}} = \ports{\PortsofPPLPS{\Phi}} \setminus \{ p_0, q_0 \}$ and $\ports{\PortsofPPLPS{\Phi'_1}} = \ports{\PortsofPPLPS{\Phi'}} \setminus \{ p_0', q_0' \}$;
\item $\portsoflink{\PortsofPPLPS{\Phi_1}} = \portsoflink{\PortsofPPLPS{\Phi}}$, $\mappriports{\PortsofPPLPS{\Phi_1}} = \mappriports{\PortsofPPLPS{\Phi}}$, $\mapleftports{\PortsofPPLPS{\Phi_1}} = \mapleftports{\PortsofPPLPS{\Phi}}$, $\paxnumber{\PortsofPPLPS{\Phi_1}} = \paxnumber{\PortsofPPLPS{\Phi}}$ and $\portsoflink{\PortsofPPLPS{\Phi_1'}} = \portsoflink{\PortsofPPLPS{\Phi'}}$, $\mappriports{\PortsofPPLPS{\Phi_1'}} = \mappriports{\PortsofPPLPS{\Phi'}}$, $\mapleftports{\PortsofPPLPS{\Phi_1'}} = \mapleftports{\PortsofPPLPS{\Phi'}}$, $\paxnumber{\PortsofPPLPS{\Phi_1'}} = \paxnumber{\PortsofPPLPS{\Phi'}}$;
\item $\edges{\Phi_1} = \edges{\Phi} \setminus \{ p_0, q_0 \}$ and $\edges{\Phi'_1} = \edges{\Phi'} \setminus \{ p_0', q_0' \}$;
\item we define the value of $\textsf{ind}_1(p)$ as follows: $$\left\lbrace \begin{array}{l} 
\textsf{ind}(p) \textrm{ if $\textsf{ind}(p) < m_0$;}\\ 
\textsf{ind}(p) - 1 \textrm{ if $m_0 < \textsf{ind}(p) < M_0$;}\\
\textsf{ind}(p) - 2 \textrm{ if $M_0 < \textsf{ind}(p)$;} \end{array} \right.$$ and the value of $\textsf{ind'}_1(p)$ as follows:
$$\left\lbrace \begin{array}{l} 
\textsf{ind'}(p) \textrm{ if $\textsf{ind'}(p) < m_0$;}\\ 
\textsf{ind'}(p) - 1 \textrm{ if $m_0 < \textsf{ind'}(p) < M_0$;}\\
\textsf{ind'}(p) - 2 \textrm{ if $M_0 < \textsf{ind'}(p)$.} \end{array} \right.$$
\end{minilist}

\subsection{The case of $\contrPLPS$}

We give here the definition of $(\Phi_1, \textsf{ind}_1), (\Phi'_1, \textsf{ind'}_1) \in \PLPSind$  of the proof of Proposition~\ref{prop : KeyProposition} (case: $\Phi \in \contrPLPS$): $(\Phi_1, \textsf{ind}_1) = \omega(\Psi_1, \textsf{ind}_2)$ and $(\Phi'_1, \textsf{ind'}_1) = \omega(\Psi'_1, \textsf{ind'}_2)$, where $(\Psi_1, \textsf{ind}_2), (\Psi'_1, \textsf{ind'}_2) \in \omegaPPLPSind$ are defined as follows:
\begin{minilist}
\item $\links{\PortsofPPLPS{\Psi_1}} = \links{\PortsofPPLPS{\Phi}}$ and $\links{\PortsofPPLPS{\Psi'_1}} = \links{\PortsofPPLPS{\Phi'}}$;
\item $\type{\PortsofPPLPS{\Psi_1}} = \type{\PortsofPPLPS{\Phi}}$ and $\type{\PortsofPPLPS{\Psi'_1}} = \type{\PortsofPPLPS{\Phi'}}$;
\item $\arity{\PortsofPPLPS{\Psi_1}}(l) = \left\lbrace \begin{array}{ll} \arity{\PortsofPPLPS{\Phi}}(l) & \textrm{if $l \not= l_0$;} \\ \arity{\PortsofPPLPS{\Phi}}(l_0-1) & \textrm{if $l = l_0$;} \end{array} \right.$ and $\arity{\PortsofPPLPS{\Psi'_1}}(l') = \left\lbrace \begin{array}{ll} \arity{\PortsofPPLPS{\Phi'}}(l_0') & \textrm{if $l' \not= l_0'$;} \\ \arity{\PortsofPPLPS{\Phi'}}(l_0'-1) & \textrm{if $l' = l_0'$;} \end{array} \right.$
\item $\ports{\PortsofPPLPS{\Psi_1}} = \ports{\PortsofPPLPS{\Phi}}$ and $\ports{\PortsofPPLPS{\Psi'_1}} = \ports{\PortsofPPLPS{\Phi'}}$; 
\item $\portsoflink{\PortsofPPLPS{\Psi_1}}(l) = \left\lbrace \begin{array}{ll} \portsoflink{\PortsofPPLPS{\Phi}}(l) & \textrm{if $l \not= l_0$;} \\ \portsoflink{\PortsofPPLPS{\Phi}}(l_0) \setminus \{ p \} & \textrm{if $l = l_0$;} \end{array} \right.$ and\\
$\portsoflink{\PortsofPPLPS{\Psi'_1}}(l') = \left\lbrace \begin{array}{ll} \portsoflink{\PortsofPPLPS{\Phi'}}(l') & \textrm{if $l' \not= l_0'$;} \\ \portsoflink{\PortsofPPLPS{\Phi'}}(l_0') \setminus \{ p' \} & \textrm{if $l' = l_0'$;} \end{array} \right.$
\item $\mappriports{\PortsofPPLPS{\Psi_1}} = \mappriports{\PortsofPPLPS{\Phi}}$ and $\mappriports{\PortsofPPLPS{\Psi'_1}} = \mappriports{\PortsofPPLPS{\Phi'}}$;
\item $\mapleftports{\PortsofPPLPS{\Psi_1}} = \mapleftports{\PortsofPPLPS{\Phi}}$ and $\mapleftports{\PortsofPPLPS{\Psi'_1}} = \mapleftports{\PortsofPPLPS{\Phi'}}$;

\item $ \paxnumber{\PortsofPPLPS{\Psi_1}} = \restriction{ \paxnumber{\PortsofPPLPS{\Phi}}}{\bigcup_{l \in \contrlinks{\PortsofPPLPS{\Phi}}} \mapauxports{\PortsofPPLPS{\Phi}}(l) \setminus \{ p \}}$ and $ \paxnumber{\PortsofPPLPS{\Psi'_1}} = \restriction{ \paxnumber{\PortsofPPLPS{\Phi'}}}{\bigcup_{l \in \contrlinks{\PortsofPPLPS{\Phi'}}} \mapauxports{\PortsofPPLPS{\Phi'}}(l) \setminus \{ p' \}}$;

\item $\edges{\Psi_1} = \edges{\Phi}$ and $\edges{\Psi'_1} = \edges{\Phi'}$;
\item $\textsf{ind}_2(q) = \left\lbrace \begin{array}{ll} \textsf{ind}(q) & \textrm{if $q \not= p$;} \\ \textit{Card}(\conclusions{\Phi})+1 & \textrm{if $q = p$;} \end{array} \right.$ and $\textsf{ind'}_2(q) = \left\lbrace \begin{array}{ll} \textsf{ind'}(q) & \textrm{if $q \not= p'$;} \\ \textit{Card}(\conclusions{\Phi'})+1 & \textrm{if $q = p'$;} \end{array} \right.$
\end{minilist}

\end{document}